\documentclass{vki}

\usepackage{lmodern,pdftexcmds,amsmath}

\usepackage{amsfonts,amsthm,bm} 

\DeclareMathAlphabet{\mathsfbi}{OT1}{\sfdefault}{bx}{sl}
\DeclareMathVersion{sfletters}
\SetSymbolFont{letters}{sfletters}{OML}{ntxsfmi}{b}{it}

\makeatletter
\newcommand{\mathbfsbilow}[1]{%
	\text{\mathversion{sfletters}$\m@th#1$}%
}

\DeclareRobustCommand{\tensor}[1]{%
	\begingroup
	\ifcat\noexpand #1\relax
	\edef\greek@test{\detokenize{#1}}%
	\edef\greek@test{\expandafter\@cdr\greek@test\@nil}%
	\edef\greek@test{\expandafter\@car\greek@test\@nil}%
	\edef\x{\the\lccode\expandafter`\greek@test}%
	\edef\y{\number\expandafter`\greek@test}%
	\ifnum\x=\y\relax
	\mathbfsbilow{#1}%
	\else
	\mathsfbi{#1}%
	\fi
	\else
	\mathsfbi{#1}%
	\fi
	\endgroup
}
\makeatother

\usepackage[most]{tcolorbox}
\usepackage{booktabs}
\usepackage{amsmath}
\usepackage{wrapfig}
\usepackage[linesnumbered,ruled,vlined]{algorithm2e}
\usepackage{algpseudocode}
\usepackage{pdflscape}
\usepackage{subcaption}

\usepackage{siunitx}
\usepackage{comment}

\usepackage{mathtools}
\tcbset{enhanced,colback=cyan!2!white,colframe=cyan!90!white,coltitle=blue!20!black,fonttitle=\bfseries}

\usepackage{listings}
\usepackage{hyperref}
\hypersetup{
	colorlinks=true,
	linkcolor=blue,
	filecolor=magenta,      
	citecolor=blue,
	urlcolor=cyan,
}

\begin{document}


\pagenumbering{roman}
\title{Statistical Methods and Modal Decompositions for Gridded and Scattered Data: \\Meshless Statistics and Meshless Data Driven Modal Analysis}
\author{Miguel A. Mendez\thanks{mendez@vki.ac.be}, Manuel Ratz, Damien Rigutto\\von Karman Institute for Fluid Dynamics}

\date{3 December 2024} 
\maketitle

Statistical tools are crucial for studying and modeling turbulent flows, where chaotic velocity fluctuations span a wide range of spatial and temporal scales. Advances in image velocimetry, especially in tracking-based methods, now allow for high-speed, high-density particle image processing, enabling the collection of detailed 3D flow fields.

This lecture provides a set of tutorials on processing such datasets to extract essential quantities like averages, second-order moments (turbulent stresses) and coherent patterns using modal decompositions such as the Proper Orthogonal Decomposition (POD).

After a brief review of the fundamentals of statistical and modal analysis, the lecture delves into the challenges of processing scattered data from tracking velocimetry, comparing it to traditional gridded-data approaches. It also covers research topics, including physics-based Radial Basis Function (RBF) regression for meshless computation of turbulent statistics and the definition of an RBF inner product, which enables meshless computation of data-driven decompositions. These include traditional methods like Proper Orthogonal Decomposition (POD) and Dynamic Mode Decomposition (DMD), as well as advanced variants such as Spectral POD (SPOD) and Multiscale POD (mPOD). We refer to this framework as "Meshless Data Driven Decomposition".

Six exercises in Python are provided. 
All codes are available at \href{https://github.com/mendezVKI/PIV_LS_2024_Signal}{this github repository.}\\

NOTE 1: This lecture is an updated version of the lecture ``Statistical Treatment, Fourier, and Modal Decomposition", which was given in the previous edition of the lecture series and can be found in \url{https://arxiv.org/abs/2201.03847}. Here the session on traditional modal decompositions has been shortened to give more space to the idea of meshless computation of statistics and data-driven decompositions.

NOTE 2: These lecture notes are now being expanted into a book, covering statistics, traditional (gridded) modal analysis, constrained RBF and meshless modal analysis. Stay tuned for more :) !
 
\vspace{20mm}

\pagenumbering{arabic}
\setcounter{page}{1}
\clearpage{\pagestyle{empty}} 

\tableofcontents
\clearpage{\pagestyle{empty}} 

\vspace{-3mm}

\section*{A note on notation and style}\label{sec_0}

\textbf{Vectors, Matrices and lists}. We use lowercase letters for scalar quantities, i.e. $a\in\mathbb{R}$. Bold lowercase letters are used for vectors, i.e. $\bm{a}\in\mathbb{R}^{n_a}$. 
The i-th entry of a vector is denoted with a subscript as $\bm{x}_i$ or with Python-like notation as $\bm{x}[i]$. Unless otherwise stated, a vector is a column vector. Thus, we omit the use of transposition when defining a vector embedded within the text of a paragraph or sentence (inline). 
\\We use upper case bold letters for matrices, e.g. $\bm{A}\in {\mathbb{R}^{n_r\times n_c}}$, with $n_r$ the number of rows and $n_c$ the number of columns. The matrix entry at the i-th row and j-th column are identified as $\bm{A}_{i,j}$ or with the Python-like notation as $\bm{A}[i,j]$. When Python notation is used, the indices begin with $0$. We use square brackets to create vectors from a set of scalars, e.g $\mathbf{a}=[a_0,a_1,\dots a_{n_a-1}]\in\mathbb{R}^{n_a}$ or matrices from a set of vectors, e.g. $\mathbf{U}=[\bm{u}_0,\bm{u}_1\dots u_{n_t-1}]\in\mathbb{R}^{n_u\times n_t}$. Note the difference between bold bold italicized $\bm{a}$ or $\bm{U}$ and upright bold $\mathbf{a}$ or $\mathbf{U}$. The first is used for quantities that are vector-valued or matrices by their nature (e.g. velocity fields or the Reynolds stresses), that is, even in their continuous form. The second is used for sampled quantities. Hence, a collection of PTV snapshots is denoted as $\bm{U}$, while the Reynolds stress tensor is denoted as $\mathbf{R}$.

\textbf{Sampling} The sampling of a continuous function is stored in a vector or a matrix. We assume uniform sampling in both space and time. For vector $\mathbf{d}(t)$ sampled the time domain $t$, considering a discretization $\mathbf{t_k}=k\Delta t$, with $f_s=1/\Delta t$ the sampling frequency and $k=[0,1,\dots n_t-1]$, we could write $\mathbf{d}[k]$ or $\mathbf{d}_k$ or $\mathbf{d}(\mathbf{t_k})$. The same is true for the space domain, although a matrix linear index must be introduced. This is important when we transform a matrix (for example, a spatial realization of a quantity) into a vector.

For example, let $p[i,j]$ be the 2D discretization of a pressure field $p(x,y)$, where the axes were discretized as $i\in[0,n_x-1]$ and $j\in[0,n_y-1]$. For this lecture, that field would be written as a single "snapshot" vector $\bm{p}\in\mathbb{R}^{n_s}$, with $n_s=n_x n_y$. The entries in this vector would be accessed with a linear matrix index, denoted in bold, i.e. $\bm{p}[\bm{i}]$. The way this accesses the data in the matrix depends on whether the flattening is performed column-wise or row-wise. For example, for a matrix $\bm{A}\in\mathbb{R}^{3\times 3}$, the column-wise and row-wise matrix indices are\footnote{Recall that here use Python-like indexing. Hence, the first entry is 0 and not 1}

$$ \mbox{column-wise }\bm{i}:\,\,\bm{A}=\begin{bmatrix}0 & 3 &6 \\1 & 4 &7 \\2 & 5 &8  \end{bmatrix} \quad \mbox{row-wise }\bm{i}:\,\,\bm{A}=\begin{bmatrix}0 & 1 &2 \\3 & 4 &5 \\6 & 7 &8  \end{bmatrix}\,,.$$

\clearpage

\section{Introduction and lecture format}\label{sec_1}

Statistical tools for describing space-time correlations, measuring turbulent stress, and identifying coherent patterns are essential in experimental fluid mechanics \cite{Pope2000,Tennekes1972,Davidson2004}. In the past decade, these tools have advanced to match rapid developments in image velocimetry and increasing spatial and temporal resolution.

A significant advancement has been the adoption of 3D tracking velocimetry over traditional correlation-based methods \citep{Schanz2016,Tan2020}, which produce scattered data and introduce new challenges for post-processing, such as calculating mean fields and Reynolds stresses. The current standard approach involves interpolating scattered data onto a uniform grid to facilitate traditional analyses. For instantaneous fields—such as those required for computing derivatives or pressure reconstruction—methods like Vic+, Vic\#, FlowFit \citep{Schneiders2016, Scarano2022, Schanz2016}, and Meshless Track Assimilation \citep{Sperotto2024b} incorporate physics-based constraints (e.g., divergence-free conditions) to ensure robust interpolation. For statistical computations, such as mean fields and Reynolds stresses, techniques like binning and ensemble PTV (EPTV, \citet{Discetti2018}) are widely used, dividing the domain into bins where local statistics are calculated \citep{Kaehler2012a}. With dense datasets, these methods often outperform cross-correlation techniques in computing Reynolds stresses \citep{Atkinson2014, Schroeder2018}.

This lecture provides an overview of statistical methods and modal decomposition techniques, covering traditional approaches for gridded data and new methods for scattered data. Although relevant literature is referenced as accurately as possible, this chapter is not intended as a comprehensive review of image velocimetry post-processing. Instead, it serves as a hands-on tutorial to guide the practical computation of key quantities, their physical significance, and their interpretation. Accordingly, the lecture is structured around a set of exercises in \textsc{Python}. Readers interested in learning \textsc{Python} for scientific computing are referred to \cite{SveinLinge2019,Langtangen2016,Johansson2018,PaulJ.Deitel2019}.

The lecture focuses on the analysis of two selected datasets, described in Section \ref{sec_3}, along with instructions for downloading or data generation. Section \ref{sec_4} provides a concise review of fundamental statistical concepts, from first- and second-order moments to ergodicity, power spectral densities, and cross-coherency. Section \ref{sec_5} addresses the statistical treatment of turbulence, covering definitions of key quantities and analysis types (pointwise statistics vs. space/time analysis) and their computations. Section \ref{sec_6} shows how to compute these quantities for gridded or scattered data, introducing the concept of Physics-Constrained Radial Basis Function (RBF) regression and its application to the processing of scattered data \cite{Sperotto2022a,Sperotto2024}. Section \ref{sec_7} then focuses on modal decompositions, beginning with the continuous formulation of Proper Orthogonal Decomposition (POD) and proceeding to its computation for gridded data using matrix factorizations and for scattered data using physics-constrained RBF. Generalizations of this methodology to other popular decompositions, such as Dynamic Mode Decomposition (DMD,\cite{Schmid,Rowley2}) and Multiscale POD (mPOD, \citet{Mendez2019,Mendez2023}), are also presented.

\section{Collecting the datasets} \label{sec_2}

We consider two test cases in this lecture, described in the following subsections.

\subsection{Transient flow past a cylinder}\label{sec_2_1}

This datasets consist of time-Resolved PIV for the flow past a cylinder of $d=\SI{5}{\milli\meter}$ diameter and $L=\SI{20}{\centi\meter}$ length in transient conditions, with varying free stream velocity. The experiments were carried out in the L10 low-speed wind tunnel of the von Karman Institute. The details of the experiment are described in \cite{Mendez2020}. The dataset can be downloaded by the provided script Get\_Cylinder\_DATA\_PIV\_PTV.py.

\begin{figure}[htbp]
	\centering    
\begin{minipage}{0.49\textwidth}
 \includegraphics[keepaspectratio=true,width=1\columnwidth]
    {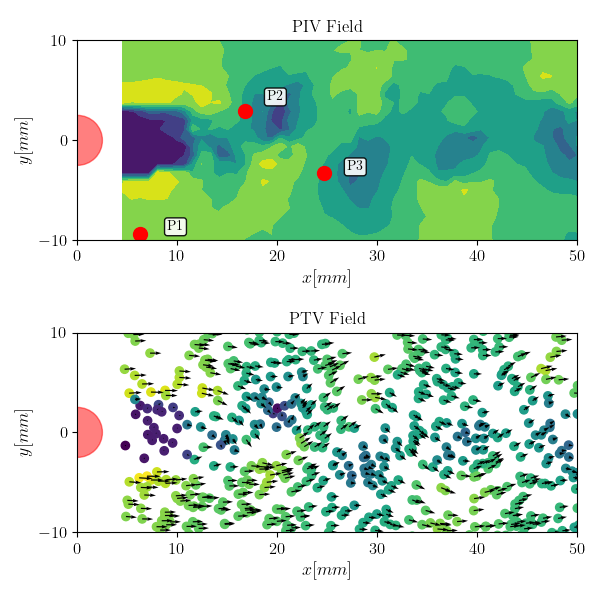}
\end{minipage}
\begin{minipage}{0.49\textwidth}         \includegraphics[keepaspectratio=true,width=1\columnwidth]{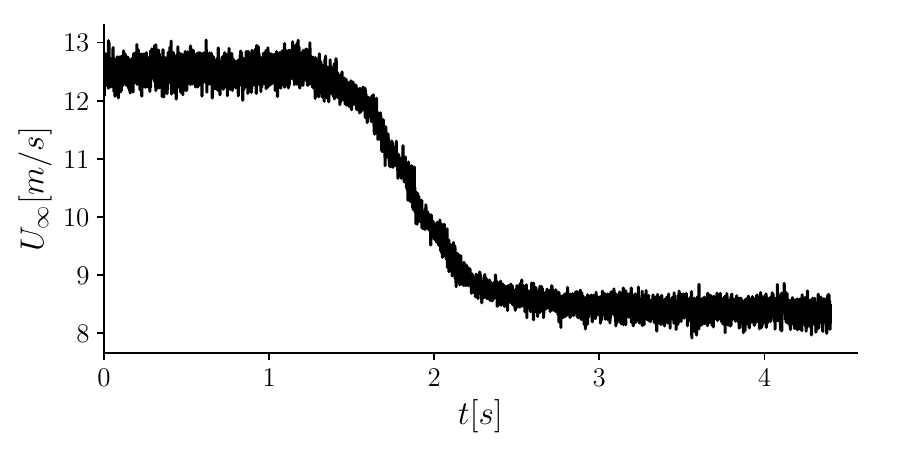}\\
\includegraphics[keepaspectratio=true,width=1\columnwidth]{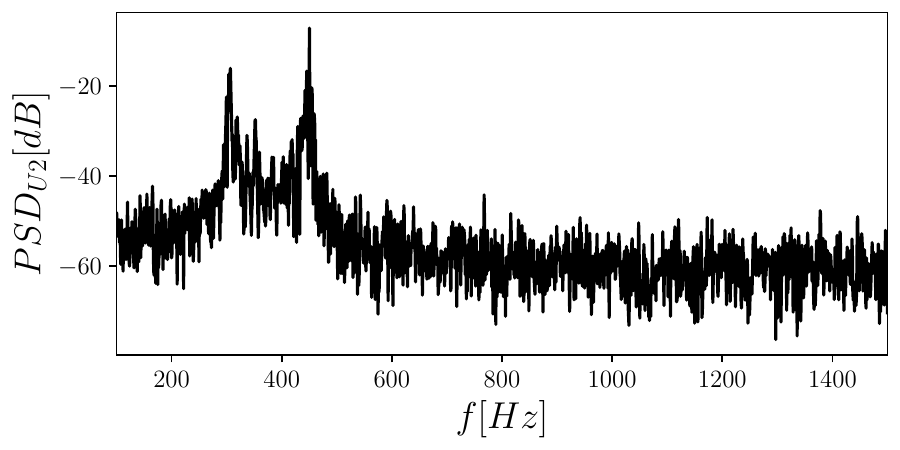}

\end{minipage}
	\caption{Right: snapshot of the velocity field from the TR-PIV measurements considered in this test case (top) and synthetic scattered data simulating a PTV velocimetry. Left: time evolution of the free stream during the experiment, collected from Probe P1 (top) and power spectral density of the signal in probe P2 (bottom).}\label{Cyl}
\end{figure}

This data set contains $n_t=13200$ velocity fields sampled at $f_s=\SI{3}{\kilo\hertz}$ over a grid of $71\times30$ points. The cylinder has a diameter $d=\SI{5}{\milli\meter}$ and the spatial resolution is approximately $\Delta x=\SI{0.85}{\milli\meter}$. The velocity of the free stream was varied from $U_\infty\approx\SI{12}{\meter\per\second}$ to $U_\infty\approx\SI{8}{\meter\per\second}$, producing an evident change in the frequency of the vortex shedding from $f\approx \SI{459}{\hertz}$ to $f\approx \SI{303}{\hertz}$. The Reynolds number is in the range $Re\in [2600,4000]$. Figure \ref{Cyl} shows a snapshot of the PIV velocity field on the top-left, together with the location of three probes P1, P2, P3, from which the signal will be analyzed in further details in the following exercises. The figure on the top right shows the evolution of the free stream velocity (probe P1).

Being time-resolved, this test case is well suited for testing the computation of power spectral densities, cross-coherency, and advection velocity between two points. Due to the change in free stream velocity and the associated variation in the vortex shedding, this also offers an excellent test case to benchmark the multiscale POD's time-frequency localization capabilities, as extensively analyzed in \cite{Mendez2020}. Figure \ref{Cyl} shows, on the bottom right, the power spectral density from the time series extracted in probe P2. This test case has also been used in other lecture series, such as in \cite{Mendez2022} and \cite{Mendez2024}, to compare the performances of various linear and nonlinear dimensionality reduction techniques.

On the other hand, this experiment did not use tracking velocimetry. The provided code instead offers an efficient way to generate synthetic scattered data from gridded data, simulating a PTV acquisition. An example of a synthetic PTV field extracted from the PIV is shown in Figure \ref{Cyl} on the bottom left. It is important to emphasize that the goal here was not to precisely replicate the PTV evaluation process since a comparison between tracking-based and cross-correlation-based velocimetry is already addressed in the other provided test case. The objective is to test the tools developed in this lecture for processing time-resolved tracking.


\subsection{A round jet flow}\label{sec_2_2}

The second dataset showcases a horizontal water jet generated by the Dantec Dynamics Educational-PIV system. The nozzle has an exit diameter of d = 5 cm and operates within an aquarium measuring 80 x 35 x 40 cm. The jet velocity is approximately \SI{20}{\milli\meter\per\second}, with the water pump set to 30\% power. Images are captured using a camera with a resolution of $1920 \times 1200$\,px at a frame rate of \SI{160}{\hertz}. The illumination is provided by a LED light source positioned at the bottom of the aquarium. As the system only supports time-resolved acquisition, a total of $10^5$ frames are recorded. These frames are down-sampled by a factor of 10 to obtain image pairs at a frequency of 16 Hz, with a time separation of 6.25 ms between frames, ensuring appropriate pixel displacement while avoiding unnecessary oversampling of the jet oscillation.

The image pairs are processed using two methods. The first is a classical PIV approach is applied using the \textit{Particle Image Reconstruction Software} (PAIRS) \citep{Astarita2004, Astarita2006, Astarita2007, Astarita2008, Astarita2009}. The second is an in-house PTV tracking approach with a custom two-pulse PTV algorithm. This is based on the super-resolution approach from \cite{Keane1995} and incorporates a predictor based on the PIV results. Particle positions were identified with an adapted version of the open-source \textit{TracTrac} code from \cite{Heyman2019}.

\begin{figure}[h]
    \centering
    \includegraphics[width=0.45\linewidth]{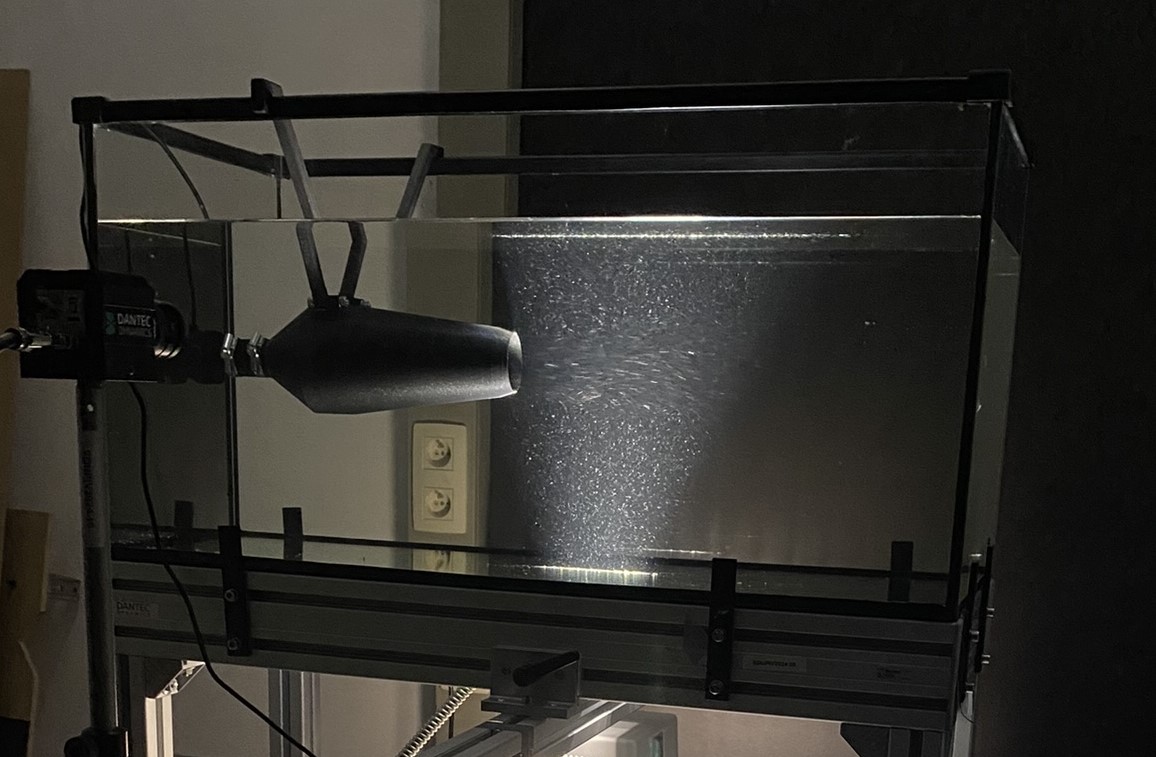}
    \includegraphics[width=0.45\linewidth]{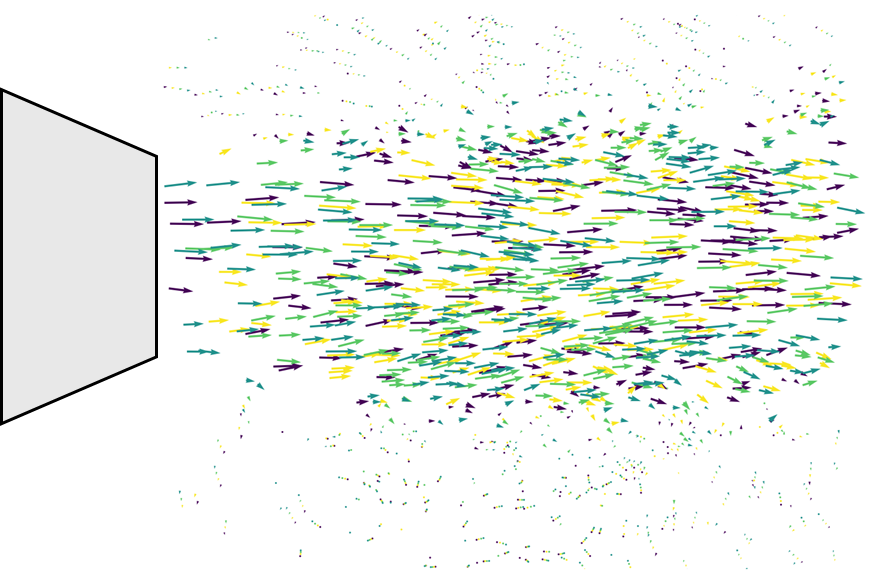}
    \caption{Picture of the jet flow setup (left). Typical velocity vectors from the PTV analysis (right)}
    \label{fig:enter-label}
\end{figure}

PIV and PTV velocity fields are obtained with the Get\_Jet\_DATA\_PIV\_PTV.py script. The data includes $n_t = 10\ 000$ time steps with a temporal resolution of $\Delta t = \SI{62.5}{\milli\second}$. For PIV analysis, velocity vectors are obtained using $64\times 64$\,px interrogation windows with \SI{75}{\percent} overlap, providing a vector pitch of $\Delta x = \Delta y = 1.83$ mm. It is worth noting that, due to limitations in illumination, the PIV vector fields contain a significant number of NaNs (Not a Numbers). For the same reason, the PTV field are much sparser near the boundaries. However, this is not considered a major limitation for the didactic purposes of this exercise. To circumvent the issues, we reduce the region of interest to a field of $50 \times \SI{100}{\milli\meter}$ in $x$ and $y$, resulting in 27 and 55 vectors in both axes. 

For PTV analysis, the algorithm extracts the velocity at the position of $n_p \approx 150$ particles for each time step in the reduced FOV. This data is provided as a large ensemble of scattered data.
The large number of images and vectors for this second data set offers an excellent dataset for statistical analysis using PIV and PTV approaches.

\section{Fundamentals of statistics} \label{sec_4}

This section reviews fundamental concepts in statics, mainly first—and second-order moments in section \ref{sec_3_1}, and then applies these to time series, with the important notion of Ergodicity in section \ref{sec_3_2}. Finally, we close with some tools for spectral analysis in \ref{sec_3_3}. 

\subsection{First and second order statistics}\label{sec_3_1}

In the statistical treatment of turbulent flows, we consider the velocity field as an infinitely dense field of \textbf{random variables}. We revisit this concept in Section \ref{sec_4}. For now, we focus on the two key properties we are primarily interested in.

Let $u$ represent a scalar random variable. For example, this could correspond to the value of one of the velocity components at a specific point in space. A random variable can be seen as a function that maps outcomes from a sample space of real numbers; we denote it as $u\in\mathbb{R}$ in this case. This is the set of all possible outcomes for $u$. We treat it as a \textbf{continuous} variable because it can take any value in $]-\infty,\infty [$, as opposed to \textbf{discrete} random variables where only a discrete set of points can be sampled.

Our continuous random variables are identified by a \textbf{probability density function} (pdf) $f_u(u)$ that describes the likelihood of different outcomes. This pdf allows to assign the probability that $u_m\leq u_n(\mathbf{t_k})<u_M$:

\begin{equation}
	\label{eq_uni}
	P\{ u_m\leq u<u_M\}=\int^{u_M}_{u_m} f_{u}(u)du\,.
\end{equation} 

Notice that, strictly speaking, we have zero probability of sampling any specific point, that is the integral in \eqref{eq_uni} tends to zero at the limit $u_M\rightarrow u_m$.

The operator from which we build our statistical treatment is the concept of expectation, which allows us to define the mean of the random variable $\mu_u$:

\begin{equation}
	\label{mu_x}
	\mu_{u}=\mathbb{E}_{\sim u}\{u\}=\int^{\infty}_{-\infty} u f_{u}(u) du\,
\end{equation}

We need this operator to compute the mean flow from the data. Note that the subscript $\sim u$ indicates the space over which the expectation is computed--in \eqref{mu_x}, this refers to the space of all possible outcomes of the random variables. In the following, we refer to this operation as \textbf{ensemble averaging}, in contrast to other types of averaging (e.g., over time or space).

To quantify the spreading of outcomes around the average, we generally use the variance $\sigma^2_u$ of the distribution, defined as 

\begin{equation}
	\label{sigma_x}
	\mbox{var}(u)=\sigma^2_{U}=\mathbb{E}_{\sim U}\{(u-\mu_{u})^2\}= \int^{\infty}_{-\infty} (u-\mu_{u})^2 f_{u}(u) du\,,
\end{equation} where $\sigma_{u}=\sqrt{\mbox{var}(u)}$ is the standard deviation. This computation involves quadratic terms in $u$, and is thus a \textbf{second order} statistic; we could continue to higher orders but this lecture will not go that far.

We are often concerned with the relation between two random variables. Let the first be $u$, with pdf $f_u(u)$, and the second be $v$, with pdf $f_v(v)$, we define the \emph{covariance} as 

\begin{equation}
\text{cov}(u, v) = \int_{-\infty}^{\infty} \int_{-\infty}^{\infty} (u - \mathbb{E}\{u\})(v - \mathbb{E}\{v\}) f_{u,v}(u, v) \, du \, dv\,
\end{equation} where $f_{u,v}(u,v)$ denotes the \textbf{joint probability density function}.
This quantity relate the variations of the two variables: if $\text{cov}(u,v)>0$, it means that the two variables tend to increase or decrease together, while $\text{cov}(u,v)=0$ means that the two variables are \textbf{uncorrelated}. This concept should not be confused with that of \textbf{independence}: we say that the variables are \textbf{independent} if $f_{u,v}(u,v)=f_u(u) f_v(v)$.

Note that $\text{var}(u)=\text{cov}(u,u)$. One can thus normalize covariances using the variances of each variable. The result is the \textbf{correlation coefficient}, ranging between -1 and 1:

\begin{equation}
\label{correlation}
R_{u,v}=\frac{\text{cov}(u,v)}{\sqrt{\text{var}(u)\text{var}(v)}}
\end{equation}

All subsequent analysis builds on these definitions, as we will discuss shortly. However, it is important to emphasize that the practical computation of these quantities always involves approximations, as we never have an analytic expression for the PDFs involved. Thus, estimating these quantities requires constructing approximations of the underlying distributions based on a large number of samples.

In practice, we rely on samples of the random variable which we collect in a vector $\mathbf{u}\in\mathbb{R}^{n_u}$. We thus must replace the notion of PDF with that of probability mass function (PMF). With this, we estimate the probability of having a specific outcome as $p(u_k)=n_k/n_u$, with $n_k$ the number of times the specific value $u_k$ occurs and $n_u$ the total number of samples corrected.
The integrals in \eqref{mu_x} and \eqref{sigma_x} should are then replaced with summations:

\begin{align}
\mu_u&\approx \mathbb{E}_{\sim \mathbf{u}}\{\mathbf{u}\}=\tilde{\mu}_u=\sum_k u_k p(u_k)=\frac{1}{n_u}\sum^{n_u}_{k=1} u_k\\
\sigma^2_U&\approx \tilde{\mbox{var}}\{\mathbf{u}\}= \sum_k (u_k-\mu_u)^2 p(u_k)= \frac{n_u}{n_u-1}\sum_k (u_k-\tilde{\mu}_u)^2 p(u_k)=\frac{1}{n_u-1}\sum^{n_u}_{k=1} (u_k-\tilde{\mu}_u)^2\,.
\end{align}

These are called, respectively, the sample mean and the sample variance. We use the $\tilde{}$ to distinguish sample quantities. Replacing the true mean with the sample mean in the definition of the sample variance requires adding a corrective term, known as Bessel's correction, to avoid bias errors (see, among others, \citet{Kay1993}). The covariance of the sample is defined similarly for two sequences of samples $\mathbf{u},\mathbf{v}\in\mathbb{R}^{n_u}$

\begin{align}
\tilde{\mbox{cov}}(\mathbf{u},\mathbf{v})=C_{UV}= \sum_k (u_k-\mu_u)(v_k-\mu_v) p(u_k,v_k)=\frac{1}{n_u-1}(u_k-\mu_u)(v_k-\mu_v)\,.
\end{align}

\subsection{Random processes, time series and ergodicity}\label{sec_3_2}

Let us now move from the concepts of random variables to the concept of \textbf{random processes}. This is the natural framework for analyzing stochastic sequences of numbers, i.e. time series (see \citet{Shumway2011} and \cite{Hyndman2021} for a comprehensive introduction).

A random process is a collection of random variables indexed by time (or space), representing a system's evolution that changes in a random manner. Focusing on time series analysis, let us $u(t)$ denote our random process; this could be the evolution of one of the velocity components at a specific location. 

Since a random process is characterized by a distribution of functions, the first and second order statistics are now functions: these are the \textbf{mean and autocovariance functions}: 

\begin{align}
\mu_u(t)&=\mathbb{E}_{\sim u(t)}\{u(t)\}=\int^{\infty}_{-\infty} u f_{u(t)}(u) du\\
\text{cov}_{u,u}(t_1,t_2)&=\mathbb{E}_{\sim u(t)}\{(u(t_1)-\mu_u(t_1))(u(t_2)-\mu_u(t_2))\}\\&=\int^{\infty}_{-\infty}\int^{\infty}_{-\infty} (u(t_1)-\mu_u(t_1))(u(t_2)-\mu_u(t_2)) f_{u(t_1),u(t_1)}(t_1,t_2)) du(t_1) du(t_2)\nonumber\,,
\end{align} having introduced the pdf of the process at a given time, $f_u(t)$, and the joint probability density function of the random variables at times $t_1$ and $t_2$ $f_{u(t_1),u(t_1)}(t_1,t_2))$. 

Note that the autocovariance function is nothing more than the covariance between the random variables obtained by sampling the random process at two different times. This is why this is also often referred to as a ``two-point statistics''. 

The variance at a given time is defined as $\sigma^2_u(t)=\text{cov}_u(t,t)$. Hence, the \textbf{auto-correlation} function can be obtained by normalizing the auto-covariance as 

\begin{equation}
R_{u,u}(t_1,t_2)=\frac{\text{cov}_u(t_1,t_2)}{{\sigma_u(t_1)\,\sigma_u(t_2)}}\,.
\end{equation}

Similarly, one can compare the degree of similarity between two different random processes. This gives the cross-covariance function between two random processes $u(t)$ and $v(t)$:

\begin{align}
\label{covariance_u_v}
\text{cov}_{u,v}(t_1,t_2)&=\mathbb{E}_{\sim u(t)}\{(u(t_1)-\mu_u(t_1))(v(t_2)-\mu_v(t_2))\}\\&=\int^{\infty}_{-\infty}\int^{\infty}_{-\infty} (u(t_1)-\mu_u(t_1))(v(t_2)-\mu_v(t_2)) f_{u(t_1),v(t_1)}(t_1,t_2)) du(t_1) dv(t_2)\nonumber\,,
\end{align} and the normalization gives the \textbf{cross-correlation}(the engine of PIV analysis!):

\begin{equation}
R_{u,v}(t_1,t_2)=\frac{\text{cov}_{u,v}(t_1,t_2)}{\sigma_u(t_1) \sigma_v(t_2)}
\end{equation}

Note that the expectation operator underpinning all these definitions is an \textbf{ensemble} operator: it averages over all possible realizations of the process \textbf{at a given time}.

As with random variables, we cannot work with these integrals directly because we cannot access the necessary probability density functions in practice. Instead, we rely on samples. While samples of a random variable can be collected in a vector, samples of a random process can be organized into a matrix, which we refer to as a \textbf{snapshot matrix}.

Let us denote by $\mathbf{U} \in \mathbb{R}^{n_t \times n_r}$ the snapshot matrix, where each column contains one of the $n_r$ realizations, each having $n_t$ samples in time: 

\[
\mathbf{U} =
\begin{bmatrix}
 u(\mathbf{t}_1, 1)  & u(\mathbf{t}_1, 2)  & \cdots & u(\mathbf{t}_{n_t}, n_r)  \\ 
\vdots & \vdots & \cdots & \cdots \\ 
u(\mathbf{t}_{n_t}, 1)  &  u(\mathbf{t}_{n_t}, 2) & \cdots &  u(\mathbf{t}_{n_t}, n_r)
\end{bmatrix}\in \mathbb{R}^{n_t \times n_r}
\]

We assume that these samples have been collected on a time list $\mathbf{t} = [t_1, t_2, \dots, t_{n_t}]$, which is not necessarily uniformly sampled (that is, ${\Delta t}_i=\mathbf{t}_{i+1}-\mathbf{t}_{i}$ is a vector of time differences). The sample mean of the process along these timestamps, denoted as $\overline{u}(\mathbf{t})$, is a vector representing the mean over the rows of $\mathbf{U}$. The sampled auto-covariance function is a matrix that collects the sample covariances across the available samples. This can be conveniently computed via matrix multiplication as:

\begin{equation}
	\label{Auto_cov_E}
	\bm{C}_{UU}(\mathbf{t_k},\mathbf{t_j})=\frac{1}{n_r-1 }\mathbf{U}'\,\mathbf{U}'^T\, \in\mathbb{R}^{n_t\times n_t}\,,
\end{equation} where $\mathbf{U}'=\mathbf{U}-\mu_U$ denotes the mean-centered snapshot matrix, i.e., having to remove the row-wise average to the matrix $\mathbf{U}$.

Note that the entry $k,j$ of this matrix is the inner product of the $k$-th and the $j$-th rows of $\mathbf{U}$ (hence all realizations collected at time $\mathbf{t}_k$ and $\mathbf{t}_j$) respectively. The fact that these are not equally spaced does not influence the computation; it only limits the pair of times that can be chosen.

The diagonal of this matrix gives the sample variance $\tilde{\sigma}^2_u(\mathbf{t})$ at the available time stamps. Normalizing \eqref{Auto_cov_E} gives the autocorrelation matrix. Writing $\bm{\Sigma}^2(t)=\text{diag}(\bm{C}_{UU})$, this can also be computed via matrix multiplication as :

\begin{equation}
\label{Auto_corr_E}
\bm{R}_{UU}(\mathbf{t_k},\mathbf{t_j})= \bm{\Sigma}^{-2}(t) \bm{C}_{UU}(\mathbf{t_k},\mathbf{t_j})
\end{equation}

The expressions to compute sample cross-covariance and sample cross-correlation of an ensemble are analogous and left as an exercise.

Two observations are now important: (1) these sample quantities are only available at the time instants $\mathbf{t}$ and (2) while uniform sampling is not required, the sampling locations must remain consistent across all samples of the process. This is the main challenge in tracking velocimetry, where samples are available in different location.

A special class of processes are \textbf{stationary processes}. A process is stationary in a strict-sense (\textbf{strong} stationarity) if the probability density function is constant over time, that is $f_{u(t)}(u)=f_{u}(u)$, hence $df_{u}/dt=0$. A process is stationary in a wide sense (\textbf{weak} stationarity) if the mean, variance, and autocovariance are constant over time (that is, first and second-order statistics). This means that all two point statistics linking the random variables at time $t_1$ and $t_2$ only depend on the time difference $\tau=t_2-t_1$. Thus we have

\begin{align}
\label{statisticS_2p}
\mu_u&=\mathbb{E}_{\sim u(t)}\{u(t)\}=\int^{\infty}_{-\infty} u f_{u}(u) du=\text{const}\\
\text{cov}_u(\tau)&=\mathbb{E}_{\sim u(t)}\{(u(t)-\mu_u)(u(t-\tau)-\mu_u)\}\\&=\int^{\infty}_{-\infty} u(t) u(t-\tau) f_{u(\tau)}(\tau)) du \nonumber\,.
\end{align}

Since these notes are solely concerned with first and second-order statistics, the distinction between weak and strong stationarity is irrelevant.

The last important concept we review in this section is that of \textbf{ergodicity}. A process is ergodic if ensemble statistics can be replaced by temporal statistics on any particular realization\footnote{ It is worth noticing that a process might be ergodic for a given statistic but not with others (e.g. a process might be ergodic in the mean but not in the autocorrelation). We will not dig deeper (see \cite{Kay1993,Oppenheim2015} for more).}. This means that the expectation over samples become (see also \citet{Lumley1,Oppenheim2015,Bruun1995}):

\begin{equation}
\label{expectation}
\mathbb{E}_{\sim u}\{u(t)\}=\int_{-\infty}^{\infty} u(t) f_{u}(u) du=\frac{1}{T}\int^T_0 u(t)dt= \mathbb{E}_{\sim T}\{u(t)\}\,,
\end{equation} where $T$ is the observation time.
The underlying idea is that all possible states allowed by the underlying pdf $f_u$ are visited for a sufficiently long time series. The proportion of time spent at each state over the observation time $T$ can be seen as the probability of that value occurring in the ensemble. Hence, the change of variables 

\begin{equation}
\label{ergodicity}
\frac{dt}{T}\approx f_u(u) du \quad  \text{as}\quad T\rightarrow \infty\,
\end{equation} implied in the equality.

For a stationary and ergodic process, the \textbf{autocovariance} and the \textbf{autocorrelation} read:

 \begin{equation}
C_{UU}(\tau) = \lim_{T \to \infty} \frac{1}{T} \int_0^T (u(t) - \mu_u)(u(t + \tau) - \mu_u) \, dt\quad\text{and}\quad R_{UU}(\tau)=\frac{C_{UU}(\tau)}{\sigma^2_u}
\end{equation} where $C_{UU}(0)=\sigma^2_u$ is the variance of the process. Similarly, the \textbf{cross-covariance} and the \textbf{cross-correlation} between two processes $u(t)$ and $v(t)$ are

 \begin{equation}
C_{UV}(\tau) = \lim_{T \to \infty} \frac{1}{T} \int_0^T (u(t) - \mu_u)(v(t + \tau) - \mu_v) \, dt\quad\text{and}\quad R_{UV}(\tau)=\frac{C_{UV}(\tau)}{\sigma_u\sigma_v}\,,
\end{equation} with $\sigma_u,\sigma_v$ the standard deviations of the two processes. 

The statistics of samples can be built by approximating integrals in time. For example, the sample cross-covariance in time between two sample sequences $\mathbf{u}(\mathbf{t})$ and $\mathbf{v}(\bm{t})$, with $\mathbf{t}\in\mathbb{R}^{n_t}$ the vector collecting the time instances where the data is available and $\Delta \mathbf{t_k}=(\mathbf{t_{k+1}}+\mathbf{t_k})/2$ the time interval representative for each sample becomes:
 
 \begin{equation}
 \label{covariance_u_v_S}
 \mbox{cov}_T(\mathbf{u},\mathbf{v})=\sum^{n_t}_{k=1} ({u}(\mathbf{t}_k)-\mu_U)({v}(\mathbf{t}_k)-\mu_V)\frac{\Delta \mathbf{t_k}}{T}\,.
 \end{equation}

 Note that, in case of equally sampled time series, with $\Delta \mathbf{t_k}=\Delta t$, one has $\Delta t/T=1/n_t$.
The computation of two-point statistics in practice is conceptually more challenging than their ensemble counterparts. Consider a sample $\mathbf{u}(\mathbf{t}) \in \mathbb{R}^{n_u}$. To compute the autocovariance at a given lag $\tau_l = \mathbf{t}_{l+1} - \mathbf{t}_{l-l}$, we need to gather many pairs $(\mathbf{u}(\mathbf{t}_j), \mathbf{u}(\mathbf{t}_{j-l}))$ and compute the time average of their product. However, this becomes problematic if the timestamps at which the samples are collected do not provide enough such pairs. Binning the possible phase lags could mitigate the problem, but we do not take that route here. 

Conversely, if the samples are collected at uniform intervals, a sequence of $n_u$ entries offers, in principle, $2n_u - 1$ possible lags, and for each lag, at least in principle, $n_u$ pairs. We say ``in principle'', because we here hit the limits of finite duration of the samples. Let us illustrate this limit in the computation of sample auto-covariance\footnote{Note that the distinction between covariances and correlations is not universally accepted. Here, we stick to a statistical definition of correlation as normalized covariance, but in numpy/scipy, for example, the function \textsc{correlate} does not normalize the output and thus provides a covariance rather than a correlation.} for a sample with uniform sampling $\mathbf{t}_k=k\Delta t$:

\begin{equation}
	\label{auto_cov_T}
	\text{cov}_{u,u}(\tau_l)=C_{UU}(\tau_l)=\mathbb{E}_{\sim t_i} \{u'(t_i) u'(t_i+\tau_l) \}=\frac{1}{n_t}\sum^{n_t}_{k=0}u'(\mathbf{t_k})u'(\mathbf{t_k}+\bm{\tau}_l)\,.
\end{equation} having used the notation $u'(\mathbf{t}_k)=u(\mathbf{t}_k)-\mu_U$. In a matrix formalism, this operation could be written as

\begin{equation}
\mathbf{C}_{UU,T}=\frac{1}{n_t} \text{circ}(\mathbf{u}')^T\text{circ}(\mathbf{u}')
\end{equation} where $\text{circ}$ is the cyclic operator that builds a Toeplitz matrix out of $n_t$ lags of a signal, hence:

\begin{equation}
\text{circ}(\mathbf{u}')=\begin{bmatrix}
 u(\mathbf{t}_1,)  & u(\mathbf{t}_2)  & \cdots & u(\mathbf{t}_{n_t})  \\ 
\vdots & \vdots & \cdots & \cdots \\ 
u(\mathbf{t}_{n_t})  &  u(\mathbf{t}_{n_t+2}) & \cdots &  u(\mathbf{t}_{2n_t})
\end{bmatrix}\in \mathbb{R}^{n_t \times n_t}
\end{equation} The problem of what to put on entries $\mathbf{t}_i$ when $i$ is larger than $n_t$. 

\begin{figure}[htbp]
	\centering
	\includegraphics[keepaspectratio=true,width=0.7 \columnwidth]
	{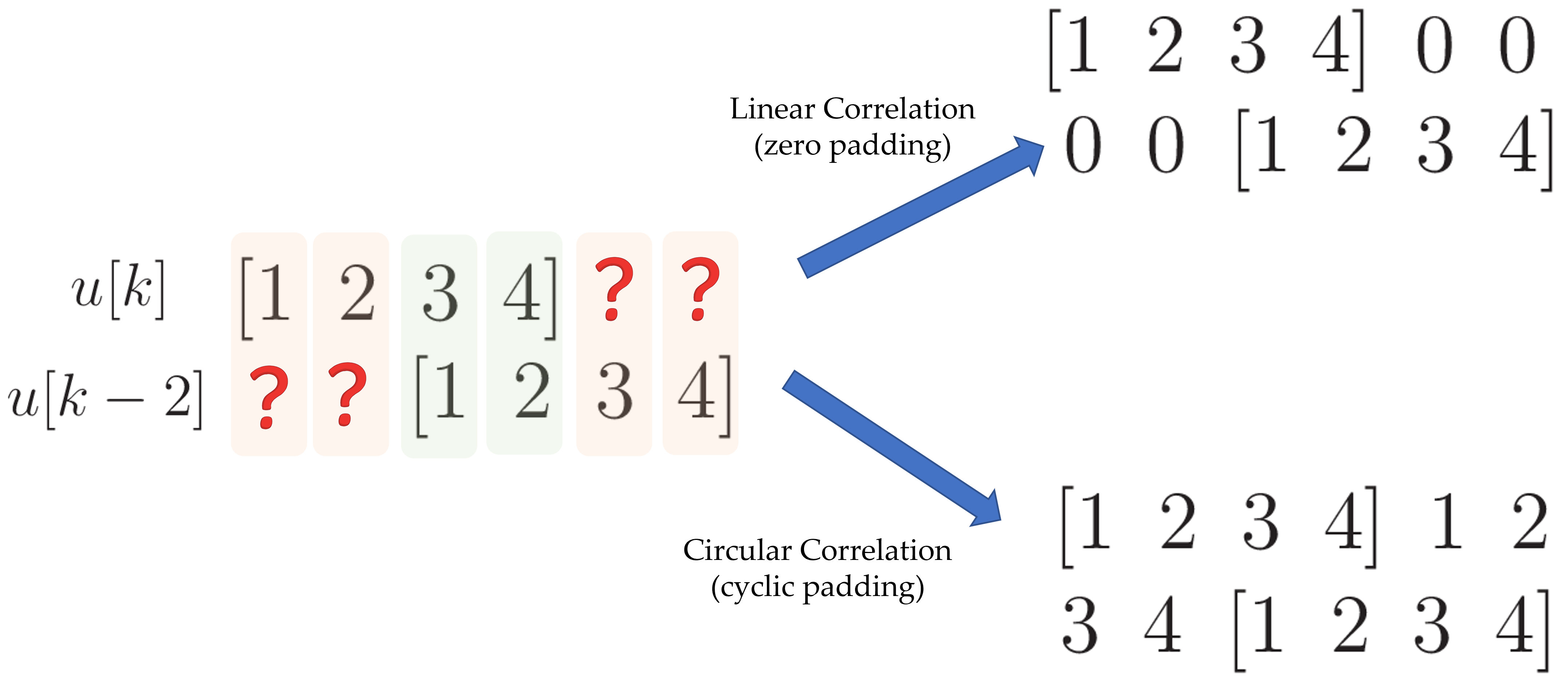}\\
	\vspace{-0.08cm}
	\caption
	{Sketch of the difference between linear and cyclic autocorrelation. The figure shows how these two definitions compute the autocorrelation entry $\tilde{R}_{UU}[-2]$ of the vector $u[k]=[1,2,3,4]$. The linear correlation pads with zero on both sides while the cyclic correlation assumes periodicity with period $n_t$.}\label{fig5}
\end{figure}

A simple example "by hand" might help illustrate where the problem is with \eqref{auto_cov_T} in the case of finite duration signals. The reader is encouraged to grab paper and pen and use \eqref{auto_cov_T} and compute the autocovariance of the time series $u(\mathbf{t_k})=[1,2,3,4]$, forgetting for the moment the mean removal. Consider a Python-like notation, hence $u(t_0)=0$ and $u(t_3)=4$. Considering $\mathbf{t}_k=k\Delta t$, we could use an index notation $u[k]$ and write $u[0]=0$ or $u[3]=4$ regardless of the $\Delta t$. Because this signal has $n_t=4$ entries, we have a total of $2n_t-1=7$ possible lags $l$. The lags in \eqref{auto_cov_T} take the form $\tau_l=l \Delta t$, and the autocorrelation is thus a vector of length $7$ which can also be indexed as $\tilde{C}_{UU}(\tau_l)=\tilde{C}_{UU}[l]$ with $l\in[-3,3]$. Figure \ref{fig5} illustrate the problem of computing $\tilde{C}_{UU}[-2]$: this is the correlation between the signal $u[k]$ and its copy shifted by $2$ entries to the right $u[k-2]$.

While it is clear that the summation will contain $3\times 1+4\times 2$, it is unclear how to deal with the entries that do not have a match in the other signal because of the finite size of the signal. On a practical level, the two classic solutions are \textbf{zero padding} and \textbf{cyclic padding}. The zero-padding leads to the \textbf{linear correlation}: the vectors are padded by zeros in all the entries lacking information. The cyclic padding leads to the \textbf{cyclic correlation}: the vectors are assumed periodic with period $n_t$. We distinguish these operators with and $L$ or a $C$, and the results for this example is:

\begin{equation}
	\tilde{C}_{UUL}[-2]=11/7 \quad 	\tilde{C}_{UUC}[-2]=22/7\,.
\end{equation}

Note that $C_{UUC}$ is periodic and for both operators we have $\tilde{C}_{UU}[l]=\tilde{C}_{UU}[-l]$. This is why one often plots only\footnote{The reader is encouraged to use \eqref{auto_cov_T} to compute the linear and the circular autocovariance of $u[k]=[1,2,3,4]$. Focusing on the `positive shifts', one has $\tilde{C}_{UUL}=[30,20,11,4]/7$ and $\tilde{C}_{UUC}=[30,24,22,24]/7$.} the last $n_t$ vectors of the autocorrelation, corresponding to the lags from $l=0$ to $l=n_t-1$. 

When the time series are `short', $\tilde{C}_{UUL}$ and $\tilde{C}_{UUC}$ differ significantly, and it is essential to clarify which is being used. `Short' here means that relevant lags (giving normalized cross-correlation of $\sim 1$) are present within a time scale that is comparable to the duration of the signal. This is the case of PIV interrogation (see \cite{Theo}), in which time is replaced by space, and the discrete-time indices are the shifts in pixels. This also explains the importance of the classic ``1/4" rule: the displacement should not exceed a quarter of the window or else the boundary problems becomes too important.

The linear (acyclic) operators are unbiased estimators \citep{Kay1993} and provide better statistical convergence to their ensemble counterparts. On the other hand, the cyclic operators are computationally more interesting because the periodic assumption enables a link with the cyclic convolution, which can be computed in the frequency domain using the Fast Fourier Transform (FFT). Briefly, it is easy to note that the cyclic convolution between two vectors $x[k]$ and $y[k]$ differs from the cyclic covariance by the flipping of the second vector (see \cite{Hayes2011}). In the frequency domain, this flipping corresponds to the conjugation of the associated Fourier transform.

To be more specific, let $X(f_n)$ be the Discrete Fourier Transform (DFT) of $x[n]$, i.e.:

\begin{equation}
	\label{Fourier_Pairs1}
	X(f_n)=\mathcal{F}\{x[k]\}=\frac{1}{n_t}\sum^{n_t}_{k=0} x[k] e^{-2 \pi f_n k\Delta t}  \leftrightarrow	x[k]=\mathcal{F}^{-1}\{X[k]\}=\sum^{n_t}_{k=0} X(f_n) e^{2 \pi f_n k\Delta t} 
\end{equation}  where $f_n=n \Delta f $, with $n=0,n_t-1$ and $\Delta f=f_s/n_t$ the frequency resolution, with $fs=1/\Delta t$ the sampling frequency. The cyclic cross-correlation between two signals $x[k]$, $y[k]$, having DFT $X(f_n)$, $Y(f_n)$ can be computed as 

\begin{equation}
	\label{cross_co_dis}
	R_{XY}=\mathcal{F}^{-1}\biggl(\mathcal{F}\{x[k]\}\overline{\mathcal{F}\{y[k]\}}\biggr)=\mathcal{F}^{-1}\bigl(X(f_n) \overline{Y}(f_n)\bigr)\,,	
\end{equation} with the overbar denoting complex conjugation. A \textsc{Python} implementation of the cyclic cross correlation is thus provided by following function:

\begin{centering}
	\begin{lstlisting}[language=Python,linewidth=15.5cm,xleftmargin=.05\textwidth,xrightmargin=.05\textwidth,backgroundcolor=\color{yellow!10}]
def R_UW_C(u,v):
  RUU=np.fft.ifft(np.abs(np.fft.fft(u)*np.fft.fft(v))).real 
  c=(RUU/len(u)-(np.mean(u)*np.mean(v)))/(np.std(u)*np.std(v))
  return c[:len(u)//2]		
	\end{lstlisting}
\end{centering}

It is possible to compute the linear operators using a circular extension if an appropriate zero padding is used before the cyclic extension (see \cite{DFT_Smith}). This is what software packages like Scipy and Numpy do when computing the 'FFT-based' cross-correlation. The linear cross-correlation using the \textit{Scipy}'s function \emph{correlate} is implemented in the following function:

\begin{centering}
	\begin{lstlisting}[language=Python,linewidth=15.5cm,xleftmargin=.05\textwidth,xrightmargin=.05\textwidth,backgroundcolor=\color{yellow!10}]
from scipy import signal
def R_UW_L(u,v):
 # Call to the scipy function correlate:
 RUU = signal.correlate(u-np.mean(u), v-np.mean(v),\
 mode='same',method='auto')/(len(u)-1)/(np.std(u)*np.std(v))
 return RUU[RUU.size//2:]		
	\end{lstlisting}
\end{centering}

The entry `method', set by default to `auto', uses the fastest option between direct and FFT-based correlation, depending on the size of the array. These functions are needed for the next exercise.

\begin{tcolorbox}[breakable, opacityframe=.1, title=Exercise 1: The statistics of an Ornstein-Uhlenbeck Process]

	Consider the Ornstein-Uhlenbeck process governed by the following stochastic differential equation (see \cite{Oksendal1998,Borodin2002,Pope2000})
	
	\begin{equation}
		\label{eqXX}
		d X_t= \kappa (\theta-X_t) dt+\sigma d W_t
	\end{equation}
	
	where the subscript $t$ here refers to a time step, $\kappa>0$ is called rate of reversion and controls how quickly the process reaches its (stationary) long-term behavior, characterized by mean $\theta$ and random fluctuation with standard deviation $\sigma$. The second term $W_t$ is a stochastic term, and it is here taken as a normal Gaussian with zero mean and unitary standard deviation, here written as $\mathcal{N}(0,1)$. The first term is often referred to as \emph{drift}, the second as \emph{diffusion}.
This process is closely related to the Langevin equation used in Lagrangian turbulence modeling \citep{thomson1987criteria,pope1994lagrangian}, where an equation similar to \eqref{eqXX} models the velocity of a particle as it moves through a turbulent flow.

\hspace{2mm} Consider a case with initial condition $X_0=0$, $\kappa=1.2$, $\theta=3$ and $\sigma=0.5$. Consider $n_t=1001$ samples, with a sampling of $\Delta t=0.01$ s. This corresponds to a physical observation time of $T=10 s$. Plot four realizations to explore the process. 
	
\hspace{2mm} The questions are two. (1) compute the ensemble mean, the ensemble standard deviation and the ensemble autocovariance as a function of time using $n_r=100$ and considering the steady state condition. (2) Study the convergence of these statistical quantities at two time steps (say at $k=10$ and $k=700$) as a function of the number of samples in the ensemble.

	\medskip
	\textbf{Solution}.  Let us begin by creating a function that produces a sample of the process, taking as input the four process parameters $\kappa,\theta,\sigma$ and vector of times $\mathbf{t_k}$. Using a simple loop, the following function does the job (see Python exercise 1):

	\begin{centering}
		\begin{lstlisting}[language=Python,linewidth=15.5cm,xleftmargin=.05\textwidth,xrightmargin=.05\textwidth,backgroundcolor=\color{yellow!10}]
import numpy as np
# Function definition
def U_O_Process(kappa,theta,sigma,t):
 n_T=len(t) 
 # Initialize the output
 y=np.zeros(n_T)
 # Define Drift and Diffusion functions in the process
 drift= lambda y,t: kappa*(theta-y)
 diff= lambda y,t: sigma
 noise=np.random.normal(loc=0,scale=1,size=n_T)*np.sqrt(dt)
 # Solve Stochastic Difference Equation
 for i in range(1,n_T):
   y[i]=y[i-1]+drift(y[i-1],i*dt)*dt+diff(y[i-1],i*dt)*noise[i]
 return y
		\end{lstlisting}
	\end{centering}
	
\hspace{2mm}The following script creates $n_r=500$ realizations and store them in a matrix $\bm{U}_n$ of size $n_t\times n_r$ (see \eqref{Auto_cov_E}). Then, it plots four randomly chosen samples, namely the numbers $r=1,10,22,55$. The four realizations are shown in Figure \ref{a)}, with the plot axis being customized (see the provided codes).

	\begin{centering}
		\begin{lstlisting}[language=Python,linewidth=15.5cm,xleftmargin=.05\textwidth,xrightmargin=.05\textwidth,backgroundcolor=\color{yellow!10}]
import matplotlib.pyplot as plt
# Initial and final time
t_0=0; t_end=10
# Number of Samples
n_t=1001
# Process Parameters
kappa=1.2; theta=3; sigma=0.5
# Create the time scale
t=np.linspace(t_0,t_end,n_t); dt=t[2]-t[1]
# Collect 500 sample and store in U_N
n_r=500; U_N=np.zeros((n_t,n_r))
for l in range(n_r):
  U_N[:,l]=U_O_Process(kappa,theta,sigma,t)
  # Plot the results 
  plt.figure(1)
  plt.plot(t,U_N[:,1])
  plt.plot(t,U_N[:,10])
  plt.plot(t,U_N[:,22])
  plt.plot(t,U_N[:,55])
		\end{lstlisting}
	\end{centering}
	
\hspace{2mm} The reader should play with the code long enough to realize that this process is characterized by a transient time of the order of $3s$ within which all signals move from zero to a stationary condition. Having arranged the data into `snapshot matrices', the computation of the time average and the temporal standard deviation can be done in one line each:

	\begin{centering}
		\begin{lstlisting}[language=Python,linewidth=15.5cm,xleftmargin=.05\textwidth,xrightmargin=.05\textwidth,backgroundcolor=\color{yellow!10}]
U_Mean=np.mean(U_N,axis=1) # Ensemble Mean
U_STD=np.std(U_N,axis=1) # Ensemble STD
		\end{lstlisting}
	\end{centering}
	
\hspace{2mm}These are essentially `row-wise' statistics. Figure \ref{b)} shows the time average together with the range $\mu_U(\mathbf{t_k})\pm \sigma_U(\mathbf{t_k})$. It appears that the standard deviation grows gently until it reaches a constant value after about $\mathbf{t_k}>4s$.

\hspace{2mm}Finally, we analyze the ensemble autocorrelation of this random process. The ensemble cross-correlation between the time steps $\mathbf{t_j}$ and $\mathbf{t_k}$ for the ensembles $\mathbf{U}_n$ and $\mathbf{W}_n$ of two random variables can be computed with the following function:
	
	\begin{centering}
		\begin{lstlisting}[language=Python,linewidth=15.5cm,xleftmargin=.05\textwidth,xrightmargin=.05\textwidth,backgroundcolor=\color{yellow!10}]
def Ensemble_Autocorr(U_N,W_N,k,j):
 n_r,n_t=np.shape(U_N)
 # Select all realizations at time t_k for U
 U_N_k=np.expand_dims(U_N[k,:],axis=0);
 # Select all realizations at time t_kj for W
 W_N_k=np.expand_dims(W_N[j,:],axis=0)
 # Note (These are row vectors)
 # Compute the average products      
 PR=U_N_k.T.dot(W_N_k)
 R_UW=np.mean(PR)/(np.std(U_N_k)*np.std(W_N_k))   
 return R_UW
		\end{lstlisting}
	\end{centering}
	
        \begin{center}%
            \captionsetup[sub]{labelformat=parens}
            \captionsetup{type=figure}
            \begin{subfigure}{0.49\textwidth}
                \includegraphics[width=\linewidth]{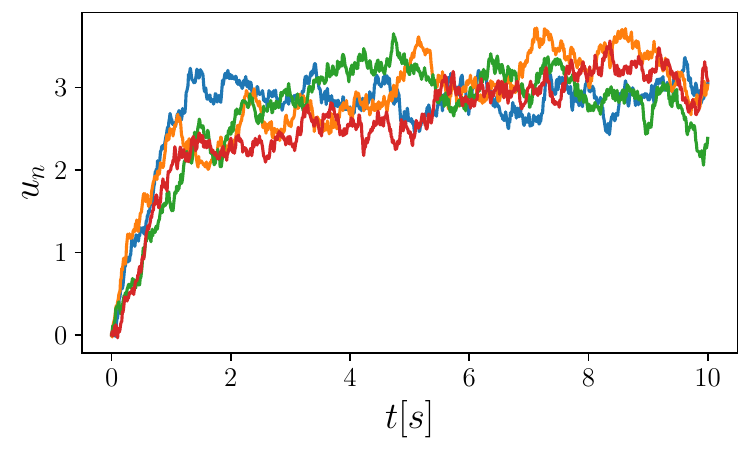}
                \caption{}
                \label{a)}
            \end{subfigure}
            \begin{subfigure}{0.49\textwidth}
                \includegraphics[width=\linewidth]{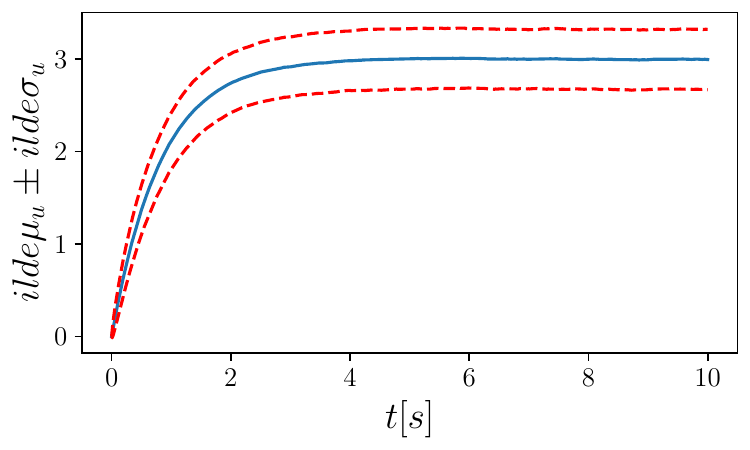}
                \caption{}
                \label{b)}
            \end{subfigure}
		\vspace{-3mm}
            \caption{Fig (a): Four randomly chosen samples of the Ornstein- Uhlenbeck process; Fig (b): Ensemble average of the process (in blue), along with the $\mu_u+\sigma$ and $\mu_u-\sigma$ curves.}
		\vspace{-0.05cm}
		\label{figEx1}
	\end{center}
	
	Having added a normalization. Then, the following script computes the autocorrelation of 100 randomly pairs, separated by a lag of $50$ samples:
	
	\begin{centering}
		\begin{lstlisting}[language=Python,linewidth=15.5cm,xleftmargin=.05\textwidth,xrightmargin=.05\textwidth,backgroundcolor=\color{yellow!10}]
# Define lag (in number of samples)
lag=50 
# Study the autocorrelation between two points at equal lags
N_S=100; R_UW=np.zeros(N_S)
# Select a 100 random points i (larger than 500)
J=np.random.randint(500,800,N_S); K=J+50
for n in range(N_S):
  R_UW[n]=Ensemble_Autocorr(U_N,U_N,J[n],K[n])
		\end{lstlisting}
	\end{centering}
	
\hspace{2mm}The resulting set of autocorrelations has a mean of $0.545$ and a standard deviation of $0.0059$: in other words, it does not matter what the exact pair $j,k$ is, as long as they differ by the same lag (in this case, 50). This is because here we are sampling in time intervals from 500 to 800, and here the process has reached its stationary condition. The reader is encouraged to repeat the exercise at an earlier interval. 
	
\hspace{2mm}Finally, we analyze the convergence of the statistics as a function of the number of realizations. The following script computes the mean and the standard deviation at $k=10$ and $k=700$ for a number of realizations that goes from $1$ to $1000$. 
	
	\begin{centering}
		\begin{lstlisting}[language=Python,linewidth=15.5cm,xleftmargin=.05\textwidth,xrightmargin=.05\textwidth,backgroundcolor=\color{yellow!10}]
n_R=np.round(np.logspace(0.1,3,num=41))
# Prepare the outputs at k=100
mu_10=np.zeros(len(n_R))
sigma_10=np.zeros(len(n_R))
# Prepare the outputs at k=700
mu_700=np.zeros(len(n_R))
sigma_700=np.zeros(len(n_R))
# Loop over all n_R's.
for n in range(len(n_R)):
 # show progress
 print('Computing n='+str(n)+' of '+str(len(n_R)))
 n_r=int(n_R[n]) # Define the number of ensembles
 U_N=np.zeros((n_t,n_r)) # Initialize the ensemble set
 for l in range(n_r): # Fill the Ensemble Matrix
   U_N[:,l]=U_O_Process(kappa,theta,sigma,t)
   # Compute the mean and the std's
   mu_10[n]=np.mean(U_N[10,:]) # Ensemble Mean
   sigma_10[n]=np.std(U_N[10,:]) # Ensemble STD
   mu_700[n]=np.mean(U_N[700,:]) # Ensemble Mean
   sigma_700[n]=np.std(U_N[700,:]) # Ensemble STD		
		\end{lstlisting}
	\end{centering}
	
\hspace{2mm}	By analyzing the vector n\_R in the script, the reader should note that some of the entries with a small number of samples are taken multiple times to show the variability in the prediction. The results are shown in Figure \ref{figEx1b}.
	
\hspace{2mm}	It can be shown that the convergence of both the mean and the standard deviation is $\propto \sqrt{n_r}$, but the convergence of the mean is $\propto \sigma_U$ while the convergence of the standard deviation is $\propto \sigma^2_U$ (see also \cite{AndreaLS}): this explains why a larger number of samples is needed to converge the second-order statistics, and why the convergence at $k=100$ is slower. The reader is encouraged to explore the provided scripts further, to observe this result in action at different points.

        \begin{center}%
            \captionsetup[sub]{labelformat=parens}
            \captionsetup{type=figure}
            \begin{subfigure}{0.49\textwidth}
                \includegraphics[width=\linewidth]{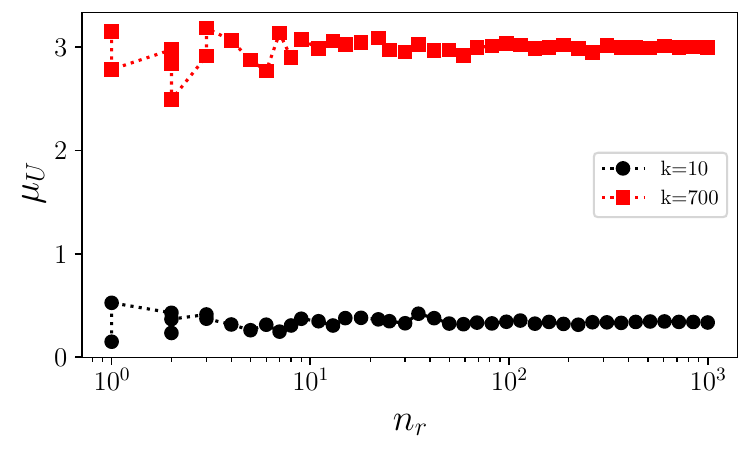}
                \caption{}
                \label{1a)}
            \end{subfigure}
            \begin{subfigure}{0.49\textwidth}
                \includegraphics[width=\linewidth]{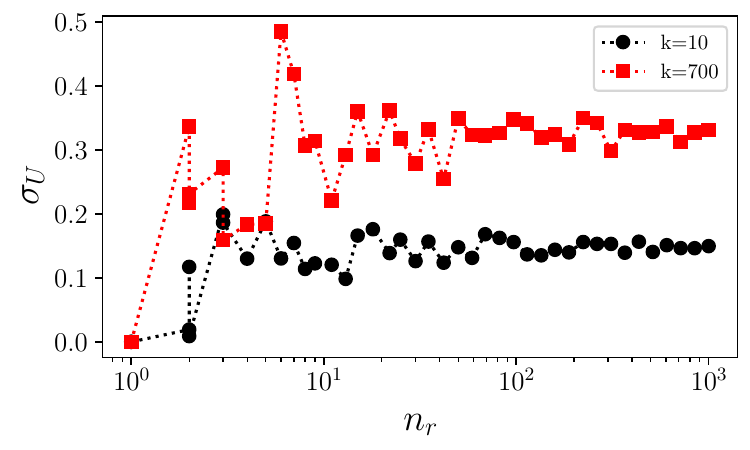}
                \caption{}
                \label{1b)}
            \end{subfigure}
		\vspace{-3mm}
            \caption{Fig (a): mean velocity convergence; Fig (b): standard deviation convergence.}
		\vspace{-0.05cm}
		\label{figEx1b}
	\end{center}
	
\end{tcolorbox}

\subsection{Power spectral densities and cross-coherency}\label{sec_3_3}

To illustrate the use of second-order statistics in spectral analysis, this section moves to the frequency representation of stochastic signals. This is slightly more involved than the frequency representation of deterministic signals. In a continuous domain, a stochastic signal is seldom square-integrable, and it thus seldom admits an ordinary Fourier transform. In the signal processing terminology, the definition of an appropriate frequency domain requires shifting the treatment from the notion of energy to the notion of power\footnote{Given an infinite duration stochastic signal $x[k]$, indexed by the integers $k$, the energy $\mathcal{E}$ and the power $\mathcal{P}$ are defined as follows:
	
	\begin{equation}
		\label{E_P}
		\mathcal{E}\{x[k]\}=\lim_{n_t\to \infty}\sum^{n_t}_{k=0} |x[k]|^2 \quad \quad \mathcal{P}\{x[k]\}=\lim_{n_t\to \infty}\frac{1}{2n_t-1}\sum^{n_t}_{k=0} |x[k]|^2\,.
\end{equation}}: stochastic signals have generally infinite energy but finite power \citep{Oppenheim2015,Hayes2011}. 

Therefore, we shall not focus on the Fourier transform of a signal but on the Fourier transform of its autocorrelation (which is square integrable). The following transform exists for all signals of interest:

\begin{equation}
	\label{R00}
	\mathbb{E}\{u^2_n(\mathbf{t_k})\}={R}_{UU}(0)=\frac{1}{2\pi} \int^{\infty}_{-\infty} S_{UU}(\omega) d\omega\,,
\end{equation} where $\omega=2\pi f$ is the pulsation, $f$ is the frequency and $S_{UU}$ is the continuous Fourier transform of the autocorrelation function. This is known as \emph{power spectral density}. Note that this is real and even in $\omega$ because the time autocorrelation $R_{UU}$ is symmetric ($R_{UU}(\tau)=R_{UU}(-\tau)$). The Fourier pairs of interest are thus 

\begin{equation}
	\label{Fourier_PAIRS}
	S_{UU}(\omega)=\int^{\infty}_{-\infty} {R}_{UU}(\tau) e^{-\mathrm{j}\omega \tau} d\tau \quad \mbox{and} \quad 	{R}_{UU}(\tau)=\frac{1}{2\pi}\int^{\infty}_{-\infty} S_{UU}(\omega) e^{\mathrm{j}\omega \tau} d\omega\,.
\end{equation}

These are also known as \textbf{Wiener-Khinchin relations}.
Similarly, the cross-spectral density and the cross-correlation are Fourier pairs:

\begin{equation}
	\label{Fourier_PAIRS}
	S_{UV}(\omega)=\int^{\infty}_{-\infty} {R}_{UV}(\tau) e^{-\mathrm{j}\omega \tau} d\tau \quad \mbox{and} \quad 	{R}_{UV}(\tau)=\frac{1}{2\pi}\int^{\infty}_{-\infty} S_{UV}(\omega) e^{\mathrm{j}\omega \tau} d\omega\,.
\end{equation}

The notion of power spectral density is important because it allows us to give a spectral representation to stochastic signals and thus to generalize the theory of linear time-invariant (LTI) systems (see \cite{MendezLS}). This theory brings powerful and simple tools for system identification, filtering and forecasting (see \cite{Oppenheim2015,FIR_Smith} for more).

Concretely, we are interested in the notion of coherency in relation to the frequency content of signals. We might want to infer, for example, how much the frequency content of two time series are linked, at least within a certain range of frequencies. This could help identify for example, if a given pattern is ``traveling'' between two probes (e.g. Probes P2 and P3 in the test case in section \ref{sec_2_1}).

We first need some definitions. Let us assume that a stochastic signal $x[k]\in \mathbb{R}^{n_t}$ is the input of a linear and deterministic system which responds with a second stochastic signal $y[k] \in \mathbb{R}^{n_t}$. Uncorrelated noise might be added to the output of this system and we only see the resulting $y_n[k]=y[k]+N[k]$. If the power of the noise is too large, we might not be able to recover any reasonable estimate of $y[k]$ and we will say that there is a poor level of \emph{coherency} between $x[k]$ and $y_n[k]$. Conversely, we might be able to identify an approximation of the underlying linear linking input and output. This link is here to be analyzed frequency by frequency.

Considering finite duration signals, the convergence problems of the Fourier representation are less stringent, and we can waive some of the formalism required for the continuous world: under the assumption of circular extension of the signals\footnote{In this section we will only consider cyclic padding. The zero-padding requires some little extra care which is not essential for this lecture (see \cite{DFT_Smith}).)}, every digital signal has a Discrete Fourier Transform (DFT). Let $X(f_n)$ and $Y(f_n)$ denote the DFT of the input $x[k]$ and the output $y[k]$ (see \eqref{Fourier_Pairs1} for the definitions).
If a linear time invariant system links $y[k]$ to $x[k]$, the output can be computed via convolution of the input with the system's impulse response. From the convolution theorem, we know that in frequency domain this is a multiplication with the transfer function of the system $H(f_n)$, i.e. the DFT of the impulse response. We thus have:

\begin{equation}
	\label{Definitions}
	y[k]=\sum^{n_t-1}_{m=0} x[k]\,h[k-m] \,\,\longleftrightarrow	\,\, Y(f_n)= H(f_n)\, X(f_n)\,.
\end{equation}  

We now introduce the discrete equivalent of \eqref{Fourier_PAIRS}. It is possible to show (see \cite{Oppenheim1996a} and eq. \eqref{cross_co_dis}) that the power spectral density of the input is

\begin{equation}
	\label{DefinitionsII}
	S_{XX}(f_n)=\sum^{n_t-1}_{k=0} R_{XX}[k] e^{-2 \pi f_n k\Delta t} =\mathcal{F}\{\mathcal{F}^{-1}\{X(f_n) \overline{X}(f_n)\}\}= X(f_n) \overline{X}(f_n)\,.
\end{equation} 

Similarly, the power spectral density of the output is $S_{YY}(f_n)=Y (f_n) \overline{Y}(f_n)$ and we can also define \emph{cross-spectral density} as $S_{YX}=X(f_n) \overline{Y}(f_n)$. We can now craft a spectral function which measures how well the output spectrum correlates with the input spectrum. This is the \emph{coherence function}:

\begin{equation}
	\label{DefinitionsIV}
	\widehat{C}_{YX}(f_n)=\frac{|S_{YX}(f_n)|^2}{S_{XX}(f_n) S_{YY}(f_n)}\,\,.
\end{equation}

At frequencies for which $Y(f_n)=H(f_n) X(f_n)$ (i.e., for which an LTI system could model the input-output relation), one has 

$$S_{YX}(f_n)=Y(f_n)\overline{X}(f_n)=H(f_n) X(f_n)\overline{X}(f_n)\rightarrow S_{YX}(f_n)=H(f_n)S_{XX}(f_n)\,.$$ 

Moreover, noticing that $S_{YY}(f_n)=|H(f_n)|^2 S_{XX}(f_n)$, we recover $\widehat{C}_{YX}(f_n)=1$. At frequencies for which no LTI system can link the two spectra, \emph{coherence} is zero. The function is undefined at $f_n$'s for which either $S_{XX}$ or $S_{YY}$ is zero.

To conclude this section, it is worth noticing that working with one single spectrum for both input and output makes little sense: an ensemble of time series leads to an ensemble of spectra. The classic approach to evaluating a stochastic signal's frequency representation involves averaging, which could be either in the ensemble or time domains. Under the assumption of ergodicity, the second is usually preferred, and the result is the well-known Welch's method \citep{Welch1967} or \emph{periodogram method}. The computation is performed by dividing the signal into successive (and overlapping) blocks, computing the DFT, and then averaging the results. In practice, a smoothing window $w[k]$ multiplies the signal in the time domain to deal with the problems arising from the (usually violated) periodicity assumption. For $S_{XX}(f_n)$, for instance, we have:

\begin{equation}
	S_{XX}(f_n)=\frac{1}{n_L}\sum^{n_L-1}_{m=1} |XW_m(f_n)|^2 \quad \mbox{with} \quad  XW_m(f_n)=\mathcal{F}\{x[k] \,w[k]\}
\end{equation} 

Here $XW_m(f_n)$ is the DFT of the windowed block $x_m[k]w[k]$, with $x_m[k]=x[k+m n_W]$, $w_m[k]$ the window function designed to gracefully taper to zero at both endpoints of the block, and $k=0,1,\dots n_w-1$ with $n_L$ denoting the number of blocks. Because the multiplication in the time domain is a convolution in the frequency domain, the method is essentially a smoothing operation on the signal's spectrum; this is why the method is sometimes also called `smoothed spectrum' and implemented in the frequency domain. 

The \textit{Python}'s \emph{scipy} package offers robust functions to compute both the power spectral densities and the spectral coherence. We use both in the next exercise.

\begin{tcolorbox}[breakable, opacityframe=.1, title=Exercise 2: Power Spectral Densities and Coherency]

Consider the time series sampled in probes P2 and P3 (see Figure \ref{Cyl}) a. (1) Study the covariance and correlation between the two signals in the first $t<1$ s of observation, i.e. when the flow is in its first stationary condition. Then, (2) analyze the power spectral densities, the cross spectral densities and the cross-coherency between the two signals.

	\textbf{Solution}. The solution to this exercise is provided in the file Exercise\_2.py. Figure \ref{scatter_EX} shows a scatter plot of the stream-wise component in P3 versus P2. The correlation coefficient between the two time series is $\rho_{1,2}=0.58$, which is rather high for a turbulent flow.  Figure \ref{figEx2b} provides the cross spectral density $S_{XY}(f_n)$ (Figure a) and the cross coherency $C_{X,Y}(f_n)$ (figure b) between the two signals.

        \begin{center}%
            \captionsetup[sub]{labelformat=parens}
            \captionsetup{type=figure}
            \includegraphics[width=0.55\linewidth]{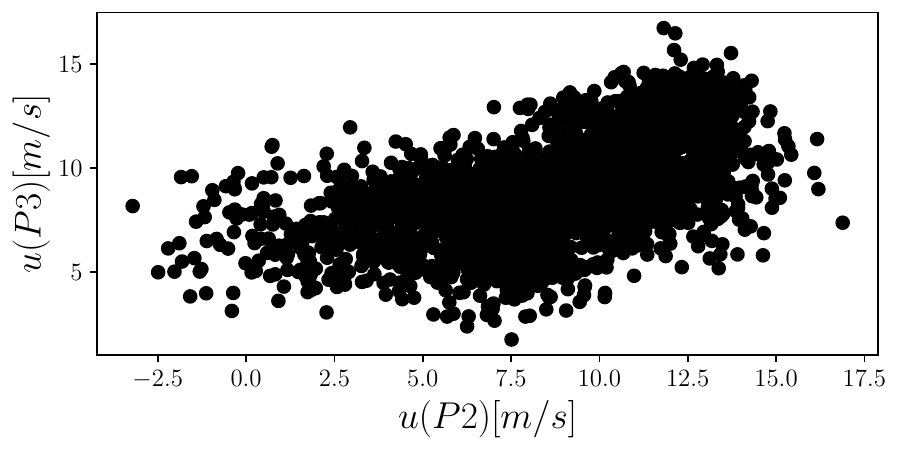}
		\vspace{-3mm}
            \caption{Scatter plot of the streamwise velocity components in probe P3 versus probe P2 (see Figure \ref{Cyl} for the location. A non-negligible correlation is visible.}
		\vspace{-0.05cm}
		\label{scatter_EX}
	\end{center}

 A strong correlation between the signals is evident near the vortex shedding frequency, which in this signal segment is approximately 440 Hz. The cross-coherency level is nearly unity, indicating that the two signals could be related within this frequency range through a linear time-invariant (LTI) system. This is due to the global nature of the oscillation mechanism in the vortex shedding, which synchronizes large portions of the flow ( in this case, the entire wake) in a coherent oscillatory behavior.

        \begin{center}%
            \captionsetup[sub]{labelformat=parens}
            \captionsetup{type=figure}
            \begin{subfigure}{0.49\textwidth}
                \includegraphics[width=\linewidth]{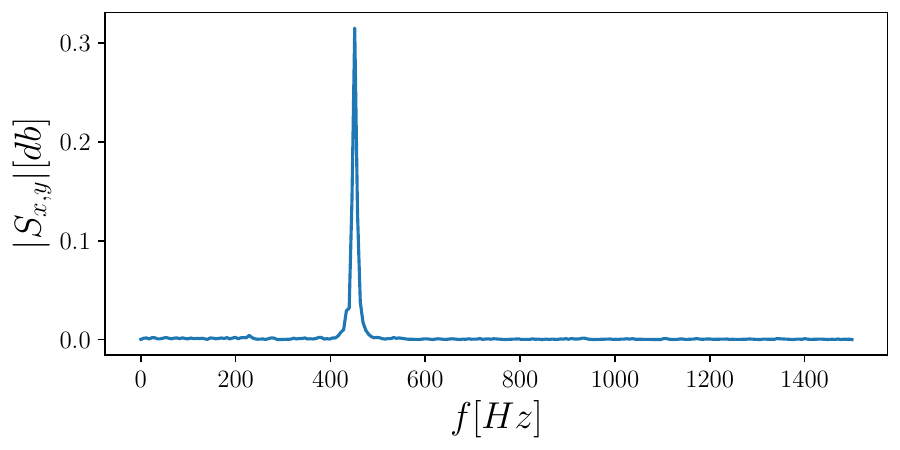}
                \caption{}
                 \label{2a)}
            \end{subfigure}
            \begin{subfigure}{0.49\textwidth}
                \includegraphics[width=\linewidth]{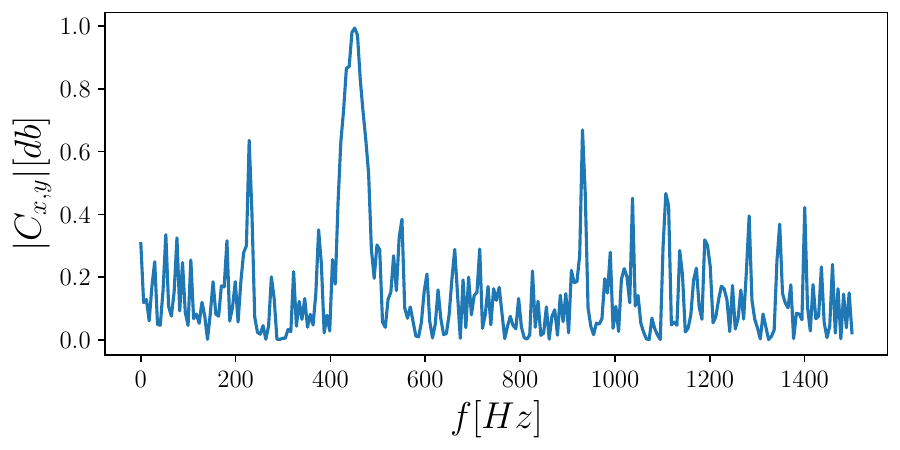}
                \caption{}
                \label{2b)}
            \end{subfigure}
		\vspace{-3mm}
            \caption{ Fig (a): Cross spectral density; Fig (b): Cross coherency between signal P1 and P2}
		\vspace{-0.05cm}
		\label{figEx2b}
	\end{center}

\end{tcolorbox}

\section{The statistical treatment of turbulence}\label{sec_5}

When analyzing turbulent flow statistics, the time series analysis tools introduced in Section \ref{sec_3_2} must be extended in two ways: (1) the random process is a velocity signal, which is a vector quantity, and (2) the random process depends not only on time but also on space. The following subsections discuss these two extensions separately.
We stress that the statistical treatment of turbulence is a vast subject, far beyond the scope of this lecture notes. The reader is referred to \cite{Pope2000,Tennekes1972,Davidson2004,Mcdonough04introductorylectures} for a comprehensive introduction to the topic and to \cite{Saarenrinne2000,Lavoie2007,Segalini2014,Scharnowski2018,Ayegba2020,Wang2021} for a discussion on the impact of measurement resolution on the main turbulence variables. This section recalls the definitions that are required to solve the exercises provided.

\subsection{Local statistics of velocity components}\label{sec_4_1}

Let $\bm{u}(\bm{x},t)$ denote the velocity field, having components $\bm{u}:=(u,v,w)$ at each location $\bm{x}:=(x,y,z)$ and time $t$. Let us introduce a more compact notation $\mathbb{E}_{E}\{a\}=\langle a\rangle_E$ to denote the expectation operator on the set of samples $E$.

In the classic Reynolds decomposition of turbulent flows, the velocity field is decomposed into the sum of an ensemble average and a fluctuating part, that is

\begin{equation}
	\label{Reynolds}
	\bm{u}(\bm{x},t)=\langle \bm{u} \rangle_{E} (\bm{x},t)+\bm{u}'(\bm{x},t) \,,
\end{equation} where $\langle \mathbf{u}\rangle=(\langle u\rangle,\langle v\rangle, \langle w  \rangle)^T$ is ensemble average field and $\mathbf{u}'=(u',v',w')^T$ is the fluctuating field. The averaging in this formulation is an \textbf{ensemble averaging} and should not be confused with the \textbf{time averaging} introduced in the following for stationary data. While this ensemble averaging acts as a smoother of the small-scale fluctuations, this should not be confused with the frequency-based filtering employed in the Large Eddy Simulation (LES) formalism.

An example might help clarify this distinction. Consider the case of a transient flow in a channel, with the flow ramping up from zero to a fully established regime within a time interval $T$. That is, let us assume that the flow rate follows the same time evolution as the stochastic process analyzed in Exercise 1. A filtering in the time domain, as in the LES formalism, would impose a cut-off frequency that removes components above a certain value, which may smooth out some of the frequencies associated with the ramp-up phase. In contrast, under an ensemble averaging approach, we would run the experiment multiple times and take averages over all realizations, as done in Exercise 1. Such ensemble averaging produces time-dependent "mean flows" which are described by the Unsteady Reynolds-Averaged Navier–Stokes (URANS) formalism. The main conceptual difficulty here is that this operation has a complex relationship with the frequency content remaining after ensemble averaging: the individual realizations could all exhibit strong gradients (thus high-frequency components) in certain regions, which would be preserved in the average. Therefore, the ensemble-averaged flow may still be time-dependent and retain high-frequency content. The key takeaway is that URANS, in general, is not equivalent to time averaging and does not inherently smooth out high-frequency fluctuations.

Particularly interesting, in an ensemble averaging formalism, are the second order statistics linking fluctuations along the different components. At each location $\bm{x}$ and for each time $t$, the covariance of the velocity component is defined as 

\begin{equation}
	\label{Reynolds_STRESS}
\bm{R}(\bm{x},t)= \begin{pmatrix}
		\langle u'^2 \rangle & \langle u' v' \rangle  & \langle u' w'\rangle\\
		\langle v' u' \rangle & \langle v'^2 \rangle & \langle v' w'\rangle\\
        \langle w' u' \rangle & \langle w'v' \rangle & \langle w'^2\rangle
	\end{pmatrix}\,,
\end{equation} having omitted the subscript E in the expectations. This matrix is known as Reynolds Stress Tensor, which arises when introducing the decomposition \eqref{Reynolds} into the Navier Stokes Equations and then ensemble averaging. The result from these operations are the URANS equations; for an incompressible flow with constant properties and negligible volume forces reads:

\begin{equation}
\label{URANS}
\rho(\partial_t \langle\bm{u}\rangle+ \langle\bm{u}\rangle\nabla \langle\bm{u}\rangle)= \nabla \langle p\rangle+\mu\nabla^2\langle\bm{u}\rangle-\rho \nabla \cdot \bm{R}\,,   
\end{equation} where $\rho$ and $\mu$ are the density and dynamic viscosity, $p$ is the pressure, and the last term models the effects of turbulence on the mean flow $\langle \bm{u} \rangle$, and its mathematically equivalent to additional stress. 

The form of the Reynolds stress gives, therefore, information about the intensity of the turbulent fluctuations and the direction across which turbulence is most prominently increasing the mean flow stresses. Turbulence intensity is usually expressed in terms of root mean square of the fluctuation $u_{rms}$, computed from the turbulent kinetic energy $\kappa=\{\bm{R}\}/2$, with $\{\}$ denoting the trace of a matrix, and a reference velocity $U$:

\begin{equation}
\mbox{TI}=\frac{u_{rms}}{U} = \frac{1}{U} \sqrt{\frac{2}{3} \kappa}\,.
\end{equation}

The Reynolds stress can be used to characterize various \textbf{states of turbulence}. The simplest state is that of \textbf{isotropic turbulence}, in which turbulent fluctuations are directionally independent and have no preferential orientation. In this case, the Reynolds stress is diagonal, with all components equal $u_{rms}=\langle u'^2\rangle=\langle v'^2\rangle=\langle w'^2\rangle$. A simple way to characterize the level of anisotropy in the flow is the anisotropy stress tensor

\begin{equation}
\bm{A}=\frac{1}{2 \kappa} \bm{R}-\frac{1}{3}\bm{I}\,,
\end{equation} where $\bm{I}$ is the identity matrix. This is identically zero in the case of isotropic turbulence, hence its Frobenious norm $||\bm{A}||_F=\sqrt{\sum_i\sum_j \bm{A}_{i,j}}$ gives a first quantitative measure of the level of anisotropy. More insights on the preferential directionality of turbulence fluctuations can be obtained by the eigenvalue decomposition of the anisotropic tensor matrix \citep{emory2014visualizing}. These can be used to compute two coordinates in a 2D map, known as an invariant map, from which various states of turbulence can be visualized. A popular map is the so-called Lumley triangle \citep{lumley1978computational}, which uses the second and third principal components of turbulence anisotropy; the coordinates in this plane are:

\begin{equation}
\mbox{II}=\lambda^2_1+\lambda_1\lambda_2+\lambda^2_2 \quad \mbox {and}\quad \mbox{III}=-\lambda_1 \lambda_2 (\lambda_1+\lambda_2)\,,
\end{equation} with $\lambda_1>\lambda_2>\lambda_3$ the ordered eigenvalues of the anisotropy tensor. Three limiting states can be found in this plane (see \cite{emory2014visualizing}): 

\begin{enumerate}
\item \textbf{One-component turbulence}: Fluctuations only exists along one direction. This occurs if $\lambda_i=[2/3,-1/3,-1/3]^T$. This condition is represented by $\bm{x}_{1C}$ in all invariant maps.

\item \textbf{Axisymmetric two-components}: Fluctuations have two leading directions with equal magnitude. This occurs if $\lambda_i=[1/6,1/6,-1/3]^T$. This condition is represented by $\bm{x}_{2C}$ in all invariant maps.

\item \textbf{Isotropic turbulence}: Fluctuations are equally relevant in all directions (spherical turbulence). As previously mentioned, the anisotropic tensor is identically zero; hence, one has $\lambda_i=[0,0,0]^T$. This condition is represented by $\bm{x}_{3C}$ in all invariant maps.

\end{enumerate}

These points can be joined to identify the boundaries of the invariant map, which also correspond to different physical behaviors. In particular:

\begin{enumerate}
\item Joining $\bm{x}_{1C}$ and $\bm{x}_{3C}$: this occurs when $0<\lambda_1<1/3$ and $-1/6<\lambda_2=\lambda_3<0$ and corresponds to an axisymmetric expansion.
\item Joining $\bm{x}_{2C}$ and $\bm{x}_{3C}$: this occurs when $-1/3<\lambda_1<0$ and $0<\lambda_2=\lambda_3<1/6$ and corresponds to an axisymmetric contraction.
\item Joining $\bm{x}_{1C}$ and $\bm{x}_{2C}$: this occurs when $\lambda_1+\lambda_3=1/3$ and $\lambda_2=-1/3$ and corresponds to the line of two components turbulence.

\end{enumerate}

\begin{figure}[h]
    \centering
    \includegraphics[width=0.6\linewidth]{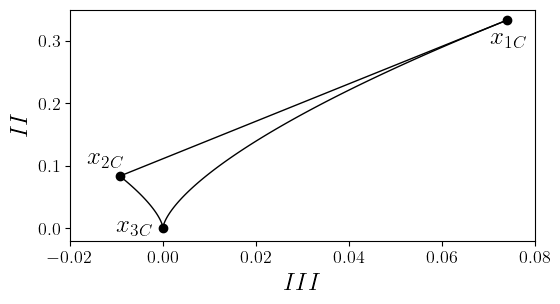}
    \caption{Lumley triangle and invariant map construction}
    \label{fig:lumley_triangle}
\end{figure}

Figure \ref{fig:lumley_triangle} plots the Lumley triangle, to which we return in the next exercise.

The properties of the Reynolds stress tensor can also be used to define realizability conditions for a turbulence model to be physical \cite{gerolymos2016algebraic}. The reader is referred to \cite{stiperski2018anisotropy,oberlack2002turbulent} for more details. 

Finally, we close with the special case in which the flow is \textbf{statistically stationary and ergodic}. In this case, as discussed in section \ref{sec_3_2}, ensemble averaging can be replaced by time averaging over a sufficiently long time series. Both the mean flow and the Reynolds stress are no longer a function of time, and all two points' statistics in time are solely functions of the time shift $\tau$ (equations \eqref{statisticS_2p}). This also implies that the autocorrelation of all velocity components at any given location vanishes at $\tau\rightarrow \infty$. For a vector-valued time series, it is convenient to define the time autocorrelation as 

\begin{equation}
R_{\bm{u}}(\bm{x},\tau)=\int^\infty_0 \bm{u}'(\bm{x},\tau)^T\bm{u}'(\bm{x},t+\tau) d\tau\,,
\end{equation} with $^T$ denoting transposition. One thus has that $R_{\bm{u}}(\bm{x},0)=||\bm{u}'(x)||_2^2$, with $||\bm{a}||_2$ the $l_2$ norm of a vector $\bm{a}$. Since $R_{\bm{u}}\rightarrow0$ for a stationary flow, it is possible to define an integral time scale as (see also \cite{Oliveira2007}) as 

\begin{equation}
	\Theta=\int^{\infty}_0\frac{R_{\bm{u}}(\bm{x},\tau)}{R_{\bm{u}}(\bm{x},0)} d\tau
	\,.
\end{equation}

It is worth stressing that this integral converges (and the definition makes sense) only if the autocorrelation function tends to zero. In practice, one should acquire a time series sufficiently long to see the autocorrelation becoming negligible.
The integral time scale measures the time after which the random process becomes uncorrelated with itself or, in a more pictorial interpretation, the time within which the variable `remembers' its history.

\subsection{Global statistics in space and time}

In Section \ref{sec_3_2}, the second-order statistics were aimed at linking two possible outcomes of the process at different times, with the outcome at each time being a random variable. In Section \ref{sec_4_1}, these were used to link the components of the vector field at a given location and a given time; again, each of these components is a random variable.  

In this section, we aim to use second-order statistics to study similarities in space and time. In space analysis, we seek to link the time series of the velocity fields sampled at different locations. We thus define the \textbf{spatial covariance function} as 

\begin{equation}
C_s(\bm{x},\bm{x}')=\int_{E} \bm{u}'(\bm{x},t)^T\bm{u}'(\bm{x}',t) f_{\bm{u}(x),\bm{u}(x')} d\bm{u}\,.
\end{equation} This definition generalizes \eqref{covariance_u_v}: the integral is carried out over all the possible set of time series occurring at location $\bm{x}$ and $\bm{x}'$ and $f_{\bm{u}(x),\bm{u}(x')}$is the (unknown) joint probability distribution. The inner product is used to obtain a scalar function from the vector-valued samples. For a stationary and ergodic process, leveraging \eqref{ergodicity}, one can exchange ensemble averaging with time averaging and obtain the \textbf{spatial covariance function}: 

\begin{equation}
\label{cov_C_function}
C_s(\bm{x},\bm{x}')=\frac{1}{T}\int_{T} \bm{u}'(\bm{x},t)^T\bm{u}'(\bm{x}',t) dt\,.
\end{equation}

A special property of turbulent flows is that of \textbf{homogeneity}: in homogeneous turbulence, the spatial covariance function solely depends on the distance between the two points considered, hence $C_s(\bm{r})=C_s(\bm{x},\bm{x}+\bm{r})$. In homogeneous and isotropic turbulence, it can be shown that it solely depends on the magnitude $||\bm{r}||$ of the shift and not its direction. The autocovariance, in this case, can be used to define an integral length scale similarly to the integral time scale introduced earlier.

A \textbf{vector-based  variant of the spatial correlation matrix} can be defined as follows

\begin{equation}
\label{cov_C_v_function}
\bm{C}(\bm{x},\bm{x'})=\begin{bmatrix}
\mbox{cov}_T(u(\bm{x}),u(\bm{x'})) & \mbox{cov}_T(u(\bm{x}),v(\bm{x'})) & \mbox{cov}_T(u(\bm{x}),w(\bm{x'})) \\
\mbox{cov}_T(v(\bm{x}),u(\bm{x'})) & \mbox{cov}_T(v(\bm{x}),v(\bm{x'})) & \mbox{cov}_T(v(\bm{x}),w(\bm{x'})) \\
\mbox{cov}_T(w(\bm{x}),u(\bm{x'})) & \mbox{cov}_T(w(\bm{x}),v(\bm{x'})) & \mbox{cov}_T(w(\bm{x}),w(\bm{x'})) \,.
\end{bmatrix}
\end{equation}

The reader should remark the similarity with the Reynolds stress tensor defined in Section \ref{sec_4_1}: indeed, the only difference is that this is now a function of two locations while the Reynolds stress tensor is defined at one location. Both are second order statistics, but $\bm{R}(\bm{x}_i)$ is a covariance matrix comparing the velocity components at location $\bm{x}_i$ while $\bm{C}(\bm{x},\bm{x'})$ is a covariance comparing velocity components at locations $\bm{x}$ and $\bm{x}'$.

In the time analysis, we seek to link the velocity fields sampled at two different times. We thus define the \textbf{temporal covariance function} as

\begin{equation}
K(t,t')=\int_{E} \bm{u}'(\bm{x},t)^T\bm{u}'(\bm{x},t') f_{\bm{u}(t),\bm{u}(t')} d\bm{u}\,.
\end{equation}

Again, this ensemble operator is a generalization of \eqref{covariance_u_v}. This time, we are considering the set of all possible fields that could occur in two time steps.
As for the previous operator, we might invoke spatial ergodicity: assuming that the domain is sufficiently large to display all the possible outcomes of the time series involved, we could replace ensemble averaging with spatial averaging to obtain:

\begin{equation}
\label{cov_K_function}
K(t,t')=\frac{1}{|\bm{\Omega}|}\int_\Omega \bm{u}'(\bm{x},t)^T\bm{u}'(\bm{x},t') d\bm{\Omega}\,,
\end{equation} where $\Omega$ is the domain under investigation and $|\Omega|$ is an appropriate measure of it; this means an area in a 2D domain or a volume in a 3D domain. For a stationary flow, this operator only depends on the time lag between the two instances considered, i.e. $K(\tau)=K(t,t+\tau)$.

In the same way, the eigendecomposition of the covariance matrix in section \ref{sec_3_2} gives information about the leading direction of turbulent mixing, the eigendecomposition of the covariance operators $C$ and $K$ give important information about the existence of coherent patterns in the data. The study of these is the subject of modal analysis in section \ref{sec_7}. Before proceeding with the computation of these operators and their sample definition, it is time to discuss the challenges in treating scattered datasets.

\section{Local statistics of a flow field}\label{sec6}

The computation of all quantities described in the previous sections for the case where data is available on a grid is textbook material. We thus start from there in section \ref{sec_6_1}. The computation for the case of scattered data requires additional work and definitions. We first briefly review the traditional approaches in section \ref{sec_6_2}. Then, we take a brief detour into the realm of physics-constrained regression in section \ref{sec_6_2}, before delving into the proposed meshless computation of statistics in section \ref{sec_6_3}.

\subsection{Gridded data}\label{sec_6_1}

The data is provided on a structured grid, denoted as $\bm{x}_{\bm{i}} = (x_m, y_n, z_l)$, with $m = 1, \dots, n_x$, $n = 1, \dots, n_y$, and $l = 1, \dots, n_z$. Thus, the grid consists of $n_p = n_x n_y n_z$ points, which we can index as $\bm{i} = 1, \dots, n_p$. For each point, it is possible to define a volume (or area, in 2D) $\Delta \Omega_{\bm{i}}$ based on the half-distance to the neighboring points. It is important to note that the grid does not need to be uniform for all the operations defined in this section. However, we assume that the \textbf{grid is fixed}, meaning that the data is collected on the same points for each snapshot. We assume that these are available on a time discretization $\mathbf{t_k}$, with $k=1,\dots n_t$. Again, this is not necessarily uniformly sampled, but we can define a time interval $\Delta t_k$ within which a given snapshot $k$ is ``representative''.

To ease computations, it is particularly convenient to store all the data in the form of snapshot matrices. For this, we reshape the velocity components at a given snapshot $k$ as a column vector. The associated snapshot matrices are thus $\mathbf{U}, \mathbf{V}, \mathbf{W}\in\mathbb{R}^{n_p\times n_t}$.

Each column collects a snapshot at time $k$ and each row contains a time series sampled at the grid point $\bm{i}$, located at $\bm{x}_i$. All statistics involving the time domain are essentially statistics along the rows of the snapshot matrices. Python offer efficient functions to compute these, as we illustrate in the following exercise.

All computations are easy as long as the statistics are computed on the set of grid points in common to all snapshots. The discrete approximation of the integrals for the average and covariances become (see section \ref{sec_3_2}): 

\begin{equation}
\langle \bm{u}\rangle(\bm{x}_i)=\sum^{n_t}_{k=1} \bm{u}(\bm{x}_i,\mathbf{t_k})\frac{\Delta t_k}{T}\quad \mbox{and} \quad 
\langle \bm{u}'_l, \bm{u}'_m \rangle(\bm{x}_i)= \sum^{n_t}_{k=1} \bm{u}_l'(\bm{x}_i,\mathbf{t_k})\bm{u}_m'(\bm{x}_i,\mathbf{t_k})\frac{\Delta t_k}{T}\,,
\end{equation} with $\Delta t_k/T=1/n_t$ in case of equally spaced samples in time. In Python, the Reynolds stresses for a 2D flow on uniformly sampled data can be computed as follows

\begin{centering}
	\begin{lstlisting}[language=Python,linewidth=16cm,xleftmargin=.05\textwidth,xrightmargin=.05\textwidth,backgroundcolor=\color{yellow!10}]
def compute_row_statistics(U, V):
 # Compute row-wise means
 mean_U = np.mean(U, axis=1)
 mean_V = np.mean(V, axis=1)    
 # Compute row-wise variances
 R_uu = np.var(U, axis=1, ddof=1)
 R_vv = np.var(V, axis=1, ddof=1)
 # Compute row-wise covariances
 R_uv = np.mean((U - mean_U[:, None]) * \
        (V - mean_V[:, None]), axis=1)    
 return mean_U, mean_V, R_uu, R_vv, R_uv
	\end{lstlisting}
\end{centering}

Further processing of the Reynolds stress matrix as described in section \ref{sec_4_1} can be carried out using standard routines as showcased in the following exercise. Finally, the covariance functions in space and time in \eqref{cov_C_function} and \eqref{cov_K_function} are now samples on $n_p$ and $n_t$ points, respectively, and thus become \textbf{covariance matrices}, build out of approximation of the integrals based on the available data. 

The (scalar) covariance matrix in space, i.e. the discrete version of \eqref{cov_C_function}, is:

\begin{equation}
\mathbf{C}_s[\bm{x}_l,\bm{x}_m]=\mathbf{C}_{l,m} =  \sum^{n_t}_{k=1} \bm{u}^T(\bm{x_l},\mathbf{t_k}) \bm{u}(\bm{x_m},\mathbf{t_k})\frac{\Delta t_k}{T} \,.
\end{equation}

Defining as $\mathbf{w}_{k,T}=\Delta t_k/T$ a set of weights, collected on a diagonal matrix $\mathbf{W}_{w,T}=\mbox{diag}(\mathbf{w}_{k,T})$, this matrix can be conveniently computed using matrix multiplication from the snapshot matrices as follows:

\begin{equation}
\label{Cov_C}
\mathbf{C}_s=\mathbf{U}\mathbf{W}_w\mathbf{U}^T+\mathbf{V}\mathbf{W}_w\mathbf{V}^T+\mathbf{W}\mathbf{W}_w\mathbf{W}^T \in\mathbb{R}^{n_p\times n_p}\,.
\end{equation}

On the other hand, the discrete version of \eqref{cov_C_v_function} reads

\begin{equation}
\mathbf{C}=\begin{bmatrix}
\mathbf{U}\mathbf{W}_{w,T}\mathbf{U}^T & \mathbf{U}\mathbf{W}_{w,T}\mathbf{V}^T & \mathbf{U}\mathbf{W}_{w,T}\mathbf{W}^T \\
\mathbf{V}\mathbf{W}_{w,T}\mathbf{U}^T & \mathbf{V}\mathbf{W}_{w,T}\mathbf{V}^T & \mathbf{V}\mathbf{W}_{w,T}\mathbf{W}^T \\
\mathbf{W}\mathbf{W}_{w,T}\mathbf{U}^T & \mathbf{W}\mathbf{W}_{w,T}\mathbf{V}^T & \mathbf{W}\mathbf{W}_{w,T}\mathbf{W}^T
\end{bmatrix} \in\mathbb{R}^{3 n_p \times 3 n_s} \,.
\end{equation}

Finally, the temporal correlation matrix \eqref{cov_K_function} becomes

\begin{equation}
\label{Cov_C}
\mathbf{K}=\mathbf{U}^T\mathbf{W}_{w,S}\mathbf{U}+\mathbf{V}^T\mathbf{W}_{w,S}\mathbf{V}+ \mathbf{W}^T\mathbf{W}_{w,S}\mathbf{W}\in\mathbb{R}^{n_t\times n_t}\,,
\end{equation} with $\mathbf{W}_{w,S}=\mbox{diag}(w_{k,S})$ and $\mathbf{w}_{k,S}=|\Delta \Omega|_k/|\Omega|$ the weights in space.

\begin{tcolorbox}[breakable, opacityframe=.1, title=Exercise 3: Statistics of a Turbulent Flow from Gridded data]

Consider the PIV data from test case \ref{sec_2_2}. We are interested in computing the (1) the mean flow, (2) the turbulent kinetic energy, (3) the Reynolds stresses and the norm of the anisotropic tensor (4) locate the turbulent states observed in the data in the Lumley triangle. What kind of turbulence is found there? 

We stress that Reynolds stresses play a crucial role in describing the mixing and momentum transfer that occur within the jet as it spreads into the surrounding fluid.

\hspace{2mm}Note that these computations require some important assumption on the flow, since only two out of the three velocity components are measured. We here assume that the flow is perfectly axisymmetric. Therefore, denoting as $(u,v,w)$ the velocity components in cylindrical coordinate along the $(x,r,\theta)$ directions (stream-wise, radial and angular), the Reynolds stress takes the form (see \cite{Pope2000})
    
\begin{equation}
	\label{Reynolds_STRESS_axis}
\bm{R}(\bm{x})= \begin{pmatrix}
		\langle u'^2 \rangle(\bm{x}) & \langle u' v' \rangle(\bm{x})  & 0\\
		\langle v' u' \rangle (\bm{x})& \langle v'^2(\bm{x}) \rangle (\bm{x})& 0\\
        0 & 0 & \langle w'^2\rangle (\bm{x})
	\end{pmatrix}\,,
\end{equation}

The circumferential symmetry ensures that all cross terms involving azimuthal components are zero. At the centerline, the radial and circumferential components become indistinguishable, allowing the assumption that $\langle v\rangle = \langle w\rangle $. While this equivalence does not generally hold at greater distances from the centerline, we assume it to be valid for the purposes of this exercise.

\vspace{2mm}

\textbf{Solution}. Once the data is organized into snapshot matrix, the computation of the mean is trivial. However, note that this datasets contains a large number of NaNs, hence a first logical check on the validity of each vector is performed. The script becomes:
    
\begin{centering}
\begin{lstlisting}[language=Python,linewidth=16cm,xleftmargin=.05\textwidth,xrightmargin=.05\textwidth,backgroundcolor=\color{yellow!10}]
# Step 1: Mean velocity field
# valid vectors
valid = np.logical_and(np.isfinite(U),
                       np.isfinite(V))
# number of valid vectors
valid_sum = valid.sum(axis=0)
# perform the average on the valid vectors only
U_mean = np.nansum(U * valid, axis=0) / valid_sum
V_mean = np.nansum(V * valid, axis=0) / valid_sum    
\end{lstlisting}
\end{centering}
    
Figure \ref{UV_mean_PIV} shows the $u$ and $v$ component of the average velocity field obtained from the PIV data.
    
    \begin{center}%
        \captionsetup[sub]{labelformat=parens}
        \captionsetup{type=figure}
        \includegraphics[width=0.55\linewidth]{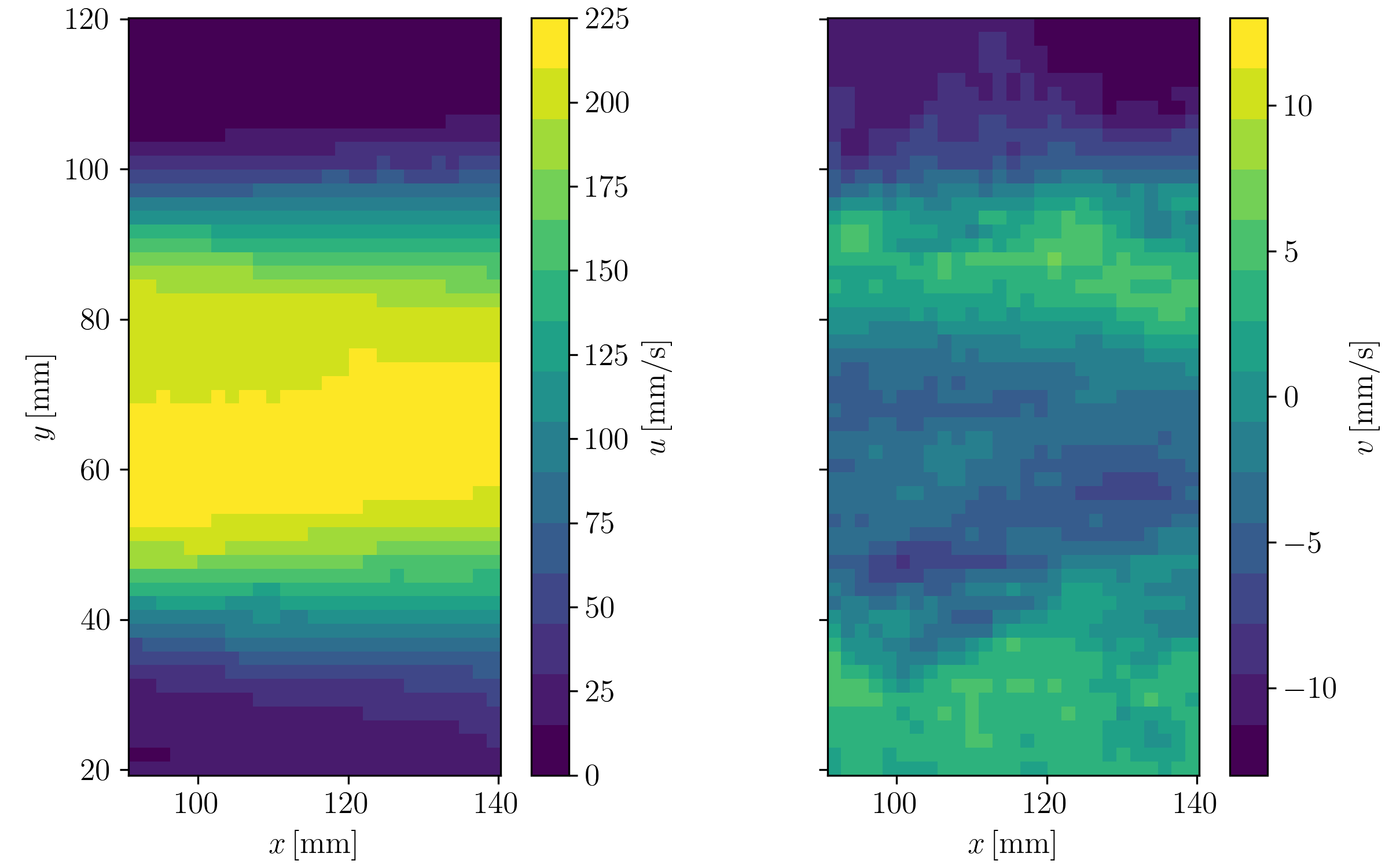}
        \vspace{-3mm}
        \caption{Average velocity field from PIV data, $u$ (left) and $v$ (right) components.}
        \vspace{-0.05cm}
        \label{UV_mean_PIV}
    \end{center}

The computation of the fluctuation component, Reynolds stress tensor, turbulent kinetic energy, anisotropic stress tensor and its norm are given by the following script:

\begin{centering}
\begin{lstlisting}[language=Python,linewidth=16cm,xleftmargin=.05\textwidth,xrightmargin=.05\textwidth,backgroundcolor=\color{yellow!10}]
# Step 2: Turbulent kinetic energy
# compute fluctuation fields
U_prime = U - U_mean[np.newaxis, :]
V_prime = V - V_mean[np.newaxis, :]
# compute average fluctuations
uu_mean = np.nansum(U_prime * U_prime * valid, axis=0) / 
    (valid_sum - 1)
vv_mean = np.nansum(V_prime * V_prime * valid, axis=0) / 
    (valid_sum - 1)
uv_mean = np.nansum(U_prime * V_prime * valid, axis=0) /
    (valid_sum - 1)
# compute mean TKE
fill_value = np.zeros_like(uu_mean)
# assemble the Reynolds stress tensor
R_ij = np.array([
    [uu_mean,       uv_mean,        fill_value  ],
    [uv_mean,       vv_mean,        fill_value  ],
    [fill_value,    fill_value,     vv_mean     ]
    ])
# compute turbulent kinetic energy
k = np.sum(np.diagonal(R_ij), axis=2) / 2
# Step 3: Anisotropic tensor
# components of the anisotropic tensor
A_ij = R_ij / (2*k[np.newaxis, np.newaxis, :]) - 
np.diag(np.full(3, 1/3))[:, :, np.newaxis, np.newaxis]
# norm of the anisotropic tensor
A_norm = np.linalg.norm(A_ij, axis = (0,1))
\end{lstlisting}
\end{centering}

Note that the python object R\_ij collects the Reynolds stress tensor in all entries of the domain (this is a tensor of size $3\times 3\times 72\times 71$). The left side of figure \ref{TKE_A_PIV} shows the turbulent kinetic energy $k$ and the turbulence intensity to give an idea of the turbulence level. The right part shows the norm of the anisotropic tensor $||\bm{A}||$ for PIV data.
    
    \begin{center}%
        \captionsetup[sub]{labelformat=parens}
        \captionsetup{type=figure}
        \includegraphics[height=0.32\linewidth]{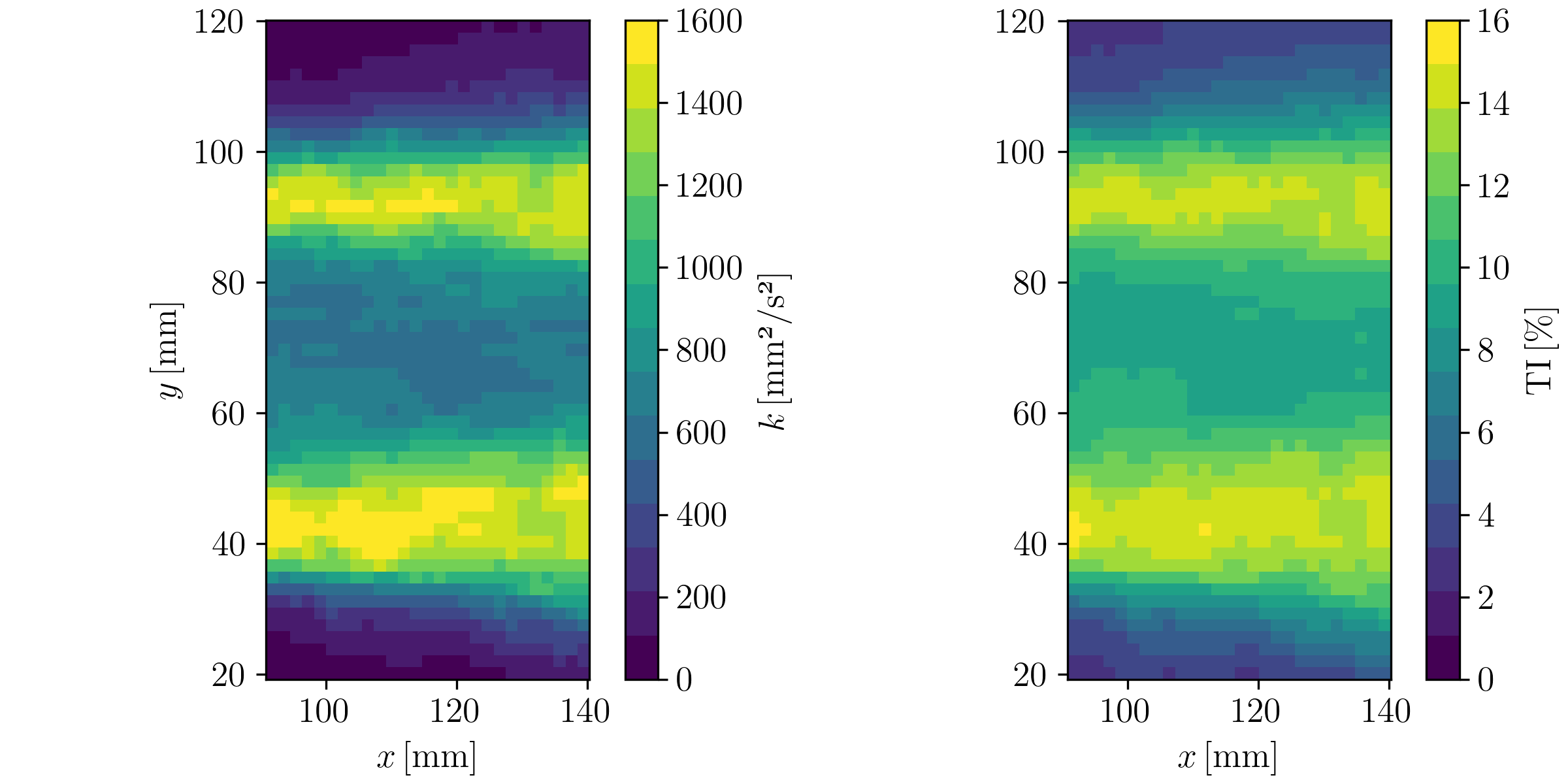}
        \includegraphics[height=0.32\linewidth]{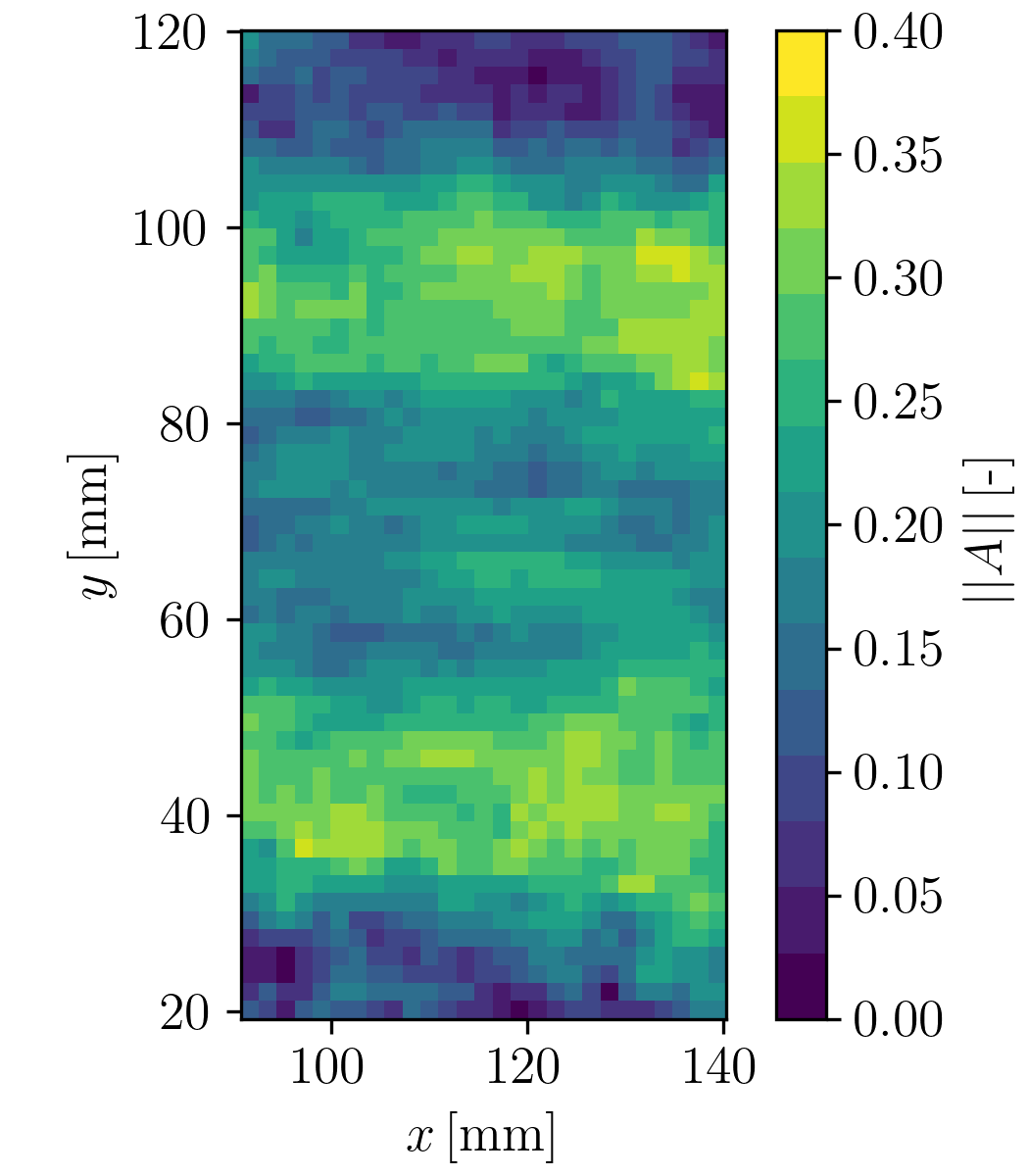}
        \vspace{-3mm}
        \caption{Average turbulent kinetic energy $k$ (left), turbulence intensity (center) and norm of the anisotropic tensor (right) from the PIV data.}
        \vspace{-0.05cm}
        \label{TKE_A_PIV}
    \end{center}

The shear layers in the flow are clearly visible. These are regions of large turbulence intensity and large anisotropicity. Traditional low fidelity turbulence models, based on the Boussinesq Hypothesis and the notion of eddy viscosity, generally have difficulties in describing anisotropicity. These model seek to capture the effects of anisotropicity in an indirect way, namely through the link to the gradients of the mean flow; but such a link does not always hold  in complex scenarios \citep{Pope2000,Davidson2004}.

\hspace{3mm}Finally, the computation required to position the dataset in the invariant map is provided below:

    \begin{centering}
\begin{lstlisting}[language=Python,linewidth=16cm,xleftmargin=.05\textwidth,xrightmargin=.05\textwidth,backgroundcolor=\color{yellow!10}]
# Step 4: Lumley triangle
# Boundaries of the invariant map
x_1C = np.array([2/3, -1/3, -1/3])
x_2C = np.array([1/6, 1/6, -1/3])
x_3C = np.array([0, 0, 0])

# II coordinate of the triangle
II_1C = x_1C[0]**2 + x_1C[0]*x_1C[1] + x_1C[1]**2
II_2C = x_2C[0]**2 + x_2C[0]*x_2C[1] + x_2C[1]**2
II_3C = x_3C[0]**2 + x_3C[0]*x_3C[1] + x_3C[1]**2

# III coordinates of the triangle
III_1C = -x_1C[0]*x_1C[1] * (x_1C[0] + x_1C[1])
III_2C = -x_2C[0]*x_2C[1] * (x_2C[0] + x_2C[1])
III_3C = -x_3C[0]*x_3C[1] * (x_3C[0] + x_3C[1])

# Number of points to draw the limiting curves
n_p = 101

# Curve from 1 to 3
x_13 = np.array([
    np.linspace(0, 2/3, n_p),
    np.linspace(0, -1/3, n_p),
    np.linspace(0, -1/3, n_p),
    ])

# Convert into II and III coordinates
II_13 = x_13[0, :]**2 + x_13[0, :]*x_13[1, :] + x_13[1, :]**2
III_13 = -x_13[0, :]*x_13[1, :] * (x_13[0, :] + x_13[1, :])

x_23 = np.array([
    np.linspace(0, -1/3, n_p),
    np.linspace(0, 1/6, n_p),
    np.linspace(0, 1/6, n_p)
    ])

II_23 = x_23[0, :]**2 + x_23[0, :]*x_23[1, :] + x_23[1, :]**2
III_23 = -x_23[0, :]*x_23[1, :] * (x_23[0, :] + x_23[1, :])

x_12 = np.array([
    np.linspace(2/3, 1/6, n_p),
    np.linspace(-1/3, -1/3, n_p),
    1/3 - np.linspace(2/3, 1/6, n_p)
    ])

II_12 = x_12[0, :]**2 + x_12[0, :]*x_12[1, :] + x_12[1, :]**2
III_12 = -x_12[0, :]*x_12[1, :] * (x_12[0, :] + x_12[1, :])

# No need for the 2 last dimensions of the array (x and y dimensions)
# A is flattened into a 3x3xn_vectors array
a_ij = np.reshape(A_ij, [3, 3, np.shape(A_ij)[2]*np.shape(A_ij)[3]]).T
# Compute eigenvalues of the anisotropic tensor
eig_vals = np.linalg.eigvals(a_ij).T
eig_vals = eig_vals - eig_vals.mean(axis=0)[np.newaxis, :]
eig_vals = np.sort(eig_vals, axis=0)[::-1, :]

# Convert to II and III coordinates
II = eig_vals[0, :]**2 + 
     eig_vals[0, :] * eig_vals[1, :] + 
     eig_vals[1, :]**2
III = -eig_vals[0, :] * eig_vals[1, :] * 
      (eig_vals[0, :] + eig_vals[1, :])

\end{lstlisting}
\end{centering}   

The resulting Lumley triangle is shown in Figure \ref{Lumley_A_PIV}. Most of the points in the far field and in the jet centerline are close to the $\mathbf{x}_{3C}$ state, hence close to anisotropic turbulene. The region in the shear layer is close to the $\bm{x}_{3C}-\bm{x}_{1C}$ corresponding to an axisymmetric contraction. This state of turbulence occurs when turbulence is ``compressed'' axially while spreading radially. As a result, the Reynolds stress tensor has nearly equal components on radial and circumferential direction and a smaller component along the axial one. This also occurs in nozzle flows.

    \begin{center}%
        \captionsetup[sub]{labelformat=parens}
        \captionsetup{type=figure}
        \includegraphics[height=0.35\linewidth]{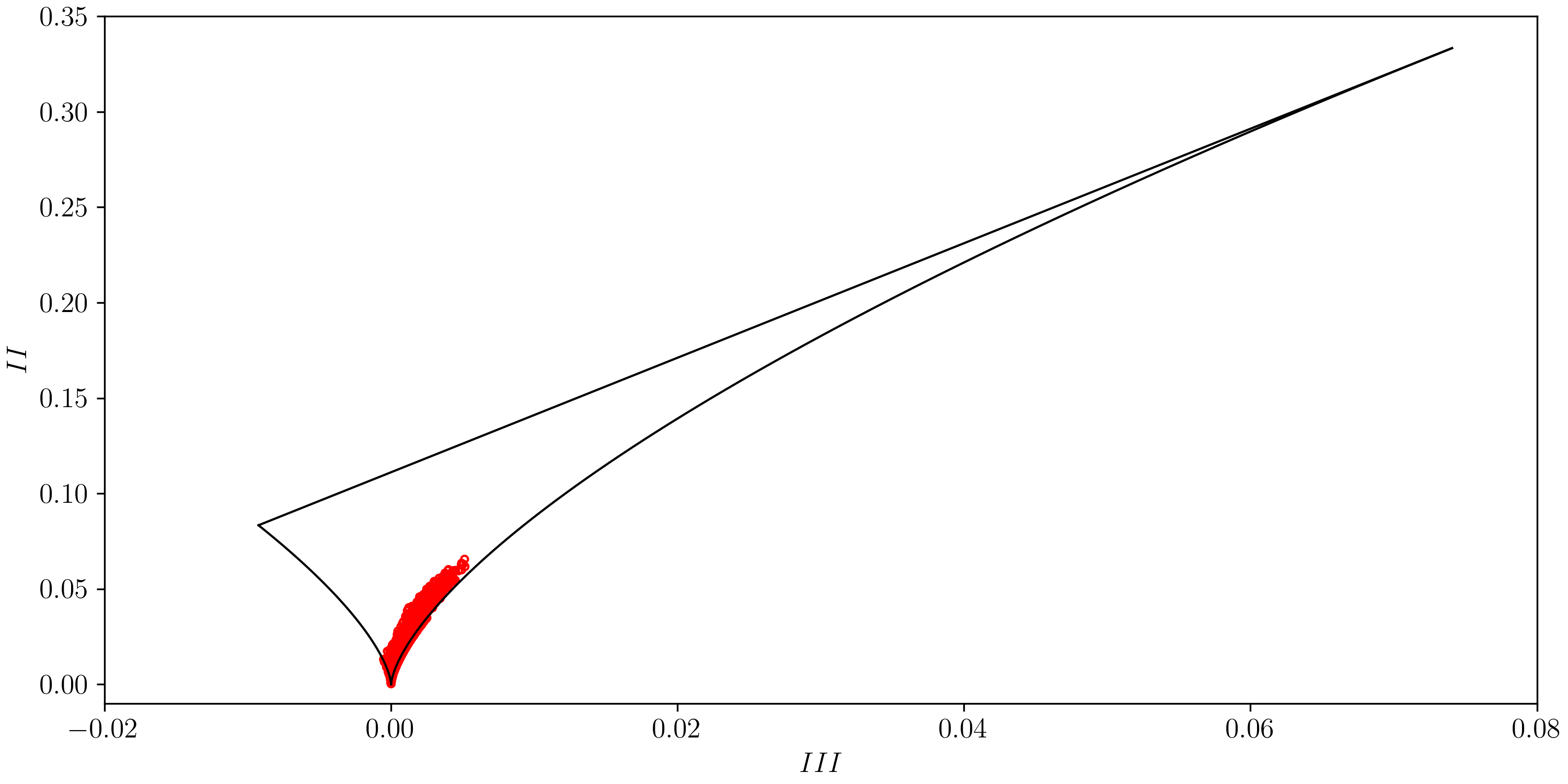}
        \vspace{-3mm}
        \caption{Lulmey triangle for the PIV data.}
        \vspace{-0.05cm}
        \label{Lumley_A_PIV}
    \end{center}

\end{tcolorbox}

\subsection{Traditional binning methods}
\label{sec_6_2_a}

The main challenge with randomly scattered data is that the velocity fields are available at different locations for each snapshot. Let $\mathbf{U}^{(j)}(\mathbf{X}^{(j)})$ denote the velocity field in snapshot $j$, with $\mathbf{X}^{(j)}=[\mathbf{x}^{(j)},\mathbf{y}^{(j)},\mathbf{z}^{(j)}]\in\mathbb{R}^{n^{(j)}_p\times 3}$ the matrix that collects the coordinate at which the velocity information is available and $n^{(j)}_p$ the number of measurement points at each snapshot. It is not rare to have that $\mathbf{X}^{(j)}\cap \mathbf{X}^{(l)}=\emptyset$ for all $j,l$ in $n_t$: in words, not two snapshots share the same sampling location within a dataset of $n_t$ snapshot. Considering for example the mean flow and recalling that this is defined as 

\begin{equation}
\langle \bm{u}\rangle(\bm{x})=\int^{\infty}_{-\infty} \bm{u}(\bm{x}) f_u(\bm{x},\bm{u}) d\bm{u}\,,
\label{eq:mean_definition}
\end{equation} with $f_u$ the joint probability function assigning a probability to a velocity vector occurring at a given location, not having sufficient samples at every location renders the problem of pdf estimation impractical.

The simplest way to compute statistics is to define a grid of "bins", that is areas where statistics are associated with a specific point called the bin center. This underpins the concept of ensemble PTV (EPTV, \citealt{Kaehler2012a}). Defining as $\bm{x_i}$ the location of a bin, one can build the following estimate for the mean flow 

\begin{equation}
\langle\mathbf{u}\rangle(\mathbf{x}_i)\approx  \frac{1}{n_{p,i}} \sum^{n_{p,i}}_{i=1} \bm{U}^{(j)}(\mathbf{x}_i)\,,
\end{equation} where $\bm{U}^{(j)}(\mathbf{x}_i)$ denotes the mapping of the PTV sample $\bm{U}^{(j)}$ onto the $i$-th bin, and $n_{p,i}$ denotes the number of measurement points available within the bin. The definition of expectation in this binned formalism can be easily extended to higher order statistics.

When a sufficiently large number of particles pass through a bin, the distribution within the bin can approximate the local flow distribution \citep{Kaehler2012a}. Binning methods differ in how they compute statistics: the "top-hat" approach assigns equal weight to all samples within a bin, while Gaussian weighting \citep{Aguei1987} prioritizes samples closer to the bin center. A significant challenge lies in the propagation of errors or unresolved gradients in the mean flow to higher-order statistics, as highlighted by \cite{Agueera2016}. This issue can be mitigated by employing local polynomial fits for the mean flow.

In these notes, we present an alternative approach recently proposed by \cite{Ratz2024}. This method is "meshless," as it avoids the need for a grid to compute derivatives, and "binless," as it eliminates the need for bins to compute local statistics. The idea is to combine the physics-constrained RBF formalism proposed in \cite{Sperotto2022a} with an ensemble trick for the regression of flow statistics. Before presenting the method in \ref{sec_6_3}, we propose a brief review of physics-constrained RBF in \ref{sec_6_2}.

\subsection{Fundamentals of (constrained) Radial Basis Functions (RBFs)}\label{sec_6_2}

Regression via Radial Basis Functions (RBFs) consists in building a target function from a set of scattered data points using a linear combination of real-valued functions that depends only on the distance from a center point. A classic RBF is the Gaussian\footnote{RBFs are usually denoted with $\phi$ or $\psi$, but in these notes, both symbols are taken for the spatial and temporal structures in modal analysis in the next section. May the readers with experience on RBF forgive us for the use of $\gamma$!}

\begin{equation}
\label{gauss}
\gamma(\bm{x}| \bm{x}_n^*, c_n)=\exp \left(-c_n^{2}\left\|\bm{x}-\bm{x}_{n}^{*}\right\|^{2}\right)
\end{equation} where $c_n$ is the {\em shape parameter} and $\bm{x}^*_k\in\mathbb{R}^{d}$ is the {\em collocation point}. The reader is referred to \cite{Fornberg2015,hoffman2018numerical,buhmann2003radial,trefethen2013approximation} for other bases and to learn more about RBFs.

The RBF approximation of a generic scalar function $f(\bm{x})$, $f:\mathbb{R}^{d}\rightarrow \mathbb{R}$ is therefore: 

\begin{equation}
\label{RBF_REG}
f(\bm{x})=\sum^{n_b}_{n=1} \bm{w}_n \gamma(\bm{x}| \bm{x}_n^*, c_n)=\sum^{n_b}_{n=1} \bm{w}_n \gamma_n(\bm{x})\,,
\end{equation} having introduced the compact notation for the n-th basis. 

In traditional RBF regression, both $c_n$ and $\bm{x}_n^*$ are pre-assigned based on the data distribution. This is typically done to ensure that the basis functions are well distributed across the domain and that each basis has sufficient data points within its region of influence\footnote{See \cite{Sperotto2022a} for a cluster-based approach to the collocation problem.}, commonly defined as the area where $\gamma_n > 0.5$. 

Once all the bases are assigned, the RBF regression consists in identifying the weights of the linear combination $\{\bm{w}_n\}^{n_b}_{n=1}$, which we here collect in a vector $\mathbf{w}\in\mathbb{R}^{n_b}$. The key advantage over more sophisticated regression methods (see \citealt{Mendez2024b} for a general introduction) is that the function approximation \eqref{RBF_REG} depends linearly on the parameters $\mathbf{w}\in\mathbb{R}^{n_b}$. Given a set of training data $\{\mathbf{x}^*_i,\mathbf{f}_i\}^{n_*}_{i=1}$, which we store in a matrix $\mathbf{X}_*\in\mathbb{R}^{n_*\times d}$ and a vector $\mathbf{f}\in\mathbb{R}^{n_*}$, these weights are usually computed as those that minimize the following cost function

\begin{equation}
\label{J}
J(\mathbf{w})= ||\mathbf{f}-\bm{\Gamma}(\mathbf{X}_*)\mathbf{w}||^2_2 + \alpha ||\mathbf{w}||^2_2\,,
\end{equation} where $\bm{\Gamma}(\mathbf{X}_*)\in\mathbb{R}^{n_*\times n_b}$ is the matrix collecting in each column the values of a given basis on the training data $\mathbf{X}_*$ and $\alpha\in\mathbb{R}^+$ is a user defined parameter controlling the weight of the penalty. In the classic probabilistic interpretation of the regression, the weights minimizing \eqref{J} provide the ``Maximum A Posteriori" (MAP) estimate, that is the most likely set of vectors combining a prior assumption\footnote{Here $\mathcal{N}(\mathbf{\mu},\bm{\Sigma})$ is a multivariate Gaussian with mean $\mathbf{\mu}$ and covariance matrix $\bm{\Sigma}$, while $\bm{I}$ is the identity matrix of appropriate size.} $\mathbf{w}\sim \mathcal{N}(\mathbf{0},\alpha^{-1}\bm{I})$ with the available data \citep{Deisenroth}. This is also known as Ridge Regression.

The minimization of \eqref{J} leads to a linear system which offers an analytic solution:

\begin{equation}
\label{Sol}
\bigl(\bm{\Gamma}^T(\mathbf{X}_*)\bm{\Gamma}(\mathbf{X}_*)+\alpha \bm{I}\bigr) \mathbf{w}=\bm{\Gamma}^T(\mathbf{X}_*) \mathbf{f}\quad \rightarrow
\mathbf{w}=\bigl(\bm{\Gamma}^T(\mathbf{X}_*)\bm{\Gamma}(\mathbf{X}_*)+\alpha \bm{I}\bigr)^{-1}\bm{\Gamma}^T(\mathbf{X}_*) \mathbf{f}\,.
\end{equation}

Nevertheless, it is worth pointing out that the inversion in \eqref{Sol} is numerically inefficient, and in practice the solution of the linear system is better carried out using Cholesky decomposition or the Conjugate Gradient (CG) method \citep{trefethen1997numerical}.

The linearity with respect to the model parameters (and the resulting quadratic dependency in \eqref{J})
facilitates the integration of \emph{quadratic penalties} and \emph{linear constraints} into the regression. Denoting these as 

\begin{equation}
||\mathbf{A}_p \mathbf{w}||^2_2\,\quad \mbox{and}\quad \mathbf{A}_c \mathbf{w}=\mathbf{c}\,,
\end{equation} respectively, with $\mathbf{A}_p\in\mathbb{R}^{n_b\times n_b}$ and $\mathbf{A}_c\in\mathbb{R}^{n_{\lambda}\times n_b}$, the penalized and constrained problem is the one that minimizes the augmented cost function:

\begin{equation}
\label{Aug}
A(\mathbf{w})=J(\mathbf{w})+\alpha_2 ||\mathbf{A}_p \mathbf{w}||^2_2 + \bm{\lambda}^T \bigl(\mathbf{A}_c \mathbf{w}-\mathbf{c}) \bigr)\,,
\end{equation} where $\bm{\lambda}\in\mathbb{R}^{n_\lambda}$ is the vector of (unknown) Lagrange multipliers required to enforce the constraints and $\alpha_2\in\mathbb{R}^+$ is an additional (user-defined) penalty parameter. The constrained RBF regression now takes the form of a traditional quadratic programming problem \citep{JorgeNocedal2006,Chong2013}: the solution gives the weights and multipliers $[\mathbf{w},\bm{\lambda}]$ minimizing \eqref{Aug}; by setting the gradient of $\eqref{Aug}$ with respect to these equal to zero, the minimization leads to a linear system 

\begin{equation}
    \label{Blocks}
    \left(  \begin{array}{cc}
      \boldsymbol{A}  & \boldsymbol{B} \\
      \boldsymbol{B^T}  & \boldsymbol{0}
    \end{array}\right) \left(\begin{array}{c}
      \boldsymbol{w}  \\
      \bm{\lambda} 
    \end{array}\right)=\left(\begin{array}{c}
      \boldsymbol{b}_1  \\
      \boldsymbol{b}_2
    \end{array}\right)\,,
\end{equation} with $\boldsymbol{A}=\boldsymbol{\Gamma}^T\boldsymbol{\Gamma}+\alpha_1 \bm{I}+\alpha_2 \mathbf{A}_p^T\mathbf{A}_p \in\mathbb{R}^{n_b\times n_b}$, $\boldsymbol{B}=\mathbf{A}_c^T \in\mathbb{R}^{n_b\times n_{\lambda}}$, $\boldsymbol{b_1}=\boldsymbol{\Gamma}^T \mathbf{f}\in \mathbb{R}^{n_b}$ and $\boldsymbol{b}_2=\mathbf{c}\in\mathbb{R}^{n_\lambda }$. The reader is referred to \cite{Sperotto2022a} and \cite{JorgeNocedal2006} for efficient methods to solve \eqref{Blocks}. 

In the context of RBF regression of tracking velocimetry, this framework allows to impose constraints such as boundary conditions (e.g. no slip or symmetry), compliance with sensor data or differential constraints such as divergece-free conditions. A detailed discussion on the constraint implementation is beyond the scopes of this introduction and the reader is referred to \cite{Sperotto2022a} for more details. An open-source library implementing this framework, currently under development at the von Karman Institute, has been released in \cite{Sperotto2024b}.

We close this section with the RBF regression of a velocity field (i.e. a vector valued function). This is here written as 

\begin{equation}
\begin{split}
\label{RBF_VEC}
\bm{u}(\mathbf{x},\mathbf{t_k}) = \begin{pmatrix}
    u(\mathbf{x}) \\
    v(\mathbf{x}) \\
    w(\mathbf{x})
\end{pmatrix} =
\sum_{n=1}^{n_b} \begin{pmatrix}
    \mathbf{w}_{u,n}(\mathbf{t_k})\, \gamma_n (\bm{x},\mathbf{t_k}) \\
    \mathbf{w}_{v,n}(\mathbf{t_k})\, \gamma_n (\bm{x},\mathbf{t_k}) \\
    \mathbf{w}_{w,n}(\mathbf{t_k})\, \gamma_n (\bm{x},\mathbf{t_k})
\end{pmatrix}
=&\sum_{n=1}^{n_b}
\begin{pmatrix}
    \mathbf{w}_{u,n}(\mathbf{t_k})\\
    \mathbf{w}_{v,n}(\mathbf{t_k})\\
    \mathbf{w}_{w,n}(\mathbf{t_k})
\end{pmatrix}\gamma_n (\bm{x},\mathbf{t_k})\\
&=
\sum_{n=1}^{n_b}\mathbf{w}_{n}(\mathbf{t_k})\gamma_n (\bm{x},\mathbf{t_k})\,.
\end{split}
\end{equation}

In other words, by taking the same bases for all the components, the RBF regression writes the vector field $\bm{u}=(u,v,w)^T$ as a linear combination of vector fields $\mathbf{w}_n=(\bm{w}_{u,n},\bm{w}_{v,n},\bm{w}_{w,n})^T$. Equation \eqref{RBF_VEC} still allows the set of bases to change from snapshot to snapshot (hence the time dependence in $\gamma_n$). 

As long as no constraints are used that involve the interaction of the components (e.g., divergence-free constraint), the regression of the weights for each component can be carried out independently, ie, setting the problem in \eqref{Sol} with $\mathbf{f}$ corresponding to the data collected for the three components of the velocity fields. Noticing that only the right-hand side is changing in the three associated systems, it is possible to use a single Cholesky factorization. Finally, note that the RBF regression \eqref{RBF_VEC} produces vector fields which are continuous in space but -- at least in the formulation in these notes-- are discrete in time.

\begin{tcolorbox}[breakable, opacityframe=.1, title=Exercise 4: Unconstrained Regression of a PTV field]

Consider the synthetic PTV field generated from the test case in Section \ref{sec_2_2}. Perform the regression of a single snapshot using the SPICY package \citep{Sperotto2024}. Use semi-random Halton points as collocation and plot the resulting circles and resulting velocity field. Then, using parallel computing, perform a regression of the first 1000 snapshots.

\textbf{Solution}. We start by loading the data and creating the SPICY object. Since the regression is unconstrained, we use a 'scalar' model which reuses the matrix factorizations for $U$ and $V$. We also save the collocation points and use them in every snapshot.
    
\begin{centering}
\begin{lstlisting}[language=Python,linewidth=16cm,xleftmargin=.05\textwidth,xrightmargin=.05\textwidth,backgroundcolor=\color{yellow!10}]
# Data loading
X_p, Y_p, U_p, V_p = np.genfromtxt(Name).T

# We initialize the Spicy object
SP = Spicy(points=[X_p, Y_p], data=[U_p, V_p],
    model='scalar', verbose=0)
# Perform the random collocation. Save these collocation points
SP.collocation(n_K=[5], method='semirandom',
    r_mM=[0.001, 50], eps_l=0.8)
collocation_path = Fol_Rbf + os.sep + 'RBFs.dat'  # Store the the RBF info
np.savetxt(collocation_path,
    np.column_stack((SP.X_C, SP.Y_C, SP.c_k)),
    delimiter='\t',fmt='%.6g')

# Assemble and solve the linear system
SP.Assembly_Regression()
SP.Solve(K_cond=1e11)

# Get the solution on a new (arbitrary) grid
n_x = 100; n_y = 50
x_reg = np.linspace(np.min(X_p), np.max(X_p), n_x)
y_reg = np.linspace(np.min(Y_p), np.max(Y_p), n_y)
X_reg, Y_reg = np.meshgrid(x_reg, y_reg)
    
U_reg, V_reg = SP.get_sol(points=[X_reg.ravel(), Y_reg.ravel()], 
             shape=X_reg.shape)
    
\end{lstlisting}
\end{centering}

The resulting collocation points are shown in Fig. \ref{collocation}. Notice that the clustering is very dense with quite large bases. The gradients in the present case are relatively tame, so it is easy to get away with that.

    \begin{center}%
        \captionsetup[sub]{labelformat=parens}
        \captionsetup{type=figure}
        \includegraphics[width=0.6\linewidth]{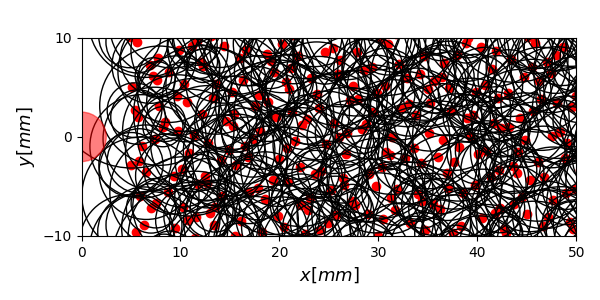}
        \vspace{-3mm}
        \caption{Example of random RBF basis collocation with shape factor controlled by the RBF value on the neighbour basis.}
        \vspace{-0.05cm}
        \label{collocation}
    \end{center}

This is already it! That is all that is needed to do a regression of data with SPICY. A few functions calls and codes result in the comparison in Fig. \ref{PTV_vs_REG} which shows side by side the tracking data used for the regression (on the left) and the RBF field on a very fine grid.

    \begin{center}%
        \captionsetup[sub]{labelformat=parens}
        \captionsetup{type=figure}
        \includegraphics[width=1\linewidth]{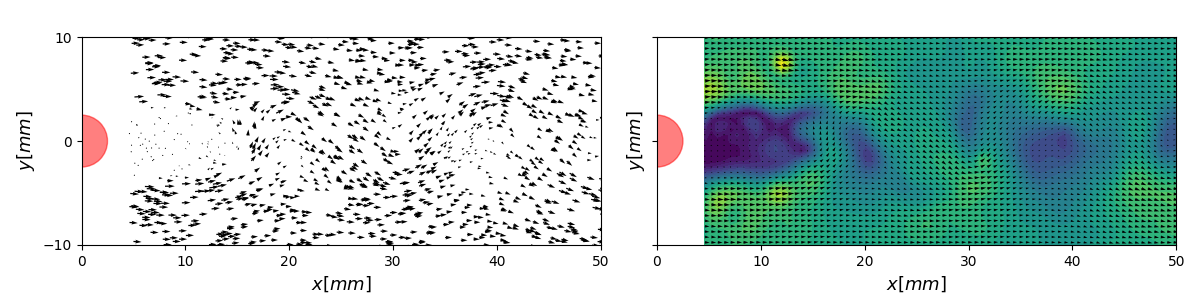}
        \caption{PTV snapshot (left) and result from the RBF regression evaluated on a fine grid (right).}
        \label{PTV_vs_REG}
    \end{center}

All the codes producing these images are in Exercise\_4.py. The code first performs one regression to generate the individual figures and then uses parallel computing to carry out the regression of 1000 snapshots, which will then be used for the exercise on modal analysis.

\end{tcolorbox}


\vspace{-8mm}
\subsection{Meshless and binless statistics of scattered data via RBF}\label{sec_6_3}

The concept of regression of instantaneous velocity fields can be extended to the regression of fields of statistical quantities, as proposed by \cite{Ratz2024}. We outline the general idea here and refer the reader to the article for a detailed derivation and example. The main idea is to introduce the RBF regression \eqref{RBF_VEC} into the ensemble expectation operator and leverage the linearity of the regression with respect to the weights.
For the mean flow, for example, \eqref{eq:mean_definition} becomes:

\begin{align}
\langle \bm{u}\rangle(\bm{x})=\int^{\infty}_{-\infty} \bm{u}(\bm{x}) f_u(\bm{x},\bm{u}) \text{d}\bm{u} = \bm{\Gamma}(\bm{x}) \int^{\infty}_{-\infty} \mathbf{w} f_w(\mathbf{w}) \text{d}\mathbf{w} = \bm{\Gamma}(\bm{x}) \langle \bm{w} \rangle\,,
\end{align} assuming that the position of the bases is maintained over all the snapshots, and having introduced the Jacobian $\text{d}\bm{u} / \text{d}\mathbf{w} = \bm{\Gamma}(\mathbf{x})$ and the probability density function $f_w(\mathbf{w}) = f_u(\mathbf{x}, \mathbf{u}) \bm{\Gamma}(\bm{x})$. The advantage is that the expectation of the weights can be estimated more easily than the one of the mean velocity field because the distribution $f_w(\mathbf{w})$ does not depend on the spatial position of the data if the velocity fields are sufficiently dense. 

A first estimate of the weight vector average $\langle \mathbf{w} \rangle$ could be obtained by averaging the weights obtained via regression of all fields:

\begin{equation}
\label{weight_averages}
    \langle \mathbf{w} \rangle \approx \mathbf{W}_A = \frac{1}{n_t} \sum_{k=1}^{n_t} \mathbf{w}_k
\end{equation}

That is, every snapshot is regressed with the same set of basis and the resulting weights are averaged. However, this requires every snapshot to be sufficiently sampled such that the regression is successful. In practice this is rarely the case because of seeding inhomogeneities. 

However, introducing \eqref{Sol} into \eqref{weight_averages} leads to interesting avenues for simplifications:

\begin{equation}
    \label{weight_mean}
    \langle \mathbf{w} \rangle \approx \mathbf{w}= \frac{1}{n_t} \sum_{k=1}^{n_t} \bigl(\bm{\Gamma}^T(\mathbf{X}^{(k)})\bm{\Gamma}(\mathbf{X}^{(k)})+\alpha \bm{I}\bigr)^{-1}\bm{\Gamma}^T(\mathbf{X}^{(k)}) \mathbf{U}^{(k)}\,,
\end{equation}where $\bm{\Gamma}(\mathbf{X}^{(k)}$ and $\mathbf{U}^{(k)}$ are the RBF matrix and velocity data in the k-th snapshot. Some of the terms in this summation could be replaced by operations on the ensemble of data points, defined as the union of all the data collected during an acquisition: 

\begin{equation}
    \mathbf{x}_E = \bigcup_{k \in 1 \dots n_t} \mathbf{X}^{(k)} \qquad \text{and} \qquad \mathbf{U}_E = \bigcup_{k \in 1 \dots n_t} \bm{u}_{k}\,.
\end{equation}

It is interesting to note that the following relations hold for the covariances and projections:

\begin{equation}
    \sum_{k=1}^{n_t} \bm{\Gamma}_k^T \bm{\Gamma}_k = \bm{\Gamma}_E^T \bm{\Gamma}_E \qquad \text{and} \qquad \sum_{k=1}^{n_t} \bm{\Gamma}_k^T \bm{U}_k = \bm{\Gamma}_E^T \bm{U}_E\,,
\end{equation} with $\bm{\Gamma}_E = \bm{\Gamma}(\mathbf{X}_E)$. These could be used in \eqref{weight_mean} to approximate the average of $n_t$ regression with one single regression on the entire ensemble of points.

I the snapshots have sufficient seeding, one could further expect the terms $\Gamma_k^T\Gamma$ to converge towards a common covariance matrix, and thus $\bm{\Gamma}_k^T \bm{\Gamma}_k = 1/n_t\, \bm{\Gamma}_E^T \bm{\Gamma}_E$. Inserting this into equation \ref{weight_mean} gives another approximation to the ensemble weight:
\begin{equation}
\label{ensamble_W}
    \langle \mathbf{w} \rangle \approx \mathbf{w}_E = (\bm{\Gamma}_E^T \bm{\Gamma}_E + \alpha \bm{I})^{-1} \bm{\Gamma}_E^T \bm{U}_E\,.
\end{equation} 

This turns the ensemble of regressions into a regression of the ensemble. While the cost in assembling the linear system is much larger, we only need to solve it once instead of $n_t$ times if each snapshot is regressed separately. Moreover, this reduces demands for seeding in the instantaneous fields. Of course, \eqref{ensamble_W} and \eqref{weight_mean} could be combined in an aggregative approach: for example computing the averages from \eqref{ensamble_W} for a set of $n_E$ ensemble and then averaging the results with \eqref{weight_mean}. 

The approach also extends effortlessly to higher order statistics since the analytical expression for the velocity field can be used to subtract the computed mean in every point of the ensemble. This field of fluctuations can then be used to compute the (scattered) expression of the field $\bm{u}^\prime_i \bm{u}^\prime_j$ which can be regressed to obtain the Reynolds stresses. Again, the reader is referred to \cite{Ratz2024} for a more detailed derivation.

\begin{tcolorbox}[breakable, opacityframe=.1, title=Exercise 5: Statistics of a Turbulent Flow from Scattered data]

Consider the PIV data from test case \ref{sec_2_2}. We are interested in computing the (1) the mean flow, (2) the turbulent kinetic energy, (3) the Reynolds stresses and the norm of the anisotropic tensor (4) locate the turbulent states observed in the data in the Lumley triangle. What kind of turbulence is found there? Use the same assumptions as for the statistics from PIV!

Hint: Using refinement regions greatly helps to reduce computational load in both the training data and the clustering.

\textbf{Solution}.  We start by pruning parts of the training data. While the seeding in the region of interest is relatively uniform, the particles outside of the jet contain more or less the same information since the flow is laminar in these regions. In contrast, the flow in the shear layer has the highest turbulence intensity so every particle counts. We start with \num{750000} from the entire dataset to illustrate the exercise. More particles are possible but this increases memory demands. We use the shapely object to define multiple refinement regions, one set of regions to prune particles and one to refine the collocation in specific areas.
    
\begin{centering}
\begin{lstlisting}[language=Python,linewidth=16cm,xleftmargin=.05\textwidth,xrightmargin=.05\textwidth,backgroundcolor=\color{yellow!10}]
refinement_1 = np.array([
        [x_min, x_max, x_max, x_min],
        [105, 115, 23, 35],
    ])
# We use shapely to define the areas
polygon_points_1 = geometry.Polygon(refinement_1.T)  
# Extract the sets of points outside the jet
X_out_glo = X[~in_polygon_1]
Y_out_glo = Y[~in_polygon_1]
U_out_glo = U[~in_polygon_1]
V_out_glo = V[~in_polygon_1]
\end{lstlisting}
\end{centering}
    
Figure \ref{refinement_points} shows the three refinement regions. Blue for outside the jet, orange for the shear layer and green for the core of the jet. We then continue by removing a percentage of particles in these areas. This is simply done with a random sample algorithm. 

    \begin{center}%
        \captionsetup[sub]{labelformat=parens}
        \captionsetup{type=figure}
        \includegraphics[width=0.3\linewidth]{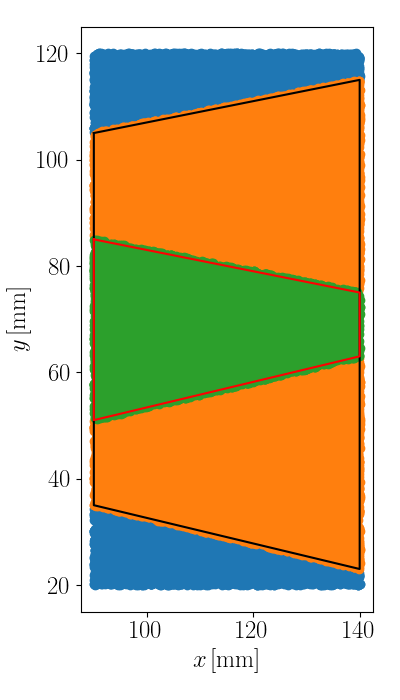}
        \caption{Refinement regions used for particles pruning}
        \label{refinement_points}
    \end{center}

\begin{centering}
\begin{lstlisting}[language=Python,linewidth=16cm,xleftmargin=.05\textwidth,xrightmargin=.05\textwidth,backgroundcolor=\color{yellow!10}]
# Here, we do the pruning in each region. These are the fractions
# of particles which are kept in each area
fraction_out = 0.6
fraction_shear = 1.0
fraction_core = 0.4

idcs_out = np.arange(X_out_glo.shape[0])
np.random.shuffle(idcs_out)
idcs_out = idcs_out[:int(X_out_glo.shape[0] * fraction_out)]
X_out = X_out_glo[idcs_out]; Y_out = Y_out_glo[idcs_out]
U_out = U_out_glo[idcs_out]; V_out = V_out_glo[idcs_out]
\end{lstlisting}
\end{centering}

We then define refinement regions in which we want to cluster more finely. These should be similar to our pruning areas. The result is shown in Fig. \ref{refinement_basis}. Note that the SPICY code plots the basis in these refinement areas if you want to take a look at it. Note that we had to give these refinement areas in normalized coordinates. Depending on your domain, rescaling the longest axis between 0 and 1 can be beneficial.

    \begin{center}%
        \captionsetup[sub]{labelformat=parens}
        \captionsetup{type=figure}
        \includegraphics[width=0.3\linewidth]{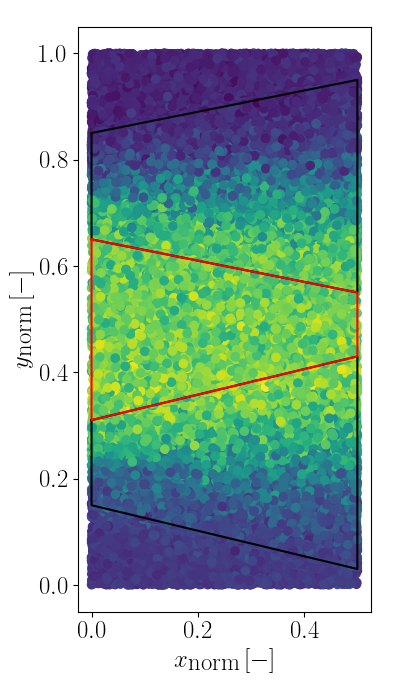}
        \caption{Refinement regions used for the clustering approach.}
        \label{refinement_basis}
    \end{center}

We continue with the use of the SPICY class \citep{Sperotto2024} to compute the regression. You are already experts from exercise 3, so we just highlight the difference of the rescaled domain and refinement.

\begin{centering}
\begin{lstlisting}[language=Python,linewidth=16cm,xleftmargin=.05\textwidth,xrightmargin=.05\textwidth,backgroundcolor=\color{yellow!10}]
# Average number of particles per basis in each refinement level
refines = [150, 150, 450, 1500]
eps_l = 0.9
SP = Spicy(
    points=[
        (X_train - x_min) / scaling,
        (Y_train - y_min) / scaling
        ],
    data=[U_train, V_train],
    basis='gauss',
    model='scalar'
    )
SP.collocation(
    n_K=refines,
    Areas=[poly_refinement_1, poly_refinement_2,
        poly_refinement_3, None],
    r_mM=[0.05, 0.8],
    eps_l=eps_l
    )
# Visualize the refined bases
SP.plot_RBFs(level=0)  # acts on first shear layer
SP.plot_RBFs(level=1)  # acts on second shear layer
SP.plot_RBFs(level=2)  # acts on core
SP.plot_RBFs(level=3)  # acts everywhere

\end{lstlisting}
\end{centering}

The resulting mean flow field is shown in Fig. \ref{ptv_mean}. The results are similar to those obtained with the gridded case in the previous exercise. However, the outcome of the RBF is a continuous function, which could be evaluated at any arbitrary grid (thus achieving super-resolution) and could provide analytic derivatives at any arbitrary point.

    \begin{center}%
        \captionsetup[sub]{labelformat=parens}
        \captionsetup{type=figure}
        \includegraphics[width=0.6\linewidth]{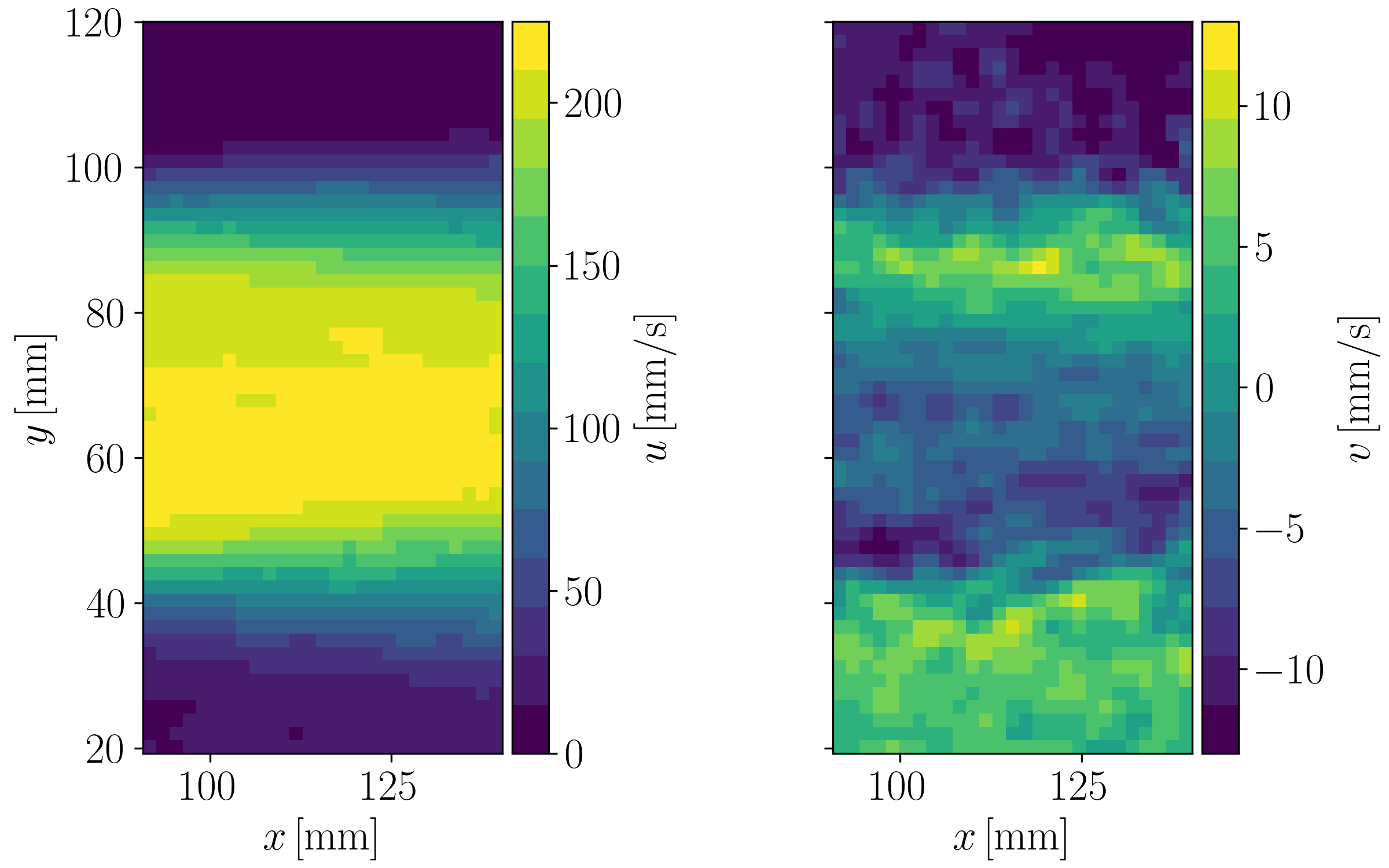}
        \caption{Resulting mean field for $U$ (left) and $V$ (right) for the meshless regression of the mean flow.}
        \label{ptv_mean}
    \end{center}

We continue by subtracting the (analytical) mean in every point and computing the products of scattered correlations which go into the second regression.

\begin{centering}
\begin{lstlisting}[language=Python,linewidth=16cm,xleftmargin=.05\textwidth,xrightmargin=.05\textwidth,backgroundcolor=\color{yellow!10}]
# Compute the mean in every point and subtract it
U_mean_train, V_mean_train = SP.get_sol(
        points=[
            (X_train - x_min) / scaling,
            (Y_train - y_min) / scaling
            ]
        )
u_train_prime = U_train - U_mean_train
v_train_prime = V_train - V_mean_train
# Compute the field of correlations
uu_train = u_train_prime * u_train_prime
vv_train = v_train_prime * v_train_prime
uv_train = u_train_prime * v_train_prime

# SPICY object for statistics
SP_stat = Spicy(
    points=[
        (X_train - x_min) / scaling,
        (Y_train - y_min) / scaling
        ],
    data=[uu_train, vv_train, uv_train],
    basis='gauss',
    model='scalar'
    )
# Here, we could also reuse the collocation points since
# they are the same
SP_stat.collocation(
    n_K=refines,
    Areas=[poly_refinement_1, 
      poly_refinement_2, poly_refinement_3, None],
    r_mM=[0.05, 0.8],
    eps_l=eps_l
    )
SP_stat.Assembly_Regression()
# We borrow the cholesky factorization from the mean flow since 
the training data is the same
# Careful! SPICY rescales your data range internally, 
so we need to adapt this
SP_stat.L_A = SP.L_A
SP_stat.b_1 = SP_stat.b_1 * SP_stat.scale_U / SP.scale_U

\end{lstlisting}
\end{centering}
  
    \begin{center}%
        \captionsetup[sub]{labelformat=parens}
        \captionsetup{type=figure}
        \includegraphics[height=0.32\linewidth]{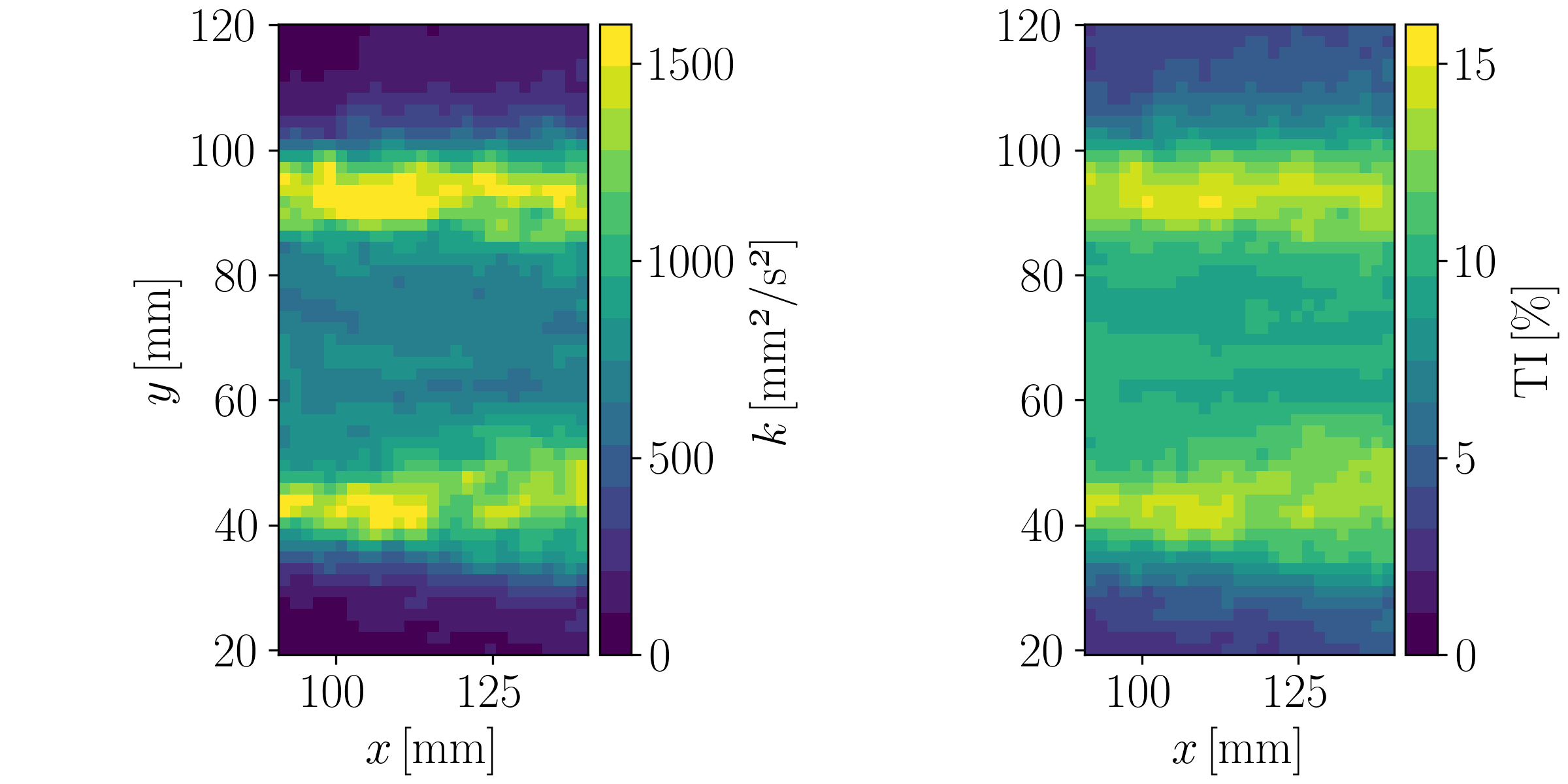}
        \includegraphics[height=0.32\linewidth]{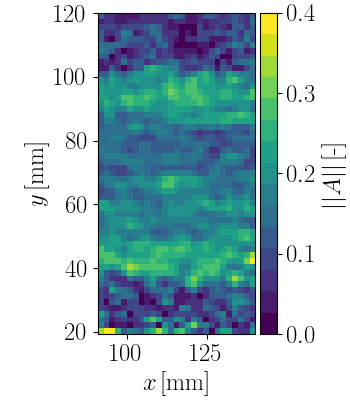}
        \caption{Average turbulent kinetic energy $k$ (left), turbulence intensity (center) and norm of the anisotropic tensor (right) from the PTV data.}
        \label{TKE_A_PTV}
    \end{center}

    \begin{center}%
        \captionsetup[sub]{labelformat=parens}
        \captionsetup{type=figure}
        \includegraphics[height=0.35\linewidth]{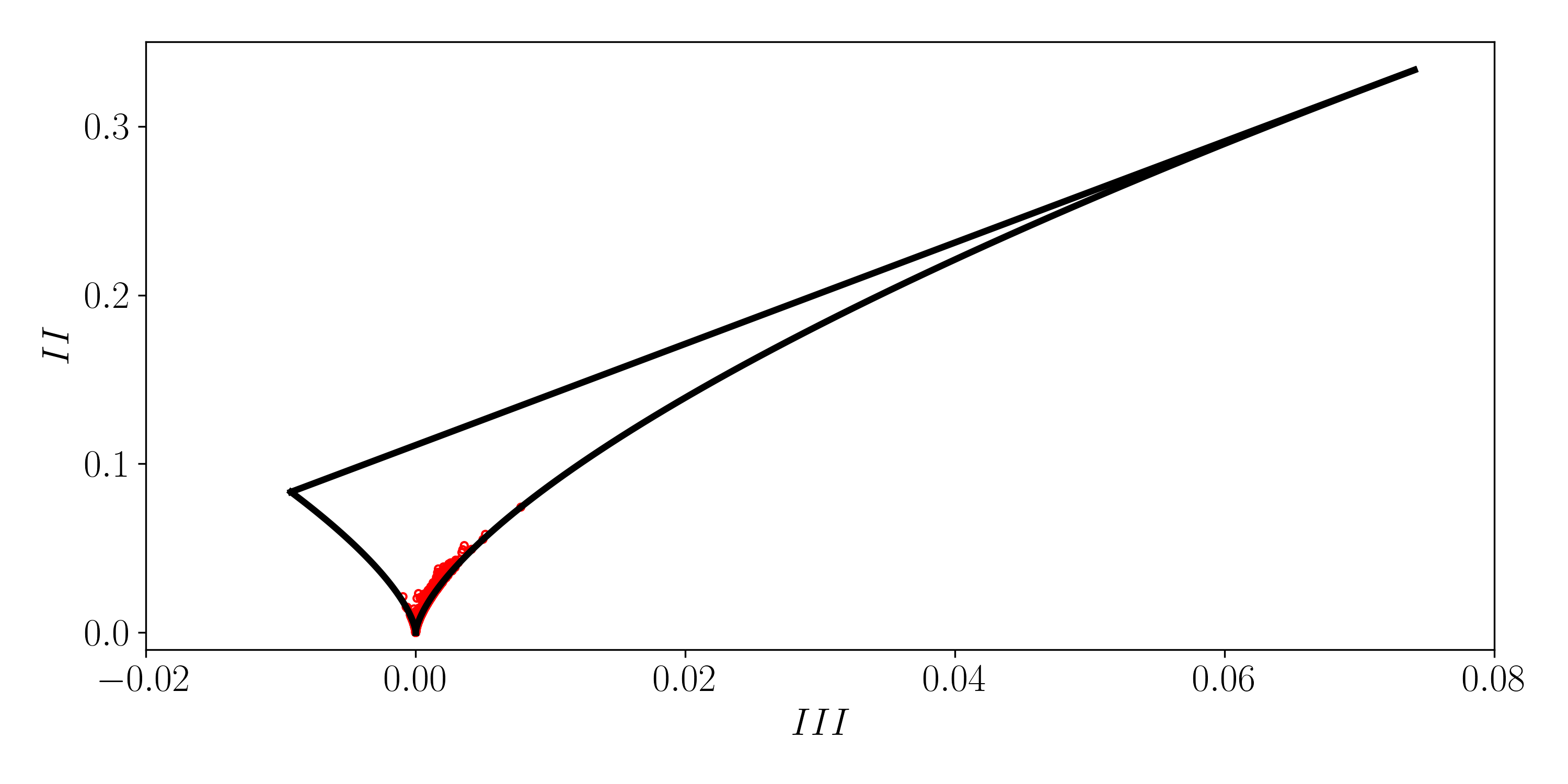}
        \caption{Lulmey triangle for the PTV data.}
        \label{Lumley_A_PTV}
    \end{center}

The rest of the computation is the same as for the PIV since we reuse the same points for plotting. The results in terms of turbulence properties agree with those of the previous exercise. However, the RBF now provides analytical functions for statistics. Moreover, we stress that this regression was carried out without physical constraints, which can significantly accelerate statistical convergence.

\end{tcolorbox}

\section{Global statistics and modal decompositions}\label{sec_7}

Data-driven modal analysis is a subset of machine learning and signal processing focusing on decomposing data as a linear combination of simpler components called modes. A broad overview of different methods is provided in \cite{Taira2017,Mendez2023}, while \cite{Mendez2024} connects these to non-linear approaches. The reader is referred to these references, along with the works cited therein, for more details. These notes focus on the implications for computations on gridded versus scattered data.

The key difference over more traditional decompositions, such as Fourier or Wavelet decompositions, is that the bases for the modes are tailored to the dataset and not defined a priori. For a velocity field, the general modal decomposition can be written as 
\begin{equation}
\label{deco}
\bm{u}(\bm{x},t)\approx\bm{u}_{n_r}(\bm{x},t)= \sum^{n_r-1}_{r=0} \sigma_r\bm{\phi}_r(\bm{x}) {\psi}_r(t)\,
\end{equation} where $\bm{\phi}_r$ and ${\psi}_r$ are the spatial and temporal structures of the $r$-th mode and $\sigma_r$ the associated amplitude. Different decompositions are obtained by setting different constraints on either $\bm{\phi}_r$ or ${\psi}_r$. In the following, we focus on the Proper Orthogonal Decomposition (POD), the fundamental decompositions derived from all other decompositions. We first start reviewing the POD for continuous data in section \ref{sec_7_1} and its implementation for gridded (section \ref{sec_7_2}) and scattered (section \ref{sec_7_3}) data using the RBF-based meshless approach. Finally, section \ref{sec_7_4} closes with some ideas to generalize the meshless approach to other decompositions.

\subsection{The (continuous) Proper Orthogonal Decomposition (POD)}\label{sec_7_1}

Let us consider the case in which both the spatial domain $\bm{x}\subseteq \Omega$ and the time domain $t\in [0,T]$ are continuous variables and hence both the "data" $\bm{u}(\bm{x},t)$ and the structures $\bm{\phi}_r(\bm{x})$ and $\bm{\psi}_r(t)$ in \eqref{deco} are continuous functions, with $\bm{\phi}_r(\bm{x})$ being flow fields. Denoting as ${\bm{u}}_{n_r}(\bm{x},t)$ an approximation of the data using $n_r$ modes in \eqref{deco}, the POD is the decompositions defined to minimize the $l_2$ error:

\begin{equation}
\label{error}
\mathcal{E}(n_r)=\frac{1}{T |\bm{\Omega}|} \int_T \int_{\bm{\Omega}} \bigl(\bm{u}(\bm{x},t)-{\bm{u}}_{n_r}(\bm{x},t)\bigr)^2dt d\bm{\Omega}\,
\end{equation}

The solution to this problem leads to the definition of spatial and temporal structures as eigenfunctions of the covariance functions \eqref{cov_C_v_function} and \eqref{cov_K_function}, respectively. These are solutions of the two Fredholm integral eigenvalue problems:

\begin{equation}
\label{Eigs_Cont}
\sigma^2_r \bm{\phi}(\bm{x})=\frac{1}{|\bm{\Omega}|}\int_{\bm{x}'\subseteq \Omega} \bm{C}(\bm{x},\bm{x}') \bm{\phi}(\bm{x}') d\bm{\Omega} \quad \mbox{and} \quad \sigma^2_r \psi(t)=\frac{1}{T}\int^{T}_{0} K(t,t') {\psi}(t') dt'\,.
\end{equation}

It is interesting to note that the integrals in \eqref{error} and in \eqref{Eigs_Cont} are based on the notion of continuous inner products. These are the inner products in space and in time, defined between functions as 

\begin{equation}
\label{Inner_PRODUCT}
\langle a(\bm{x}) b(\bm{x}) \rangle_\Omega=\frac{1}{|\bm{\Omega}|}\int_{\bm{x}'\subseteq \Omega} a(\bm{x}') b(\bm{x}') d\bm{x}'\quad \mbox{and}\quad \langle a(t) b(t) \rangle_T=\frac{1}{T} \int^T_{0} a(t') b(t') dt' \,.
\end{equation}

Note that these were also used in the definition of functions \eqref{cov_C_v_function} and \eqref{cov_K_function}.  The main implication in deriving algorithms for computing POD on gridded or scattered data is in how the integrals in the inner products and the definitions of the covariance functions are approximated.

\subsection{The POD of gridded data}\label{sec_7_2}

Consider the case of grid data as introduced in section \ref{sec_6_1}. The simplest and most popular approach to approximate the inner product in space in \eqref{Inner_PRODUCT} is the midpoint rule, hence:

\begin{equation}
\label{inner_Space}
\langle a(\bm{x}) b(\bm{x}) \rangle_\Omega= \frac{1}{|\bm{\Omega}|}\int_{\bm{x}'\subseteq \Omega} a(\bm{x}) b(\bm{x}) d\bm{x}'\approx  \sum^{n_s-1}_{j=0} a(\bm{x}_j)b(\bm{x}_j) \frac{d\Omega_j}{|\bm{\Omega}|}=\mathbf{b}^T \bm{W}_{s,w}\mathbf{a}\,, 
\end{equation} having introduced a weight matrix similar to equation \eqref{Cov_C} and the sample vectors $\mathbf{a},\mathbf{b}\in\mathbb{R}^{n_t}$ collecting samples of the functions $a(\bm{x}), b(\bm{x})$ on $n_s$ points.

In this setting, the space eigenvalue problem in \eqref{Inner_PRODUCT} becomes a matrix eigenvalue problem:

\begin{equation}
\label{eigen_C}
(\mathbf{C} \mathbf{W}_{s,w}) \bm{\phi}_r = \sigma^2_r \bm{\phi}_r,
\end{equation} where $\bm{\phi}_r\in\mathbb{R}^{n_s}$ is the sampled r-th eigenfunctions in space. The reader is referred to \cite{Sun2015,Kumar2012} for more quadrature approaches to approximate the integral. Similarly, the approximation for the inner product in time in \eqref{Inner_PRODUCT} becomes

\begin{equation}
\label{inner_time}
\langle a(t) b(t) \rangle_T=\frac{1}{T} \int^T_{0} a(t) b(t) dt'\approx \sum^{n_t-1}_{j=0} a(t_j)b(t_j) \frac{\Delta t_j}{T}=\mathbf{b}^T \bm{W}_{t,w}\mathbf{a}\,, 
\end{equation} and the time eigenvalue problem in \eqref{Inner_PRODUCT} becomes

\begin{equation}
\label{eigen_K}
(\mathbf{K} \mathbf{W}_{t,w}) \bm{\psi}_r = \sigma^2_r \bm{\psi}_r\,,
\end{equation} with $\psi_r\in\mathbb{R}^{n_t}$ is the r-th sampled eigenfunction in time.

By definition, the eigenvectors obtained in \eqref{eigen_C} and \eqref{eigen_K} are orthonormal according to the weighted inner product in \eqref{inner_Space} and \eqref{inner_time}. The main implication is that it is easy to compute $\phi_r's$ from $\psi_r's$ and vice versa. The discrete version of \eqref{deco} now becomes a matrix factorization:

\begin{equation}
\label{discrete_deco}
\bm{u}(\bm{x}_i,\mathbf{t_k})=\sum^{n_r-1}_{r=0} \sigma_r \bm{\phi}_r(\bm{x}_i)\psi^T_r(\mathbf{t_k})\,,
\end{equation} having reshaped each snapshot of the velocity field and each spatial structure of the modes into a column vector. The (weighted) time inner product of the expansion \eqref{discrete_deco} by $\psi_r$ reads:

\begin{equation}
\label{discrete_deco_PRO}
\langle\bm{u}(\bm{x}_i,\mathbf{t_k}),\psi_r(\mathbf{t_k})\rangle_{T}=\Bigg\langle \sum^{n_r-1}_{r=0} \sigma_r \bm{\phi}_r(\bm{x}_i)\psi^T_r(\mathbf{t_k}),\psi_r(\mathbf{t_k})\Bigg\rangle_{T}=\sigma_r \bm{\phi}_r(\bm{x_i})\,,
\end{equation} so the both the inner decomposition \eqref{discrete_deco} can be completed. Note that, in the case $\bm{W}{s,w}$ and $\bm{W}_{t,w}$ become identity matrices (i.e., the data is uniformly sampled in space and time), the POD can be computed as the Singular Value Decomposition of the snapshot matrix (see \cite{Dawson2023} for more details). The most popular POD algorithm, due to \cite{Sirovich1987}, computes the POD by first assembling and solving the eigenvalue problem in \eqref{eigen_K}, and then computing the spatial structures via the projection in \eqref{discrete_deco_PRO}.

\subsection{The meshless POD of scattered data}\label{sec_7_3}

Consider now the case where all the $n_t$ datasets have been written as linear combinations of RBFs as \eqref{RBF_VEC}. We thus have a finite set of continuous velocity fields in $\bm{x}\in\Omega\subset\mathbb{R}^d$. The inner products in time are discrete, while those in space are continuous. Therefore, following the traditional snapshot POD approach, the eigenvalue in time remains discrete as in \eqref{eigen_K}. Still, both the definition of the temporal correlation matrix and the projection to identify the spatial structures must change. Building adaptations of these takes us to meshless POD, introduced recently by \cite{Tirelli2024}. A similar idea is known in functional analysis as Functional Principal Component Analysis
\citep[FPCA,][]{ramsay2005principal,
Ramsay1997, Wang2016, Hall2006}. FPCA generalizes the traditional PCA to functional data, usually taking the form of continuous functions over time. FPCA uses discrete inner products to build a continuous eigenvalue problem, while the proposed approach uses a continuous inner product to build a discrete eigenvalue problem. A brief discussion on the difference between the two is provided in \cite{Tirelli2024}.

The idea is to use the RBF expansion within the definition of the temporal correlation matrix. This provides an analytical expression for the integrand in \eqref{cov_K_function}, which can be evaluated at any arbitrary point, enabling location-based quadrature methods, such as the Gauss-Legendre quadrature. However, in these notes, we take an alternative approach: we derive an explicit expression for the temporal correlation matrix by substituting the RBF expansion \eqref{RBF_VEC} into the definition of the temporal covariance function \eqref{cov_K_function} and formulating the result in terms of the RBF weights. Using the index $l\in\{u,v,w\}$ to span the velocity components and the RBF weight vectors for each velocity component, the result reads:

\begin{equation}
\begin{split}
\mathbf{K}_{i,j}&=\frac{1}{|\bm{\Omega}|}\int_\Omega \bm{u}'(\bm{x},t_i)^T\bm{u}'(\bm{x},t_j) d\bm{\Omega}=\frac{1}{|\bm{\Omega}|}\int_\Omega \sum^{3}_{l=1}\bm{u}_l'(\bm{x},t_i)\bm{u}_l'(\bm{x},t_j) d\bm{\Omega}\\
&=\frac{1}{|\bm{\Omega}|}\int_\Omega \sum^{3}_{l=1}\biggl(\sum_{n=1}^{n_b}\mathbf{w}_{l,n}(\mathbf{t_i})\gamma_n (\bm{x},\mathbf{t_i})\biggr)\biggl(\sum_{m=1}^{n_b}\mathbf{w}_{l,m}(\mathbf{t_j})\gamma_m (\bm{x},\mathbf{t_j})\biggr) d\bm{\Omega}\\
&=\sum^{3}_{l=1}\sum_{n=1}^{n_b}\sum_{m=1}^{n_b}\mathbf{w}_{l,n}(\mathbf{t_i})\mathbf{w}_{l,m}(\mathbf{t_j})\biggl(\frac{1}{|\bm{\Omega}|}\int_\Omega \gamma_n (\bm{x},\mathbf{t_j})\gamma_m (\bm{x},\mathbf{t_j})d\bm{\Omega}\biggr) \\
&=\mathbf{w}^T_{u}(\mathbf{t}_i) \mathbf{I}(\mathbf{t}_i,\mathbf{t}_j)\mathbf{w}_{u}(\mathbf{t}_j)+\mathbf{w}^T_{v}(\mathbf{t}_i) \mathbf{I}(\mathbf{t}_i,\mathbf{t}_j)\mathbf{w}_{v}(\mathbf{t}_j)+\mathbf{w}^T_{w}(\mathbf{t}_i) \mathbf{I}(\mathbf{t}_i,\mathbf{t}_j)\mathbf{w}_{w}(\mathbf{t}_j)\,,
\end{split}
\label{eq:meshless_K}
\end{equation} having introduced the matrix:

\begin{equation}
\label{eq:integral_pod}
\mathbf{I}_{m,n}(\mathbf{t}_i,\mathbf{t}_j)=\frac{1}{|\bm{\Omega}|}\int_\Omega \gamma_n (\bm{x},\mathbf{t_i})\gamma_m (\bm{x},\mathbf{t_j})d\bm{\Omega}\,\in\mathbb{R}^{n_b\times n_b}.
\end{equation}

If the basis functions differ between snapshots, this matrix will depend on each specific pair of snapshots $(i, j)$. However, if a common basis is used across all snapshots, only a single matrix, which can be precomputed, is needed. This significantly reduces the computational cost of evaluating \eqref{eq:meshless_K}.

The proposed approach is numerically less accurate than the quadrature-based approach described in \cite{Tirelli2024}. However, it is computationally cheaper and better adapts to complex geometries.

Finally, the calculation of the spatial structures following \eqref{discrete_deco_PRO} leads to the expansion of the RBF of the spatial structures $\phi_r(\bm{x})$:

\begin{equation}
\label{RBF_Pro}
\sigma_r \phi_r(\bm{x})=\langle \bm{u}(\bm{x},\mathbf{t}_k) ,\psi_r(\mathbf{t}_k)\rangle_T=\sum^{n_b}_{n=1}\langle\mathbf{w}_n(\mathbf{t}_k),\psi_r(\mathbf{t}_k) \rangle _T\,\gamma_n(\bm{x})\,.
\end{equation}

\begin{tcolorbox}[breakable, opacityframe=.1, title=Exercise 6: Meshless POD vs Gridded POD]

Consider the PIV data from test case \ref{sec_2_2}, resampled to produce scattered data as in PTV. As for exercise 2, consider only the portion of the data with $\mathbf{t}_k$, that is, during the first stationary condition. 
The PIV dataset is sufficiently dense to provide a good decomposition (see \cite{Mendez2020}); The scope of this exercise is to test the meshless POD and see if the obtained modes agree with the gridded ones. 

For the comparison, we first perform the traditional (grid based) POD of the PIV dataset following Section \ref{sec_7_2}. We compute modal amplitudes, spatial structures and the frequency spectra of the temporal structures for the first three modes. Then, the computation for the meshless approach will be repeated, and the results will be compared. The reader is encouraged to decrease the seeding density in the data generation to see when the decomposition fails.

\textbf{Solution}.

The POD of gridded data is relatively straightforward since it only involves matrices and matrix multiplications. To perform it, we use the open-source code \href{https://github.com/mendezVKI/MODULO/tree/master}{MODULO} \citep{Poletti2024}. The code requires the data matrix $\bm{D}$ of all the stacked snapshots as input. This can be loaded using parallel computing and then passed to MODULO as follows:

\begin{centering}
\begin{lstlisting}[language=Python,linewidth=16cm,xleftmargin=.05\textwidth,xrightmargin=.05\textwidth,backgroundcolor=\color{yellow!10}]
# Function to load the data
def load_piv_data(file_name):
    data = np.genfromtxt(file_name)[1:, :]
    return data

# Parallel processing to load the files
num_workers = 4
with ThreadPoolExecutor(max_workers=num_workers) as executor:
    results = np.array(list(executor.map(lambda file_name:\
        load_piv_data(Fol_Piv + os.sep + file_name), file_names)))

# Reshape into size (n_s, n_t)
D = np.transpose(results, axes=(0, 2, 1)).reshape(n_t, n_s).T

# Import the Modulo package and perform decomposition
from modulo_vki import ModuloVKI
modu = ModuloVKI(data=np.nan_to_num(D), n_Modes=1000)
Phi_grid, Psi_grid, Sigma_grid = modu.compute_POD_K()

\end{lstlisting}
\end{centering}

Note that MODULO computes the POD decomposition in one line (line 18). We compute 50 modes even if only 3 are asked to observe the decomposition convergence. Figure 19a) shows the convergence of the modes, that is the normalized amplitude as a function of the mode index. Figure 19b) shows the frequency content on the leading modes.

\begin{center}%
    \captionsetup[sub]{labelformat=parens}
    \captionsetup{type=figure}
    \begin{subfigure}{0.49\textwidth}
        \includegraphics[width=\linewidth]{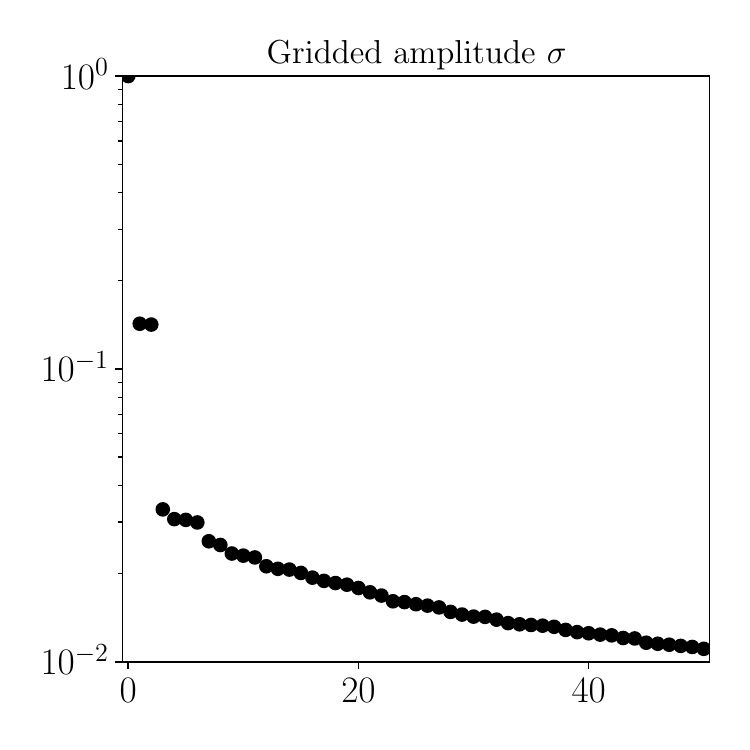}
        \caption{}
        \label{5a}
    \end{subfigure}
    \begin{subfigure}{0.49\textwidth}
        \includegraphics[width=\linewidth]{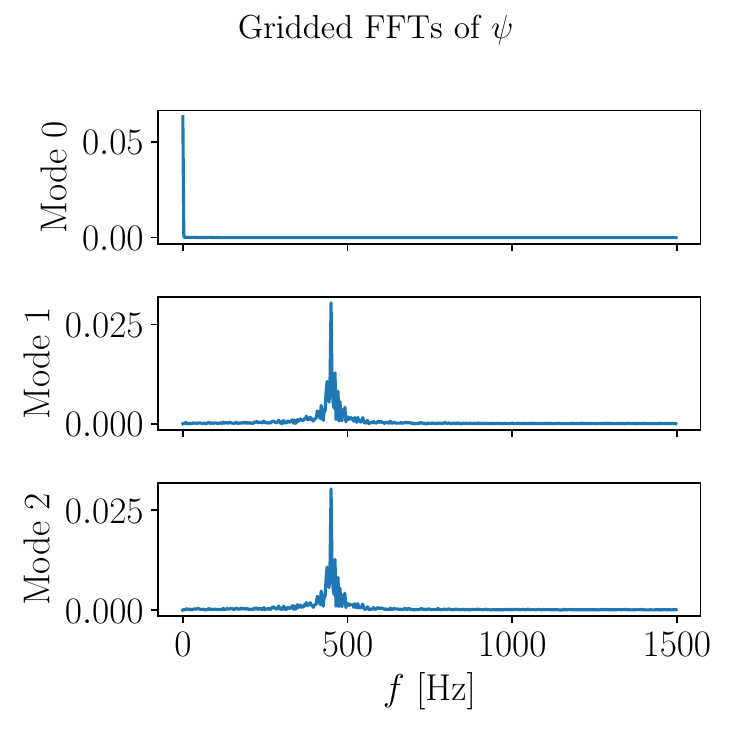}
        \caption{}
        \label{5b}
    \end{subfigure}
    \begin{subfigure}{0.7\textwidth}
        \includegraphics[width=\linewidth]{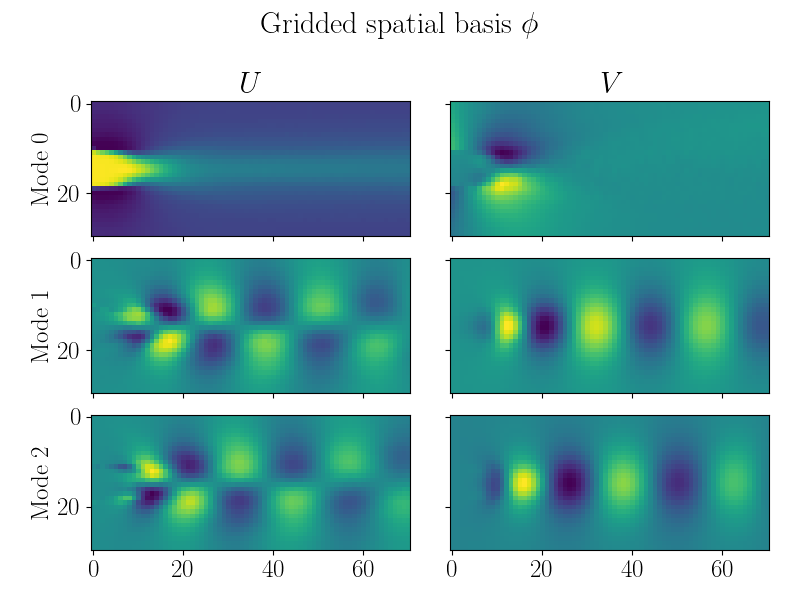}
        \caption{}
        \label{5c}
    \end{subfigure}
\vspace{-3mm}
    \caption{Results from the traditional POD of PIV data. Fig (a): Energy $\sigma$ of the first 50 POD modes; Fig (b): spectra of the temporal modes $\psi$ Fig (c): contour of the spatial structure of the first 3 POD modes (U component on the left, V component on the rigth).}
\vspace{-0.05cm}
\label{figEx5_grid}
\end{center}

Figure 19c) shows the contour plots of the velocity fields associated with the first three modes, i.e. $\bm{\phi}_1(\bm{x})$, $\bm{\phi}_2(\bm{x})$, $\bm{\phi}_3(\bm{x})$. 

\hspace{3mm}Since the data is in stationary conditions and has a solid statistical convergence, the first mode corresponds to the mean flow (note that this is generally not the case for the POD; see \cite{Mendez2019}). The second and third modes are paired to describe the traveling wave pattern in the shedding (see \cite{Mendez2020}). These modes have the same amplitude but are in quadrature in space and time (see \cite{BarreiroVillaverde2021} for methods to identify traveling patterns)
The leading frequency in both modes corresponds to the vortex shedding frequency. Combining the leading wavelength on the spatial structures with the leading frequencies in the temporal one, it is possible to estimate the advection speed of these vortices.

Moving onto the meshless POD, we first need to build the matrix $\mathbf{I}_{m,n}$ \eqref{eq:integral_pod}, which requires an integral. Having used the same RBF basis for all snapshots, this matrix must be evaluated only once. For simplicity, we sample it on a domain and use a summation to save the computational cost of a more precise integration rule; note, however, that this integral could be carried out with the Montecarlo method in case of more complex geometries and is the only step that requires considerations on the shape of the domain.

The definition of integrand, as a function of the RBF parameters, is provided by the function \textit{func} while the integration is carried out using a nested for loop. This could be significantly accelerated using parallel computing; however, given the small cost of the operation, we keep it as a simple nested loop.

\begin{centering}
\begin{lstlisting}[language=Python,linewidth=16cm,xleftmargin=.05\textwidth,xrightmargin=.05\textwidth,backgroundcolor=\color{yellow!10}]

#%% Meshless POD, computation of I matrix
# Input folder of the RBF weights
Fol_Rbf = 'RBF_DATA_CYLINDER'
weight_list = sorted([file for file in os.listdir(Fol_Rbf) 
if 'RBF' not in file])

# Function for the integrand in equation 77
def func(x, y, x_c_n, x_c_m, y_c_n, y_c_m, c_n, c_m):
    return np.exp(-c_n**2 * ((x-x_c_n)**2 + (y-y_c_n)**2)) * \
        np.exp(-c_m**2 * ((x-x_c_m)**2 + (y-y_c_m)**2))

# load the RBF data
X_C, Y_C, c_k = np.genfromtxt(Fol_Rbf + os.sep + 'RBFs.dat').T
n_b = c_k.shape[0]

# Integration domain (for classic integration)
x_integrate = np.linspace(Xg.min(), Xg.max(), 151)
y_integrate = np.linspace(Yg.min(), Yg.max(), 61)
X_integrate, Y_integrate = np.meshgrid(x_integrate, y_integrate)
X_integrate = X_integrate.ravel()
Y_integrate = Y_integrate.ravel()

# We compute a single matrix I since the RBFs do not change 
in between time steps, only their weights.
# This allows to save time because we use a single I matrix 
instead of 1000

I_meshless = np.zeros((n_b, n_b))
for m in tqdm(range(n_b), mininterval=1, desc='Filling I matrix'):
    for n in range(0, n_b):
        I_meshless[m, n] = func(X_integrate, Y_integrate, 
        X_C[n], X_C[m], Y_C[n], Y_C[m], c_k[n], c_k[m]).sum()
# Divide by the area to normalize the integral
area = (Xg.max() - Xg.min()) / (Yg.max() - Yg.min())
I_meshless = I_meshless / area

\end{lstlisting}
\end{centering}

Since $\mathbf{I}_{m,n}$ is a constant in between all snapshots, we just need to multiply it with the individual weights at $\mathbf{t}_i$ and $\mathbf{t}_j$ to get the correlation matrix $\mathbf{K}$ \eqref{eq:meshless_K}. The necessary weights of the first 1000 snapshots were already computed in exercise 3. We here use parallel computing to prepare the matrix K while taking advantage of its symmetry:

\begin{centering}
\begin{lstlisting}[language=Python,linewidth=16cm,xleftmargin=.05\textwidth,xrightmargin=.05\textwidth,backgroundcolor=\color{yellow!10}]
# Function to compute a single element of the covariance matrix
def compute_K_element(i, j, w_U_all, w_V_all, I_meshless):
    w_U_i = w_U_all[i]; w_V_i = w_V_all[i]
    w_U_j = w_U_all[j]; w_V_j = w_V_all[j]
    # Calculate the element of the covariance matrix K[i, j]
    K_value = w_U_i.T @ I_meshless @ w_U_j +
    w_V_i.T @ I_meshless @ w_V_j
    return (i, j, K_value)

# Load all weights ahead of time to avoid repeated I/O
w_U_all = []; w_V_all = []

# Data Reader
for i in tqdm(range(len(weight_list)), desc='Loading weights'):
    w_U_i, w_V_i = np.genfromtxt(Fol_Rbf + os.sep 
    + weight_list[i]).T
    w_U_all.append(w_U_i)
    w_V_all.append(w_V_i)

# Stack weights for easy indexing
w_U_all = np.stack(w_U_all)
w_V_all = np.stack(w_V_all)

# Number of snapshots
n_t = len(weight_list)

# Use joblib to parallelize the double loop calculation
results = Parallel(n_jobs=-1)(
    delayed(compute_K_element)(i, j, w_U_all, w_V_all, I_meshless)
    for i in tqdm(range(n_t), desc="Computing covariance matrix")
    for j in range(i + 1)  
    #(Only compute for j <= i since K is symmetric)
)

# Create an empty covariance matrix
K_meshless = np.zeros((n_t, n_t))

# Fill in the covariance matrix with the results from 
the parallel computation
for i, j, value in results:
    K_meshless[i, j] = value
    K_meshless[j, i] = value  # Since K is symmetric


\end{lstlisting}
\end{centering}

To minimize the I/O overhead, all weights are first loaded and appended into two lists. The function ``compute\_K\_element'' implements \eqref{eq:meshless_K}, and lines 28-32 distribute the job to various processors.
The last step builds the matrix $\mathbf{K}$.

The eigendecomposition of $\mathbf{K}$ can now be computed as in the gridded case to produce the amplitudes and temporal bases (see the provided codes). Finally, to compute the spatial basis, we compute the projections from \eqref{RBF_Pro}:S

\begin{centering}
\begin{lstlisting}[language=Python,linewidth=16cm,xleftmargin=.05\textwidth,xrightmargin=.05\textwidth,backgroundcolor=\color{yellow!10}]

# The Gamma matrix is the same at every step
Gamma = Phi_RBF_2D(Xg.ravel(), Yg.ravel(), X_C, Y_C,\
c_k, basis='gauss')
for i in range(n_modes):
    weights_U_projected = np.squeeze(
        Psi_meshless[:, i][np.newaxis, :].dot(w_U))
    weights_V_projected = np.squeeze(
        Psi_meshless[:, i][np.newaxis, :].dot(w_V))

    Phi_meshless[:nxny, i] = Gamma.dot(weights_U_projected)\
        / Sigma_meshless[i]
    Phi_meshless[nxny:, i] = Gamma.dot(weights_V_projected)\
        / Sigma_meshless[i]
    
\end{lstlisting}
\end{centering}

We conclude by plotting the same quantities as in the gridded case in Figure 20. The agreement with the grid-based POD is remarkable, although the RBF configuration used in this exercise encounters some challenges in regions with sizeable mean flow gradients. A more refined RBF collocation strategy could improve accuracy in these areas. Nevertheless, for this exercise, the comparison is enough to demonstrate the validity of the approach. It is worth emphasizing that this method can be easily applied to more complex geometries, and the resulting modes are analytic in space, offering super-resolution and enabling the analytic computation of derivatives.

\begin{center}%
    \captionsetup[sub]{labelformat=parens}
    \captionsetup{type=figure}
    \begin{subfigure}{0.49\textwidth}
        \includegraphics[width=\linewidth]{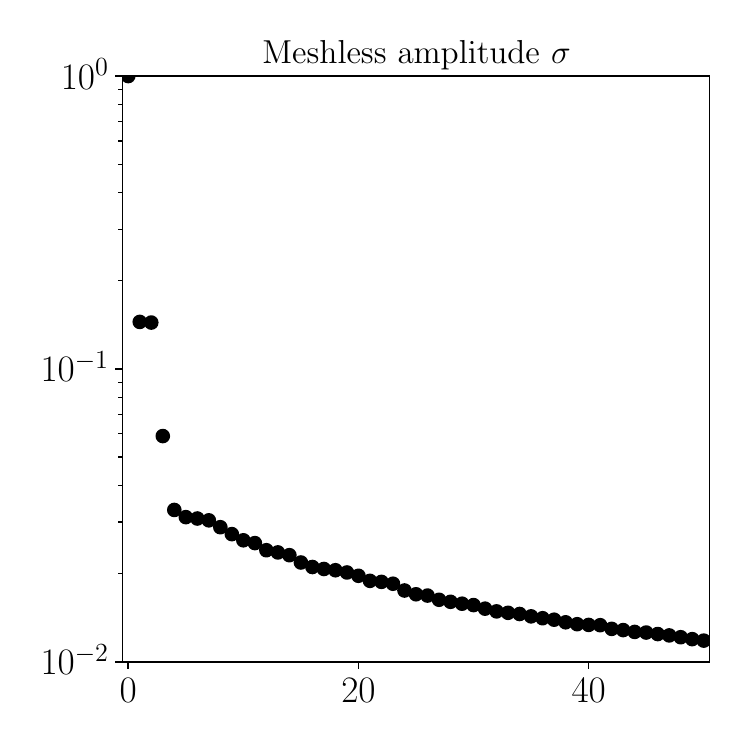}
        \caption{}
        \label{54}
    \end{subfigure}
    \begin{subfigure}{0.49\textwidth}
        \includegraphics[width=\linewidth]{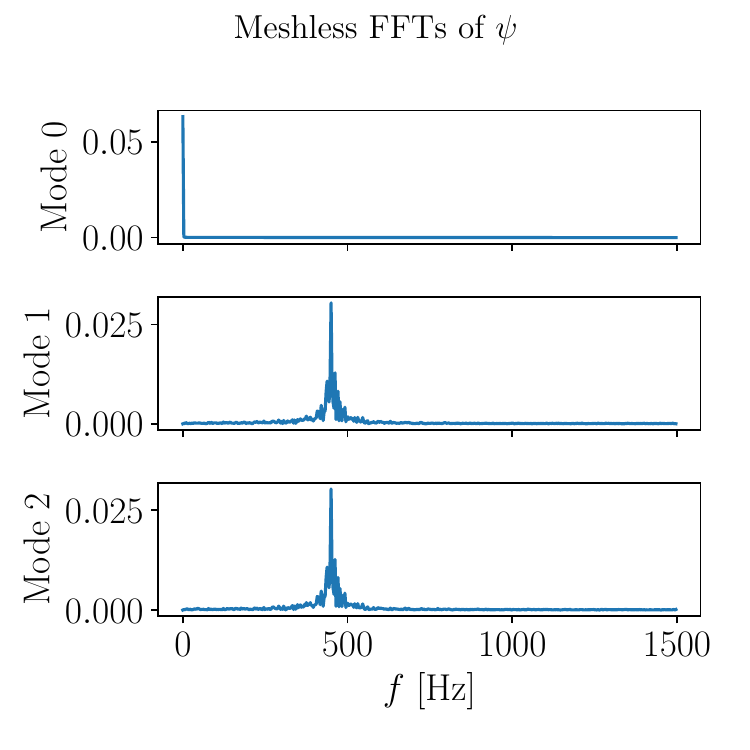}
        \caption{}
        \label{5e}
    \end{subfigure}
    \begin{subfigure}{0.7\textwidth}
        \includegraphics[width=\linewidth]{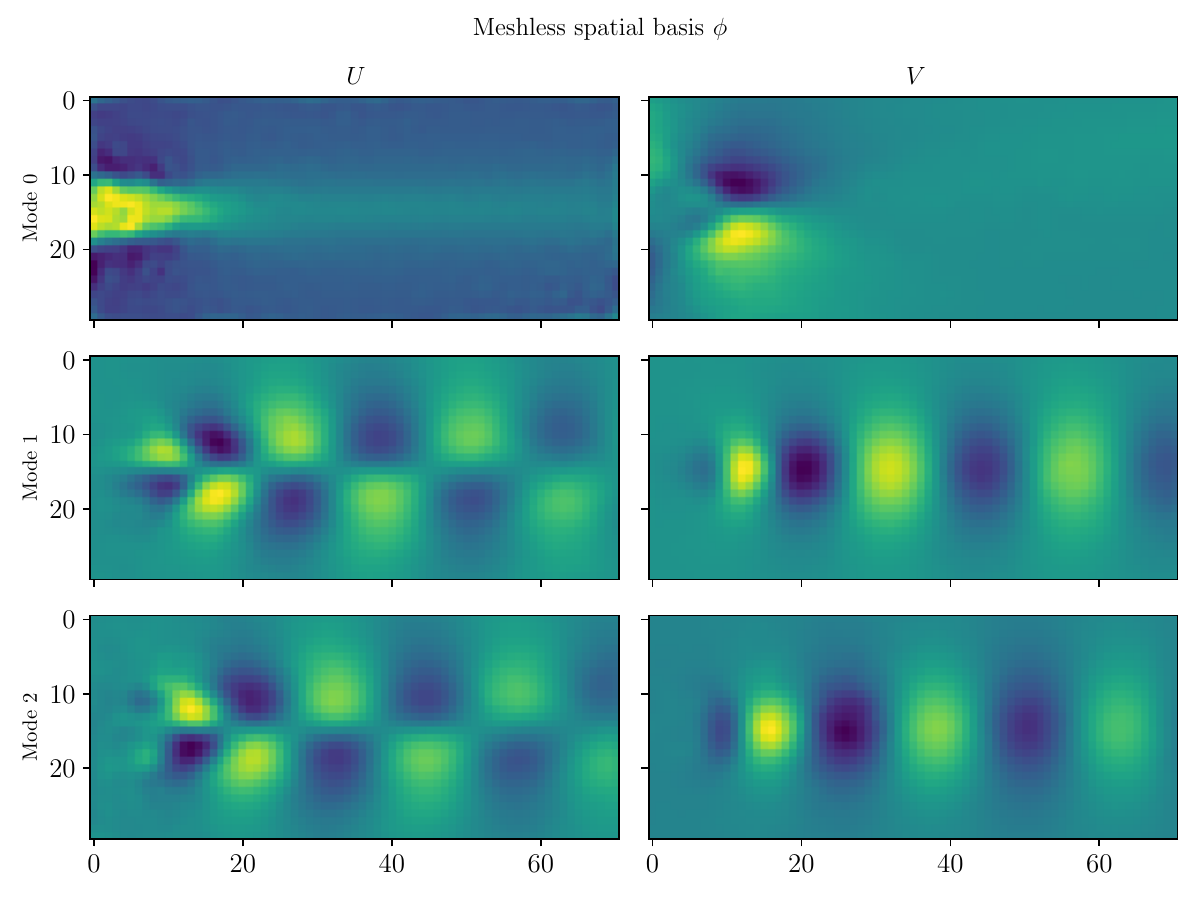}
        \caption{}
        \label{5f}
    \end{subfigure}
    \vspace{-3mm}
    \caption{Meshless POD.	Fig (a): Energy $\sigma$ of the first 50 POD modes; Fig (b): spectra of the temporal modes $\psi$}
    \vspace{-0.05cm}
    \label{figEx5_meshless}
\end{center}

\end{tcolorbox}

\subsection{Generalizations beyond the POD}\label{sec_7_4}

All linear decompositions are based on inner products in space and time \citep{Mendez2023}. These solely differ in how the temporal \textbf{or} the spatial structures are computed. Therefore, using the inner products introduced in the previous section in the traditional decompositions produces the meshless formulation for scattered data. Below, we briefly comment on how this could be done for the most common Spectral POD by \cite{sieber_paschereit_oberleithner_2016}, the Spectral POD by \cite{towne_schmidt_colonius_2018}, the Multiscale POD by \cite{Mendez2019} and the Dynamic Mode Decomposition by \cite{Schmid}. 

The reader is referred to \cite{Mendez2024} for a comprehensive introduction to all these decompositions, all available in the MODULO software package MODULO \citep{Poletti2024}.
The meshless implementation of all these decompositions is a work in progress!

\subsubsection{The meshless version of \cite{sieber_paschereit_oberleithner_2016}'s Spectral POD}\label{sec_7p5}

\cite{sieber_paschereit_oberleithner_2016}'s Spectral POD is a variant of the traditional POD that aims to improve the spectral purity of the POD for the case of stationary data. The idea is to apply a diagonal filter to the temporal covariance matrix $\mathbf{K}$ that forces the temporal structure of the POD to have a narrower spectrum. This allows for circumventing pathological conditions in datasets where the energy optimality is more a limitation than an advantage, for example, if features characterized by vastly different scales (i.e., frequency content) have comparable energy contributions to the data. In these cases, the POD modes tend to mix different features in the same modes, which limits the identification of the features.

The derivation of the meshless SPOD is straightforward because this decomposition adds only one additional step to the traditional POD: the filtering operation. This is carried out on the matrix $\mathbf{K}$, which would simply be replaced by the same meshless version used in the meshless POD. Similarly, the projection for the spatial structures remains identical to that for the meshless POD.

\subsubsection{The meshless version of \cite{towne_schmidt_colonius_2018}'s Spectral POD}\label{sec_7p6}

\cite{towne_schmidt_colonius_2018}'s Spectral POD generalizes Welch's method to modal decompositions. The idea is to first compute the modal discrete Fourier Transform (DFT) over different chunks of the data in time. Then, the average frequency content in each chunk is replaced by a POD carried out on a snapshot matrix obtained by looking at the evolution of the DFT over the chunks. The result is a large set of harmonic modes.

To obtain the meshless version of this decomposition, one must combine the meshless version of the DFT with the meshless POD. The meshless DFT is obtained by combining the RBF expansion in space with the Fourier Transform of the RBF's weights. The meshless POD can then be applied for each leading frequency to obtain the SPOD modes.

\subsubsection{The meshless version of \cite{Mendez2019}'s Multiscale POD}\label{sec_7p8}

\cite{Mendez2019}'s Multiscale POD combines the concept of Multiresolution Analysis (MRA) with POD. The mPOD basis is optimal for a given partition of the frequency domain, effectively constraining the frequency content of each mode to a specific range (scale). This frequency decomposition uses Wavelets or filter banks, ensuring the modes remain orthogonal in the time domain. Since MRA is applied to the temporal correlation matrix $\mathbf{K}$, the derivation of meshless mPOD follows the same procedure as the meshless POD, with the MRA steps remaining unchanged.

\subsubsection{The meshless version of \cite{Schmid}'s DMD}\label{sec_7p9}

\cite{Schmid}'s Dynamic Mode Decomposition consists of decomposing the data as a linear combination of complex exponentials. This implies approximating the data with a linear dynamical system and fitting its eigenvalues using a least-square approach. Since fitting the linear dynamical system directly to the dataset (that is, to all the time series at each location) is computationally prohibitive, the idea is to fit it on a subset of leading POD modes. Given the complex harmonics that best describe the evolution of the POD modes, it is possible to compute the associated spatial structures via projection in space. The meshless DMD, therefore, can be built from the meshless POD with least-square regression of complex exponentials on the temporal structures. An alternative and more expensive approach would be to fit the complex exponentials to the time evolution of the RBF coefficients. 

\section{Conclusions and Outlook}\label{sec_6}

That was a long journey. These notes began with traditional tools for computing statistics and modal decompositions of gridded data and explored novel methods for scattered data. As fluid dynamicists, we are primarily concerned with first- and second-order statistics, used to relate random variables in vastly different contexts: the components of a vector field at a given location or at various times and locations. The covariance matrices obtained in these contexts carry different physical meanings. The covariance matrix built from velocity field components at a given location corresponds to the Reynolds stress, whose eigendecomposition provides insights into the nature of turbulence. The covariance functions (or matrices) constructed from random fields in space and time reveal coherent patterns in the data, with their eigendecomposition leading to the Proper Orthogonal Decomposition.

The main challenge with scattered data from tracking velocimetry is not the lack of a grid itself but the changing sampling locations across snapshots, which complicates the evaluation of the ensemble operator. Radial Basis Functions (RBFs) provide an effective way to manage these datasets and compute the necessary integrals. They also make it possible to add physics constraints and generate analytic fields for both instantaneous snapshots and statistical fields. While the tools introduced for gridded data are well established, the meshless formalism discussed here is still a research topic, and readers are encouraged to join this research effort. The exercises and codes provided should offer a helpful starting point. Given the increasing use of tracking velocimetry in experimental fluid mechanics and the growing importance of meshless particle-based methods in computational fluid dynamics (e.g., Smoothed Particle Hydrodynamics, SPH, or Lagrangian Differencing Dynamics, LDD), meshless computations, meshless POD, and, more broadly, meshless decompositions are poised to become central themes in the near future.

\bibliographystyle{apalike}


\bibliography{chapter_x}
\end{document}